%% file: main.tex
\documentclass[fullpage,twocolumn]{article}
\pdfoutput=0

\usepackage{algorithm}
\usepackage{algpseudocode}
\usepackage{amsfonts,amssymb,amsmath,amsthm}
\usepackage{cite} 
\usepackage[breaklinks,colorlinks,linkcolor=blue,citecolor=blue,urlcolor=black]{hyperref}
\usepackage[hyphenbreaks]{breakurl}
\usepackage{paralist}
\usepackage{times}
\usepackage{xspace}
\usepackage{graphicx}
\usepackage{svg}
\usepackage{listings}

\long\def\com#1{}

\long\def\xxx#1{}

\long\def\arxiv#1#2{#2}			

\newcommand{\ie}{{\em i.e.}\xspace}
\newcommand{\eg}{{\em e.g.}\xspace}

\def\tlc#1#2#3{$\mathrm{TLC}(#1,#2,#3)$}
\def\qsc{QSC\xspace}
\newcommand{\msg}{\ensuremath{m}\xspace} 
\newcommand{\tm}{\ensuremath{t_\msg}\xspace} 
\newcommand{\ack}{\ensuremath{a}\xspace} 
\newcommand{\ta}{\ensuremath{t_\ack}\xspace} 
\newcommand{\tw}{\ensuremath{t_w}\xspace} 
\newcommand{\lts}{\ensuremath{s}\xspace} 
\newcommand{\NN}{\ensuremath{\mathbb{N}}\xspace} 

\begin{document}

\title{Threshold Logical Clocks for \\
	Asynchronous Distributed Coordination and Consensus \\
	{\large\em preliminary work-in-progress; may become
		\href{https://bford.info/book/}{part of a book}}}

\author{Bryan Ford \\
	Swiss Federal Institute of Technology in Lausanne (EPFL)}

\maketitle

\input{abs}

\tableofcontents

\input{intro}

\input{bg}
\input{tlc}

\input{app}

\input{cons}
\input{byz}

\input{dkg}

\input{time}

\input{causal}

\input{arch}

\xxx{
\input{stamp}

\input{opt}

\input{proto}

}
\input{analys}
\input{eval}

\input{rel}

\input{concl}

\bibliographystyle{plain}
\arxiv{
\bibliography{os,net,sec,theory,soc}
}{
\bibliography{main}
}

\section*{Appendices}
\appendix

\arxiv{
\lstloadlanguages{Go,Promela}
}{}

\input{go}
\input{spin}

\xxx{
\input{tlc_properties}
}

\end{document}

%% file: abs.tex
\begin{abstract}

Consensus protocols for asynchronous networks
are usually complex and inefficient,
leading practical systems to rely on synchronous protocols.
This paper attempts to simplify asynchronous consensus
by building atop a novel {\em threshold logical clock} abstraction,
which enables upper layers to operate as if on a synchronous network.
This approach yields an asynchronous consensus protocol for fail-stop nodes
that may be simpler and more robust than Paxos and its leader-based variants,
requiring no common coins and achieving consensus
in a constant expected number of rounds.
The same approach can be strengthened against Byzantine failures
by building on well-established techniques such as
tamper-evident logging and gossip, accountable state machines,
threshold signatures and witness cosigning,
and verifiable secret sharing.
This combination of existing abstractions and threshold logical clocks
yields a modular, cleanly-layered approach to building
practical and efficient Byzantine consensus,
distributed key generation,
time, timestamping, and randomness beacons, and other critical services.

\xxx{summarize experimental results}

\end{abstract}

%% file: intro.tex
\section{Introduction}

Consensus protocols tend to be delicate and complex,
despite numerous attempts to simplify or reformulate
them~\cite{lamport01paxos,boichat03deconstructing,ongaro14search,renesse15paxos,howard15raft}.
They become even more complex and fragile when we want them
to tolerate
Byzantine node failures~\cite{castro99practical,kotla09zyzzyva,clement09making,clement09upright,bessani14state,aublin15bft,yin19hotstuff},
and/or asynchronous network conditions~\cite{cachin01secure,cachin05random,correia06consensus,correia11byzantine,mostefaoui14signature,miller16honey,duan18beat,abraham19vaba}.
Because relying on synchrony assumptions and timeouts
can make consensus protocols vulnerable to 
performance attacks~\cite{clement09making,amir11byzantine}
and routing-based attacks~\cite{apostolaki16hijacking},
we would prefer to allow for both adversarial nodes
{\em and} an adversarial network.

This paper explores a new approach to asynchronous consensus
that decomposes the handling of {\em time}
from the consensus process itself.
We introduce TLC, a new {\em threshold logical clock} protocol,
which synthesizes a virtual notion of time on an asynchronous network.
Other protocols, including consensus protocols,
may then be built more simply atop TLC
as if on a synchronous network.
This layering is thus conceptually related
to Awerbuch's idea of {\em synchronizers}~\cite{awerbuch85complexity},
but TLC is designed to operate in the presence of failed or Byzantine nodes.

TLC is inspired in part by Lamport clocks~\cite{lamport78time,raynal92about},
vector clocks~\cite{fischer82sacrificing,liskov86highly,mattern89virtual,fidge91logical,raynal92about},
and matrix clocks~\cite{wuu84efficient,sarin87discarding,ruget94cheaper,drummond03reducing,raynal92about}.
While these classic notions of virtual time
label an unconstrained event history
to enable before/after comparisons,
TLC in contrast labels {\em and} constrains events
to ensure that a threshold of nodes in a group progress through logical time
in a quasi-synchronous ``lock-step'' fashion.
In particular, a TLC node reaches time step $\lts+1$
only after a threshold of all participants has reached time $\lts$
and a suitable threshold amount of round-trip communication
has demonstrably occurred since then.
A particular protocol instance \tlc{\tm}{\tw}{n} is parameterized by
message threshold $\tm$,
witness threshold $\tw$, and
number of nodes $n$.
This means that to reach time $\lts+1$, a node $i$ must have received
messages broadcast at time $\lts$ by at least $\tm$ of the $n$ nodes,
and $i$ must have seen each of those $\tm$ messages
acknowledged by at least $\tw$ of the $n$ nodes.
In a Byzantine environment,
TLC ensures that malicious nodes cannot advance their clocks
either ``too fast'' (running ahead of honest nodes)
or ``too slow'' (trailing behind the majority without catching up).

We find that it becomes simpler to build other useful protocols
atop TLC's logical notion of time,
such as threshold signing,
randomness beacons,
and consensus.
To explore TLC's usefulness for this purpose,
we develop an approach to consensus we call
{\em que sera consensus} or {\em \qsc}.
In \qsc,
the participants each propose a potential value to agree on
(\eg, a block in a blockchain),
then simply ``wait'' a number of TLC time steps,
recording and gossiping their observations at each step.
After the appropriate number of logical time steps have elapsed,
the participants decide independently on the basis of public randomness
and the history they observed
whether the consensus round succeeded and, if so, which value was agreed on.
This ``propose, gossip, decide'' approach relates to
recent DAG-based blockchain consensus
proposals~\cite{lewenberg15inclusive,baird16hashgraph,popov18tangle,danezis18blockmania},
which reinvent and apply classic principles
of secure timeline entanglement~\cite{maniatis02secure}
and accountable state machines~\cite{haeberlen07peerreview,haeberlen10accountable}.
The approach to consensus we propose attempts to clarify and systematize
this direction in light of existing tools and abstractions.

To handle network asynchrony, including adversarial scheduling,
our observation is that it is sufficient to associate random {\em tickets}
with each proposed value or block for symmetry-breaking,
while ensuring that the network adversary cannot {\em learn}
the random ticket values until the communication pattern
defining the consensus round has been completed and indelibly fixed.
In a Paxos-equivalent version of the consensus protocol
for $n=2f+1$ well-behaved, ``fail-stop'' nodes (Section~\ref{sec:cons}),
we ensure this {\em network adversary obliviousness} condition
simply by encrypting each node's self-chosen ticket 
(\eg, via TLS~\cite{rfc8446}), keeping it secret from the network adversary
until the consensus round's result is a {\em fait accompli}.

To tolerate $f$ Byzantine nodes colluding with the network adversary,
as usual we need
$n=3f+1$ nodes total~\cite{pease80reaching,schneider90implementing}.
We rely on gossip and transferrable authentication (digital signatures),
and treat all participants as accountable state machines
in the PeerReview framework~\cite{haeberlen07peerreview,haeberlen10accountable}
to handle equivocation and other detectable misbehavior by faulty nodes.
We use threshold public randomness~\cite{canetti93fast,cachin05random,syta17scalable}
via secret sharing~\cite{shamir79share,stadler96publicly,schoenmakers99simple}
to ensure that the adversary can neither learn nor bias
proposal ticket values until the round has completed.

\xxx{ rename "fat consensus" QSCW for ``witnessed QSC'';
plus 3-step or 4-step thin consensu QSC3 and QSC4 respectively? }

These tools simplify the construction of
asynchronous Byzantine consensus protocols.
QSC3 (Section~\ref{sec:cons})
builds on the TLC protocol configured
with the message and witness thresholds $\tm = \tw = 2f+1$,
\ie, \tlc{2f+1}{2f+1}{3f+1}.
\xxx{
The {\em thin consensus} protocol (Section~\ref{sec:thin})
requires no TLC acknowledgments,
\ie, \tlc{2f+1}{0}{3f+1}.
This approach incurs only $O(n)$ messages per time step,
but requires more time steps and less trivial reasoning.
}
The protocol is attractive for its simplicity and clean layering,
and for the fact that it requires no common coins or trusted dealers.


%% file: bg.tex
\xxx{ 
\section{Background}

Briefly summarize existing building blocks...

\subsection{Logical clocks}

Lamport clocks, vector and matrix clocks, entangled timelines, ...

\subsection{Accountable state machines}

PeerReview, AVM, ...
limitation: hard to deal with secrets; doesn't solve consensus

Summarize: each node maintains its own log of all nondeterministic inputs
affecting its state: mainly the record of messages it received in the order it
[claims to have] received them, but also the nondeterministic inputs the node
itself injects into the system when permitted: e.g., the contents of proposed
values or blocks for consensus.  Also the random elements of signatures or
zero-knowledge proofs… 

Apart from each node’s nondeterministic input logs, all other aspects of the
node’s state and behavior is strictly deterministic and may be (and is)
verified independently by all other nodes. …

\subsection{Secret sharing}

SS, VSS, PVSS, recent optimization...

\subsection{Threshold cryptographic randomness}

Cachin, RandHound, etc.

\subsection{Threshold cryptographic signatures}

CoSi, BLS, ...

Not necessarily for consensus itself, but useful to make results of consensus
efficiently offline-verifiable via structures such as SkipChains[]…

}

%% file: tlc.tex
\section{Threshold Logical Clocks (TLC)}
\label{sec:tlc}

We now introduce TLC and explore its properties informally,
emphasizing simplicity and clarity of exposition.
For now we consider only the non-Byzantine situation
where only the network's scheduling of message delivery,
and none of the participating nodes themselves,
may exhibit adversarial behavior.
We leave Byzantine node failures to be addressed later
in Section~\ref{sec:byz}.

\subsection{TLC as a Layer of Abstraction}

In the tradition of layered network
architectures~\cite{zimmermann80osi,clark90architectural},
TLC's main purpose is to provide a layer
that simplifies the construction of interesting higher-lever protocols atop it.
Building atop a fully-asynchronous underlying network, in particular,
TLC offers a coordinating group of nodes
the abstraction of a simple synchronous network
in which time appears to advance for all participants in lock-step
through consecutive integer {\em time-steps} ($1, 2, 3, \dots$).
TLC's synchronous network abstraction is analogous to that
provided by Awerbuch's {\em synchronizers},
except that TLC tolerates a threshold number
of faulty nodes that may be unavailable and/or compromised.

The contract TLC offers upper-layer protocols on participating nodes
may be summarized concisely as follows:

\begin{itemize}
\item	I, TLC, will give you, the upper-layer protocol,
	an integer clock that {\em measures} time,
	by counting rounds of communication
	that network connectivity permits among the group members
	while tolerating a threshold number of
	unreachable and/or malicious nodes.

\item	I will {\em pace} your communication with the group
	by notifying you when logical time advances,
	which is when you may broadcast your next message.

\item	For reference in formulating your next broadcast,
	I will make available a record  of {\em history},
	which will contain
	a potentially incomplete subset of the messages that all nodes broadcast
	in recent time steps.

\item	This record of history will tell you not just what {\em you} saw
	in recent time steps,
	but also exactly what prior messages {\em other nodes} had seen
	by the moment of each recorded event in the history.

\item	I will ensure that the recorded history
	{\em includes} messages from at least a threshold number of nodes
	at each past logical time step it records.

\item	Optional: I will ensure that a threshold number of messages
	in each time step
	were {\em seen} by a threshold number of nodes
	before that time step completes.
\end{itemize}

We expand on these rules and explore how TLC implements them
in the sections below.

\subsection{Time Advancement in Basic TLC}

TLC assumes at the outset that we have
a well-defined set of participating nodes,
each of which can send messages to any other participant
and can correctly authenticate messages received from other participants
(\eg, using authenticated TLS~\cite{rfc8446}).
Further, in addition to the number $n$ of participants and their identities,
TLC requires a {\em message threshold}, $t_m$,
as a configuration parameter defining
the number of other participants a node must ``hear from''
during one logical time-step before moving on to the next.
For simplicity we will assume that $1 < t_m < n$.

At any moment in real-world time,
TLC assigns each node a {\em logical time step}
based on its communication history so far.
Like Lamport clocks~\cite{lamport78time,raynal92about}
but unlike vector or matrix clocks~\cite{fischer82sacrificing,liskov86highly,wuu84efficient,sarin87discarding,raynal92about},
TLC represents logical time steps
as a single monotonically-increasingly integer $\lts \in \NN$
with global meaning across all $n$ participating nodes.
Lamport clocks give individual nodes, or arbitrarily-small groups of nodes,
unconstrained freedom to increment their notion of the current logical time
to reflect events they locally observe, or claim to have observed.
TLC instead constrains nodes so that they must coordinate
with a threshold $t_m$ of nodes in order to ``earn the privilege''
of creating a new event and incrementing their notion of the logical time.

At the ``beginning of time'' when an instance of the TLC protocol starts,
all nodes start at logical time-step $s=0$,
and have the right to broadcast to all participants
a single initial message labeled with $s = 0$.
On reaching this and every subsequent time-step $s$,
each node $i$ then waits to receive messages labeled step $s$
from at least $t_m$ distinct participants, including $i$ itself.
At this point, node $i$ has earned the right
to advance its logical time to step $s+1$,
and consequently broadcasts a single message labeled $s+1$
before it must wait again.

Node $i$ does not care {\em which} set of $t_m$ participants
it received step $s$ messages from
in order to meet its threshold and advance to $s+1$.
Critical to tolerating arbitrary (adversarial) network scheduling,
$i$ simply takes {\em the first} threshold set of messages to arrive,
regardless of which subset of participants they came from,
then moves on.

\begin{figure}
\begin{center}
\includegraphics[width=1\columnwidth]{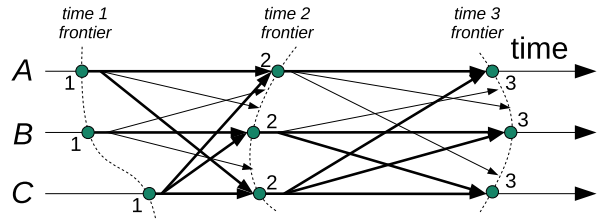}
\end{center}
\caption{Illustration of basic threshold logical clocks without witnessing.
Each node waits to receive a threshold number of messages
at each logical time step: 2 of 3 including its own in this example.
Darker arrows indicate messages that enable nodes to progress;
the rest arrive “too late” to contribute to time advancement.
}
\label{fig:tlc-basic}
\end{figure}

Figure~\ref{fig:tlc-basic}
illustrates this process for a single three-node example
with a message threshold $t_m=2$.
This TLC configuration requires each node
to collect one other node's step $s$ message in addition to its own
before advancing and broadcasting its step $s+1$ message.

Different nodes may start at different real-world wall-clock times,
and the network may arbitrarily delay or reorder
the delivery of any node's message to any other.
This implies that different nodes may reach a given logical time-step
at vastly different wall-clock times than other nodes.
We refer to the varying sets of real-world times
that different nodes arrive at a given logical step $s$
as the {\em time frontier} of step $s$.
Since each node advances its logical clock monotonically,
the time frontier for each successive step $s$
divides real time into periods ``before'' and ``after'' step $s$
from the perspective of any given node $i$.
A given moment in real time may of course occur before $s$
from one node's perspective but after $s$ for another node.

\subsection{Causal Propagation of Messages}
\label{sec:tlc:causal}

To simplify reasoning about logical time
and the protocols we wish to build on it,
we will assume that any TLC implementation ensures that
knowledge propagates ``virally'' according to a causal ordering.
For example, suppose node $A$ sends message $1_A$ at step $1$,
node $B$ receives it before moving to step $2$ and broadcasting message $2_B$,
and node $C$ in turn receives message $2_B$.
In this case, message $1_A$ is {\em causally before} $2_B$.
We will assume that the underlying network or overlay protocol,
or the TLC implementation itself,
ensures that node $C$ learns about message $1_A$
either before or at the same time as $C$ learns about $2_B$:
\ie, in causal order.

One way to ensure causal message propagation
is conceptually trivial, if impractically inefficient.
Each node $i$ simply includes in every message it sends
a record of $i$'s entire {\em causal history}:
\eg, a complete log of every message $i$ has ever received directly
or heard about interactly from other nodes.
There are more practical and efficient ways
to ensure causally-ordered message delivery,
of course:
Section~\ref{sec:arch:causality},
will employ standard gossip and vector time techniques for this purpose.
For now, we will simply take 
causally-ordered message propagation for granted
as if it were a feature of the network.

\subsection{Viral Advancement of Logical Time}

A consequence of the threshold condition for time advancement,
combined with causally-ordered message propagation,
is that not just messages but also {\em time advancement events}
propagate virally.

Suppose, for example, that node $i$ is waiting at logical time-step $s$
while another node $j$ has advanced to a later step $s' > s$
arbitrarily far ahead of $i$.
If $i$ receives the message $j$ broadcast at step $s'$,
then this delivery causes $i$ to ``catch up'' instantly to step $s'$.
This is because, due to causal message propagation,
$i$ obtains from $j$ not just $j$'s step $s'$ broadcast
but also, indirectly, the $t_m$ threshold set of messages $j$ used
to advance from $s$ to $s+1$,
those that $j$ used to advance to $s+2$, etc., up through step $s'$.

\subsection{Information Propagation in TLC}

The basic TLC protocol outlined above makes it easy to reason about
the information that flowed {\em into} a node
leading up to a particular time step $s+1$.
Because any node $i$ had to obtain a threshold $t_m$ of step $s$ messages,
either directly or indirectly,
in order to advance to $s+1$ at all,
this trivially implies that $i$'s ``view'' of history at step $s+1$
will contain at least a $t_m/n$ fraction of all messages from step $s$,
as well as at all prior steps.

To build interesting protocols atop TLC, however,
we will need to be able to reason similarly about information flowing
{\em out of} a particular node into other nodes after some step $s$.
In particular, after a node $i$ broadcasts its step $s$ message,
how can we tell how many nodes have received that message
by some future logical time,
say $s+1$?
The adversarial network schedule ultimately determines this, of course,
but it would be useful if we could at least {\em measure} after the fact
the success (or lack thereof) of a given message's propagation to other nodes.
For this purpose, we enhance TLC to support {\em witnessed} operation.

\subsection{Threshold Witnessed TLC}
\label{sec:tlc:wit}

One way we can determine
when a message broadcast by a node has reached other nodes
is by requiring the node to collect delivery confirmations proactively,
as a new prerequisite for the advancement of logical time.
We might, for example,
require each node to transmit each of its broadcasts to every other node
and await TCP-like acknowledgments for its broadcast.
If we require a node to confirm message delivery to {\em all} other nodes,
or even to any pre-defined set of other nodes, however,
this would present denial-of-service opportunities to the adversarial network,
which could arbitrarily delay the critical message or acknowledgment deliveries.

To tolerate full network asynchrony,
we must again invoke threshold logic,
this time to confirm a message's delivery to {\em any subset} of participants
meeting some threshold,
without caring {\em which} specific subset confirms delivery.
Confirming message delivery to a threshold of participants
is the basic  purpose of a threshold {\em witnessing} protocol
such as CoSi~\cite{syta16keeping}.
Threshold witnessing is useful, for example,
in proactively ensuring the public transparency
of software updates~\cite{ford16apple,nikitin17chainiac}
or building scalable cryptographically-trackable
blockchains~\cite{kokoris16enhancing,kokoris17omniledger,kokoris19secure}.

Threshold witnessing may be secured against Byzantine behavior using
cryptographic multisignature or threshold signing
schemes~\cite{desmedt89threshold,rabin98simplified,shoup00practical,boneh18compact,drijvers19security,bagherzandi08multisignatures}.
Since we are assuming no Byzantine nodes for now, however,
simple acknowledgments suffice for the moment in TLC.

We introduce a new {\em witness threshold}
configuration parameter $t_w$ to TLC.
A TLC protocol instance is thus now parameterized
by message threshold $t_m$, witness threshold $t_w$,
and total number of nodes $n$.
We will label such a TLC configuration
\tlc{t_m}{t_w}{n} for brevity.
In practice we will typically pick $t_w$ either to be equal to $t_m$,
or to be zero, reducing to unwitnessed TLC as described above.
We separate the message and witness thresholds, however,
because they play orthogonal but complementary roles.

\begin{figure}
\begin{center}
\includegraphics[width=1\columnwidth]{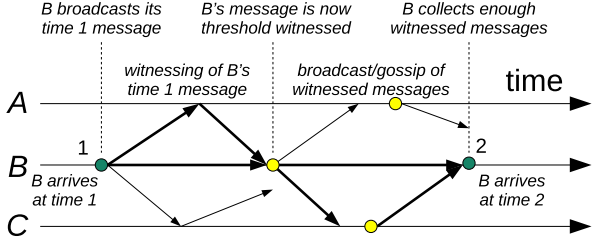}
\end{center}
\caption{Illustration of one witnessed TLC time-step
	from the perspective of one particular node $B$.
	Darker arrows indicate messages on the ``critical path''
	enabling node $B$ to make progress.}
\label{fig:tlc-witnessed}
\end{figure}

These threshold parameters establish
a two-part condition for a node to advance logical time.
To get from step $s$ to $s+1$,
each node must collect not just $t_m$ messages
but $t_m$ {\em threshold witnessed} messages from step $s$.
Each threshold message must have been witnessed
by at least $t_w$ participants
before it can ``count'' towards $t_m$.

To create a threshold witnessed message,
each node $i$ first broadcasts its ``bare'' unwitnessed step $s$ message $m$,
and begins collecting {\em witness acknowledgments} on $m$
from participants serving as witnesses.
Another node $j$ that receives $m$ in step $s$
simply replies with an acknowledgment that it has witnessed $m$.
Upon collecting a $t_w$ threshold of witness acknowledgments within step $s$,
node $i$ broadcasts an assertion that $m$ has been threshold witnessed.
Only upon receiving this threshold witness confirmation
may any node count $m$ towards its message threshold $t_m$
required to advance to step $s+1$.
Figure~\ref{fig:tlc-witnessed} illustrates this process
in a simple 3-node configuration.

Suppose a node $i$ broadcasts an unwitnessed message $m$ for step $s$,
and another node $j$ receives $m$ not in step $s$,
but {\em after} having advanced to a later step $s' > s$.
In this case, receiving node $j$ considers $m$ to be ``too late'' for step $s$,
and declines to witness $m$ for step $s$.
Instead, $j$ replies with the information $i$ needs
to ``catch up'' to the most recent step $s'$ that $j$ is aware of.
If too many nodes receive $i$'s message $m$ too late,
this may make it impossible for $m$ ever to be threshold witnessed --
but $i$ can still advance its logical time with the information $j$ provided
in lieu of a witness acknowledgment for step $s$.

Due to network scheduling,
a node $i$ may receive $t_m$ threshold witnessed messages of {\em other} nodes,
and hence satisfy the conditions to advance time,
before $i$ has obtained a threshold $t_w$ of witness acknowledgments
to its own step $s$ message.
In this case, $i$ simply abandons its collection of witness acknowledgments
for its own message and moves on,
using only other nodes' threshold witnessed messages and not its own
as its basis for advancing time.
This rule preserves the property that time advancement advances virally,
as discussed above,
and ensures that a lagging node can ``catch up'' instantly to the rest
upon receiving a message from a recent time-step.

With witnessed TLC,
we now have a convenient basis for reasoning about information flow
both {\em into} and {\em out of} a node at a given time-step.
As before,
we know that to reach step $s+1$ any node $i$
must have collected information --
and hence be ``caught up'' on the histories of --
at least $t_m$ nodes as of step $s$.
With witnessed TLC,
we additionally know by construction that
any node's step $s$ message that is threshold witnessed by step $s+1$
has propagated ``out'' to and been seen by at least $t_w$ nodes by $s+1$.
Finally, because only threshold witnessed messages count
towards the $t_m$ threshold to advance time,
we know that by the time any node reaches step $s+1$
there are at least $t_m$ threshold witnessed messages from step $s$.

\subsection{Majoritarian Reasoning with TLC}
\label{sec:tlc:maj}

So far we have developed TLC
in terms of {\em arbitrary} thresholds $t_m$ and $t_w$
without regard to any specific choice of thresholds.
But many interesting protocols, such as consensus protocols,
rely on {\em majoritarian} logic:
\ie, that a participant has received information from,
or successfully delivered information to,
a majority of participants.

For this reason,
we now explore the important special case of TLC
configured with majority thresholds:
\ie, $t_m > n/2$ and $t_w > n/2$.
To tolerate Byzantine faults, 
Section~\ref{sec:byz} will adjust these thresholds
to ensure majorities of {\em correct}, non-Byzantine nodes --
but the fundamental principles remain the same.

Configured with majority thresholds,
TLC offers two key useful properties:
time period delineation and two-step broadcast.
We develop these properties next.

\begin{figure}
\begin{center}
\includegraphics[width=1\columnwidth]{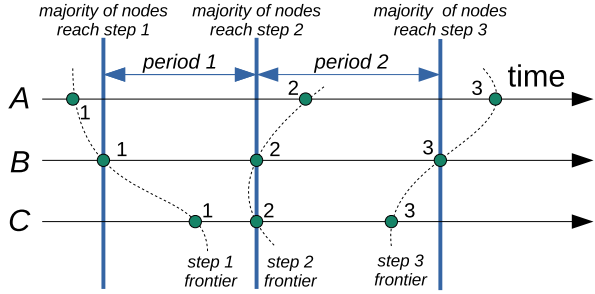}
\end{center}
\caption{Global time periods demarked by the moments a majority of
	correct nodes reach a threshold time $t$.}
\label{fig:time-periods}
\end{figure}

\subsection{Global time period delineation}
\label{sec:tlc:periods}

Even though different TLC nodes reach a given time step at varying real times,
majoritarian TLC nevertheless divides not just logical
but also {\em real} wall-clock time
into a well-defined quasi-synchronous succession of real time periods.
The start of each global time period may be defined by
the moment in real time that {\em a majority of} nodes
first reaches a given logical time step $s$.
Figure~\ref{fig:time-periods} illustrates this principle,
with real time delineated into successive time periods,
each starting the moment the {\em first two} of the three nodes
have advanced to a given time step.

Because each node's logical clock advances monotonically,
and a majority of nodes must reach step $s$
before a majority of nodes can reach $s+1$,
these majoritarian time periods likewise advance monotonically.
These time periods in principle create the effect of
a purely synchronous ``lock-step'' system,
but with time periods ``self-timed'' by 
the progress of underlying network communication.

Even though these majoritarian time periods are easy to define in principle,
we face a practical challenge in protocol design.
Without precisely-calibrated real-time clocks, which we prefer not to assume,
an individual node will rarely be able to tell whether it has advanced
to logical time step $s$ before, or after, other participants.
This implies in turn that no node can realistically be expected to know
or determine precisely when a previous time period ends and the next begins.
In the Figure~\ref{fig:time-periods} example,
although globally there is a clear and well-defined ``fact of the matter''
regarding the moment each majoritarian time period begins and ends,
a node will be unable to tell whether it advanced to step $s$
before majoritarian time period $s$ started (\eg, $1_A$),
after period $s$ started ($1_C$),
or happened to be the ``critical moment''
that launched period $s$ ($1_B$).

Despite this limitation in the knowledge of any given node,
this majoritarian delineation of real time into periods
gives us important tools for reasoning conservatively about
when any particular message could, or could not,
have been formulated and sent.
Consider in particular a given time period $s$,
starting the moment a majority of participants reach step $s$
and ending the moment a majority of participants reach $s+1$.
We can be sure that:

\begin{enumerate}
\item	No node can advance to step $s+1$,
	or send a message labeled $s+1$,
	before the prior global time period $s$ has started.
	Such a node $i$ would have had to collect a majority $t_m$
	of step $s$ messages to meet its condition to advance logical time,
	but no majority of step $s$ messages can be available to $i$
	before a majority of nodes has actually reached step $s$.

\item	After global time period $s$ has ended and $s+1$ begun,
	no node can formulate or successfully threshold witness
	any new message for step $s$.
	Getting a step $s$ message threshold witnessed
	would require a majority of nodes to provide witness acknowledgments
	for step $s$.
	But after period $s+1$ begins,
	a majority of nodes has ``moved on'' to $s+1$
	and stopped providing witness acknowledgments for step $s$,
	leaving only an inadequate minority of nodes
	that could potentially witness new messages for step $s$.
\end{enumerate}

Looking at an illustration like Figure~\ref{fig:time-periods},
one might reasonably ask whether the wandering time frontiers,
representing each node's advancement to a given step $s$,
can ``cross'' over not only the majoritarian time period $s$ boundary,
but also the time period boundaries before ($s-1$) and/or after ($s+1$).
The above two guarantees in a sense answer this question in the negative,
effectively keeping all nodes approximately synchronized with each other,
plus or minus at most one logical time step.

The first property above trivially ensures that no node can reach step 2 
before global time period 1 has begun,
can reach step 3 before period 2 has begun, etc.
Thus, no node can ``race ahead''
of the majority's notion of the current logical time
by more than one time step.

And although communication patterns such as denial-of-service attacks
could cause a particular node to ``lag'' many time-steps behind the majority
in terms of real time,
the second property above guarantees that such a lagging node
cannot actually produce any effect,
{\em observable via threshold witnessed messages},
after period $s$ has ended and $s+1$ begun.
Any new messages the lagging node might produce after period $s+1$ has begun
will effectively be ``censored'',
by virtue of being unable ever to be threshold witnessed.
The lagging node will once again be able to send threshold witnessed messages
when, and only when, it ``catches up'' to the current global time period.

\com{ explain global time periods delimited by moments
	when $f+1$ correct nodes reach time $t$.
Three key properties:
1. No message can be certified for time $t$ before period $t$ starts,
because at least $f+1$ correct nodes would have to witness that message
and they will not do so before they reach time $t$; and
2. No message formulated after period $t$ ends (and $t+1$ starts)
can be certified for time $t$,
because at least $f+1$ correct nodes would have to witness it
including at least one of the correct nodes that has reached $t+1$
and hence stopped witnessing messages for time $t$.
3. Every message $m$ certified in time $t$ will be known to
at least $2f+1$ correct nodes by time $t+1$,
and by {\em every} correct node by time $t+1$.
This is because the set of $f+1$ correct nodes that certified $m$ in period $t$
must overlap with the set of $f+1$ certified messages from time $t+1$
that any node $i$ collected in order to reach time $t+1$,
and that overlapping node will causally forward its time $t$ knowledge of $m$
to every node that uses its certified message from time $t+1$
to advance to time $t+2$.
(But not every node will necessarily know that $m$ was certified.)
}

\subsection{Two-step semi-reliable broadcast}
\label{sec:tlc:broadcast}

Another key property we obtain from majority message and witness thresholds
is a guarantee that a majority of the messages sent at any time step $s$
will be known to {\em all} participants by step $s+2$.
TLC thus implicitly provides {\em two-step broadcast}
at least for a majority, though not all,
of the messages sent at any time step.

To see why this is the case,
consider that in order for any node to advance to step $s+1$,
it must collect a majority $t_m$ of threshold witnessed messages from step $s$.
Each of these messages must have been seen by a majority $t_w$ of nodes
in order to have been successfully threshold witnessed.
To reach step $s+2$, in turn,
each node must collect a majority $t_m$ of threshold witnessed messages
from step $s+1$.
The majority of nodes that witnessed
any threshold witnessed message $m$ from step $s$
must overlap, by at least one node $i$,
with the majority of nodes that any other node $j$ collects messages from
in order to reach $s+2$.
This intersection node $i$ effectively serves as a conduit
through which $j$ is guaranteed to learn of message $m$ transitively
through causal knowledge propagation,
even if $j$ itself did not directly witness $m$ during step $s$.

Since the real time at which nodes reach step $s+2$
is determined by the network's arbitrary communication schedule,
this two-step broadcast property can make no guarantees
about when {\em in real time} any node
actually learns about threshold witnessed message $m$ from step $s$.
A minority of nodes might lag many time steps behind the majority,
and learn about $m$ only when they eventually ``catch up'' and resynchronize.
By the time period delineation properties above, however,
no lagging node will be able to have any {\em effect} on the majority,
observable through threshold witnessed messages,
before catching up with the majority.
If the lagging node catches up at step $s+2$ or later,
it learns about threshold witnessed message $m$ from step $s$,
through causal propagation,
in the process of catching up.

\xxx{ illustrate with figure}

It is important to keep in mind
that this two-step broadcast property applies only to
the ``lucky'' majority of messages that were threshold witnessed in step $s$,
however.
A minority of messages that other participants
{\em tried} to send in step $s$ may never be threshold witnessed
before too many nodes advance to $s+1$ and the ``gate closes'' on step $s$.
These unlucky step $s$ messages might be seen by some participants,
but TLC can make no guarantee that all, or any particular number,
of participants will ever see them.
Further, the adversarial network gets to decide which messages
are in the lucky majority that are threshold witnessed and broadcast,
and which are unlucky and potentially lost to history.
Messages that fail to achieve threshold witnessed status during a time step
may be considered casualties of network asynchrony.

Another subtle but important caveat
with two-step broadcast in TLC
is that even if message $m$ is threshold witnessed in step $s$
and broadcast to all nodes by $s+2$,
this does not mean that all nodes will
{\em know that $m$ was threshold witnessed} by $s+2$.
Suppose a node $i$ receives and acknowledges
the bare, unwitnessed version of $m$ during step $s$, for example,
thereby contributing to the eventual threshold witnessing of $m$.
Node $i$ might then, however,
advance to steps $s+1$ and $s+2$ on the basis of
{\em other} sets of threshold witnessed messages not including $m$,
without ever learning that $m$ was fully threshold witnessed.
In this case,
while $i$ has indeed, like all nodes, seen at least
a bare unwitnessed version of $m$ by step $s+2$,
only some nodes may know by $s+2$
that $m$ was successfully threshold witnessed.
This subtlety will become important later in Section~\ref{sec:cons:universal}
as we build consensus protocols atop TLC.

\xxx{ optional: Lazy, On-Demand TLC }

%% file: app.tex
\section{Building Basic Services on TLC}
\label{sec:app}

Before we tackle asynchronous consensus in Section~\ref{sec:cons},
we first briefly sketch several classic distributed services
not requiring consensus that are easy and natural to build
using TLC for pacing.
While these services may of course be built without TLC,
the threshold logical clock abstraction makes it simple
for such distributed services to operate atop fully asynchronous networks
in self-timed fashion as quickly as network communication permits.

\subsection{Network Time Services}
\label{sec:app:time}

Even in asynchronous distributed systems
that we do not wish to be {\em driven by} wall-clock time or timeouts,
it is still important in many ways
to be able to {\em tell time} and interact properly with the wall clock.
We first discuss three basic time-centric services
and how they might benefit from asynchronous operation atop TLC:
clock synchronization, trusted timestamping, and encrypted time capsules.

\subsubsection{Clock Initialization and Synchronization}
\label{sec:app:time:sync}

Time services such as NTP~\cite{mills91internet,rfc5905},
by which networked devices around the world synchronize their clocks,
play a fundamental role in the Internet's architecture.
Without time services,
all devices' real-time clocks gradually drift,
and can become wildly inaccurate after power or clock failures.
Drifting or inaccurate device clocks can undermine the functioning
of real-time systems~\cite{kopetz11real-time} and
wireless sensor networks~\cite{zhao04wireless,sundararaman05clock}.
Security protocols often rely on devices
having roughly-synchronized
clocks~\cite{davis96kerberos,vratonjic11inconvenient,topalovic12towards},
otherwise becoming vulnerable to attacks
such as the replay of expired credentials, certificates,
or outdated software with known exploits~\cite{nikitin17chainiac}.

While a correct sense of time is critical to the reliability and security
of today's networked devices,
numerous weaknesses have been found in traditional time
services~\cite{roosta07time,ganeriwal08secure,malhotra16attacking,malhotra16broadcast,malhotra17security}.
The fact that clients typically rely entirely on a {\em single} NTP time server
(\eg, the nearest found on a list)
is itself an inherent single-point-of-failure weakness.
Using GPS as a time source~\cite{lewandowski91gps,dana97global},
while ubiquitous and accurate under normal conditions,
is less trustworthy as GPS spoofing
proliferates~\cite{psiaki16gnss,robinson17using,c4ads19above}.
A networked device might achieve a more secure notion of the time
by relying on a group of independent time servers rather than just one,
thereby avoiding any single point of failure or compromise.

TLC represents a natural substrate atop which to build
such a {\em distributed time service} or {\em beacon}.
One simple approach is for each each server in a TLC coordination group
to publish a log (or ``blockchain'') of current-time records,
one per TLC time-step.
Each successive record indicates the server's notion
of the record's publication time,
ideally measured from a local high-precision source such as an atomic clock.
Each published record is both digitally signed by the server,
and witness cosigned by other coordination group members~\cite{syta16keeping},
thereby attesting to clients jointly that the server's notion of time
is consistent to some tolerance.
Clients may still follow just one time server at a time
as they currently do (\eg, the closest one),
but protect themselves from major inaccuracy or compromise
of their time source by verifying the witness cosignatures as well.
We address later in Section~\ref{sec:arch:realtime}
the important detail of allowing witnesses
to validate proposed time records in an asynchronous setting
without introducing arbitrary timeouts or tolerance windows.

The precision a distributed time beacon can provide
will naturally depend on factors such as
how widely-distributed the participating servers are
and how reliable and predictable their mutual connectivity is.
Building a distributed time beacon atop TLC offers the potential key benefit
of adapting automatically to group configurations and network conditions.
A time beacon composed of globally-distributed servers
could offer maximum independence and diversity, and hence security,
at the cost of limited precision due to the hundreds-of-milliseconds
round-trip delays between group members.
Such a widely-distributed service could offer client devices
a coarse-grained but highly-secure ``backstop'' reference clock
ensuring that the device's clock
cannot be off by minutes or hours
even if more-precise time sources are unavailable or subverted.
Another complementary time beacon running the same TLC-based protocol,
but composed of servers colocated in a single city or data center
with a low-latency interconnect,
would automatically generate much more frequent, high-precision time reports,
while still avoiding single points of failure
and degrading gracefully during partial failures or attacks.

\subsubsection{Trusted Timestamping and Notarization}
\label{sec:app:time:stamp}

A closely-related application is a {\em digital timestamping} service,
which not only tells the current time,
but also produces {\em timestamps} on demand
attesting that some data known to the client existed at a certain time.
Standards such as the Time-Stamp Protocol~\cite{rfc3161,ansi16trusted}
allow clients to request a signed timestamp
on cryptographically hashed data from a trusted timestamping authority.
Such an authority is again a single point of failure, however,
motivating recently-popular decentralized approaches to timestamping,
such as collecting content hashes into a Merkle tree~\cite{merkle79secrecy}
and embedding its root
in a Bitcoin transaction~\cite{nakamoto08bitcoin,szalachowski18towards},
or collectively signing each root~\cite{syta16keeping}.

An asynchronous distributed timestamping service,
whose period and timestamp granularity is self-timed
to the best allowed by group configuration and prevailing network conditions,
represents a natural extension to a TLC-based time service.
Each server in such a group might independently collect
client-submitted content hashes into Merkle trees,
publishing a signed and witness cosigned tree each TLC time step,
as in the CoSi time service~\cite[Section V.A]{syta16keeping}.
In addition, a newly-started or long-offline device
can bootstrap its internal clock with strong freshness protection,
preventing an upstream network attacker from back-dating its notion of time,
by initializing its clock according to a witness cosigned timestamp
it requests on a freshly-generated random nonce.

\subsubsection{Encrypted Time Capsules}
\label{sec:app:time:vault}

A third classic time-related service with many potential uses
is a cryptographic {\em time vault} or {\em time capsule},
allowing clients to encrypt data so that it will become decryptable
at a designated future time.
In games, auctions, and many other market systems, for example,
participants often wish to encrypt their moves or bids from others
until a designated closing time to guard against
front running~\cite{czernik18frontrunning,eskandari19transparent}.
Time-lock puzzles~\cite{rivest96time} and
verifiable delay functions~\cite{boneh18verifiable,wesolowski19efficient}
represent purely cryptographic proposals to achieve this goal,
but cryptographic approaches face inherent challenges
in accurately estimating the future evolution
of, and market investment in, computational and cryptanalytic puzzle-solving
power~\cite{mahmoody11time,conner-simons19programmers}.

Another approach to time capsules more compatible with TLC
relies on a time service that holds a master key
for identity-based encryption
(IBE)~\cite{shamir84identity,boneh01identity,waters05efficient}.
Clients encrypt their messages
to a virtual ``identity'' representing a particular future time.
The time service regularly generates and publishes
the IBE private keys representing these ``time identities'' as they pass,
allowing anyone to decrypt any time-locked ciphertext
after the designated time passes.
Threshold secret-sharing~\cite{shamir79share,stadler96publicly,schoenmakers99simple}
the IBE master key among the group
avoids single points of failure or compromise.
The asynchronous setting presents the challenge that clients
seemingly must predict the future rate
at which the time capsule service will operate,
and hence the granularity at which it will publish time-identity keys,
a detail we address later in Section~\ref{sec:arch:release}.

\subsection{Public Randomness Beacons}
\label{sec:app:random}

Like time,
trustworthy public randomness has become an essential ``utility''
needed by numerous applications.
Lotteries and games need random choices
that all participants can trust to be fair and unbiased,
despite the many times such trust has been breached
in the past~\cite{fienberg71randomization,weissman15how,song18guess,forgrave18man}.
Governments need public randomness to choose a sample of ballots
to select jury candidates~\cite{robinson50bias},
to audit election results~\cite{calandrino07machine,lindeman12gentle},
and experimentally,
to involve citizens in policy deliberation
through sortition~\cite{fishkin05experimenting,dowlen08political}.
Large-scale decentralized systems such as blockchains need public randomness
to ``scale out'' via sharding~\cite{luu16secure,kokoris17omniledger}.

The many uses for public randomness have inspired
beacons such as NIST's~\cite{nist19reference}.
Concerns about centralized beacons
being a single point of compromise~\cite{beacon14se},
however,
again motivate more decentralized approaches to public
randomness~\cite{cachin05random,clark10beacon,lenstra15random,bonneau15bitcoin,syta17scalable}.
Threshold-secure approaches~\cite{cachin05random,syta17scalable,kozlov19league}
are naturally suited to being built on TLC,
which can pace the beacon to produce fresh random outputs dynamically
as often as network connectivity permits,
rather than at a fixed period.

%% file: cons.tex
\section{Que Sera Consensus (\qsc)}
\label{sec:cons}

We now explore approaches to build consensus protocols atop TLC,
using a series of strawman examples to address the key challenges
in a step-by-step fashion for clarity.
This series of refinements will lead us to QSC3,
a randomized non-Byzantine (\ie, Paxos-equivalent) consensus protocol.
We leave Byzantine consensus to Section~\ref{sec:byz}.

Although the final QSC3 protocol this section arrives at is quite simple,
the reasoning required to understand and justify it subtle,
as with any consensus protocol.
A desire to clarify this reasoning motivates
our extended, step-by-step exposition.
Expert readers may feel free to
skip to the final solution
summarized in Section~\ref{sec:cons:solution} if desired.

\xxx{
spin-the-bottle illustration? 
Background, need for randomness to
achieve both safety and liveness… 
}

\subsection{Strawman 0: multiple possible histories}
\label{sec:cons:histories}

As a starting point,
we will not even try to achieve consensus reliably on a single common history,
but instead simply allow each node to define and build
its own idea of a {\em possible history},
independently of all other nodes.
For convenience and familiarity,
we will represent each node's possible history as a {\em blockchain},
or tamper-evident log~\cite{schneier99secure,crosby09efficient}
in the form popularized by Bitcoin~\cite{nakamoto08bitcoin}.

At TLC time-step 0, we assume all $N$ nodes start building
from a common {\em genesis block}
that was somehow agreed upon manually.
At each subsequent time-step,
each node independently formulates and {\em proposes} a new block,
which contains a cryptographic hash or {\em back-link} to the previous block.
Thus, node $i$'s block 1 contains a hash of the genesis block,
node $i$'s block 2 contains a hash of node $i$'s block 1, and so on.
The main useful property of this structure
is that the blockchain's entire history is identified and committed by
knowledge of the {\em head}, or most recent block added.
It is cryptographically infeasible to modify any part of history
without scrambling all the hash-links in all subsequent blocks
including the head, thereby making any modification readily detectable.

\begin{figure}
\begin{center}
\includegraphics[width=1\columnwidth]{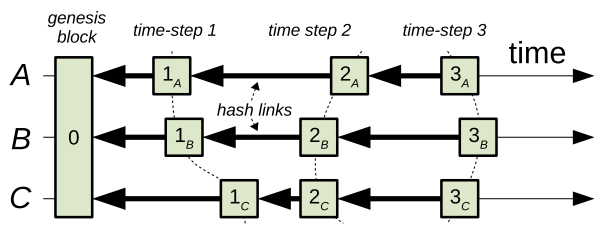}
\end{center}
\caption{Illustration of Strawman 0: each of the $N$ nodes
independently builds its own {\em possible history},
each represented by a blockchain with one block per TLC time-step.
}
\label{fig:cons-chains}
\end{figure}

Figure~\ref{fig:cons-chains} illustrates three nodes
building three independent blockchains in this way.
The real (wall-clock) time at which each node reaches a given TLC time-step
and proposes the corresponding block on its blockchain
may vary widely across nodes due to network delays,
but TLC serves to pace all nodes' advancement of time
and keep them logically in lock-step despite these delays.

If we assume each node's proposed block at a given time-step
contains a set of transactions submitted by clients, as in Bitcoin,
then even this strawman protocol can provide a limited notion of ``consensus.''
If a client submits some transaction $T$
to {\em all} $N$ nodes
(\eg, ``Alice pays Bob 1 BTC''),
and the client succeeds in getting $T$ embedded in each node's history,
then the client can consider $T$ to be ``committed.''
This is because regardless of which of the $N$ posssible histories
we might choose to believe,
all of them contain and account for transaction $T$.

However, if a ``double-spending'' client manages
to get $T$ onto some nodes' blockchains
and gets a conflicting transaction $T' \ne T$ onto others
(\eg, Alice pays Charlie the same 1 BTC),
then we will forever be uncertain whether Bob or Charlie now holds the 1 BTC
and unable ever to resolve the situation.
Thus, we need a way to break the symmetry
and enable some competing histories to ``win'' and others ``lose'' --
the challenge we tackle next.

\com{
At the end of each time period,
there can be up to $N$ conflicting bloickchain heads.
Each node, upon reaching time $t+1$,
will be aware of a threshold number of those possible heads --
but of course will have no basis for predicting which of those, if any,
of those will ``survive'' or ``win'' in the long term.
Since each node might well choose simply
to keep building its own local blockchain and ignore the others forever,
this strawman thus fails to ensure that the group
ever reaches any state we would clearly recognize as ``consensus'' --
except that limiting the possibility space to $N$ possible competing blockchains
is at least arguably better than an unlimited number of possibilities.
}

\subsection{Strawman 1: genetic consensus}
\label{sec:cons:genetic}

Introducing randomness makes it surprisingly simple to create
a working, if non-ideal, consensus protocol.
Suppose we modify the above strawman such that at each time-step,
one randomly-chosen node chooses to adopt and build on the blockchain
of a randomly-chosen neighbor instead of its own.
This node's prior history is thus dropped from the space of possibilities,
and effectively replaced with the node's newly-adopted view of history.

\begin{figure}
\begin{center}
\includegraphics[width=1\columnwidth]{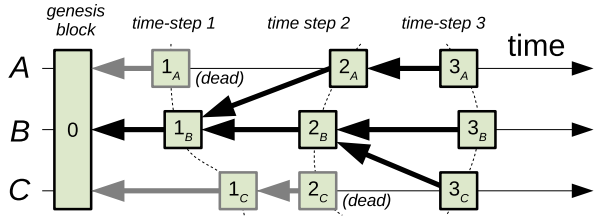}
\end{center}
\caption{Illustration of Strawman 1,
in which each of the $N$ nodes occasionally
choose to adopt a random neighbor's blockchain in favor of their own,
as if one history ``genome'' had ``died'' while another ``reproduced.''
}
\label{fig:cons-genetic}
\end{figure}

Consider the simple example in Figure~\ref{fig:cons-genetic}.
At TLC step 1, each node independently builds on the genesis block as before.
At step 2, however, node $A$ randomly chooses
to build on $B$'s prior blockchain instead of its own.
Similarly, at step 3, 
node $C$ chooses to adopt $B$'s blockchain.
While we still have three competing heads and hence competing histories
(namely $3_A$, $3_B$, and $3_C$),
nevertheless they happen to share a common prefix,
namely block $1_B$.
Because all future time-steps must build on
one of these three possible histories, all sharing this common prefix,
we can consider the common prefix (block $1_B$) to be {\em committed} --
even if we can't (yet) say anything about the more recent blocks.
This situation is directly analogous to the common event
of a temporary fork in Bitcoin,
where two miners mine competing blocks at about the same time,
deferring resolution of the conflit for later.
The main difference is that we pace the ``mining'' of blocks in our protocol
using TLC instead of via proof-of-work.

Whenever one node adopts another's blockchain,
any transactions that had existed only in the node's prior blockchain
become lost or in effect ``aborted.''
All transactions on the adopted blockchain, in contrast,
become more likely to survive long-term because they are represented redundantly
in the (new) history of one additional node,
and become correspondingly more likely to propagate further
via future adoption events.
If we ever happen to observe that through this random history-adoption process,
a particular transaction of interest has propagated
to all $N$ nodes' view of history,
then we can consider that transaction to be definitely ``committed.''
But will every transaction reach such a state of being
definitely either ``committed''
(by virtue of being on all nodes' views of history)
or ``aborted''
(by virtue of being on none of them)?

Given time, the answer is definitely yes.
This is because from the perspective of a particular transaction
that any node first introduces in a block on its local blockchain,
that transaction's subsequent propagation or elimination
corresponds to a Moran process~\cite{moran58random,nowak06evolutionary},
a statistical process designed to model genetic drift
in a population constrained to a fixed size
(\eg, by natural resource limits).
A node's adoption of another's blockchain corresponds to
the successful ``reproduction'' of the adopted blockchain,
coincident with the ``death'' of the replaced blockchain.
We might view all the transactions in the adopted blockchain's view of history
to be the ``genome'' of the successfully-reproducing blockchain,
whose constituent blocks and transactions
become more likely to survive with each successful reproduction.

This process is a variation on the
P\'olya urn model~\cite{johnson77urn,mahmoud08polya,propp15polyas},
where we view each competing blockchain (or the transactions on them)
as colored balls in an urn.
From this perspective, we view one node's adoption of another's blockchain
as the removal of a pair of colored balls from the urn,
where we duplicate one, discard the other,
and return the two duplicates to the urn.
With time, this process guarantees that any particular transaction
in any particular blockchain's ``genome'' is eventually either lost (aborted)
or propagated to all other individuals (committed).
If all nodes' blockchains have the same ``genetic fitness''
or likeliness to reproduce,
then a transaction first introduced in a block on any one node
has a uniform probability of $1/N$ of eventually being ``committed'' this way.

Of course, this strawman has several obvious limitations.
$1/N$ is not a great probability
of a proposed transaction being successfully committed.
We must wait a considerable time before we can know a transaction's
commit/abort status for certain.
And we must monitor {\em all} nodes' blockchains --
not just a threshold number of them --
in order to reach absolute certainty of this commit/abort status.
However, this strawman does illustrate how simple it can be in principle
to achieve {\em some} notion of ``consensus'' through a simple random process.

\subsection{Strawman 2: a genetic fitness lottery}
\label{sec:cons:fitness}

We can speed up the above ``genetic process'' in two ways,
which we do now.
First, we can simply increase the global rate of death and reproduction,
by requiring several -- even {\em all} -- nodes
to replace their history at each time-step
with a randomly-chosen node's prior history.
TLC's lock-step notion of logical time facilitates this process.
At each step $s$ each node proposes and announces a new block,
then at $s+1$ each node chooses {\em any} node's step $s$ block at random
to build on in its step $s+1$'s proposal.
Thus, each node's proposal will survive even just one round
only if some node (any node) happens to choose it to build on.

The second, more powerful way we can accelerate the process --
and even achieve ``near-instant'' genetic consensus --
is by using randomness also to break the symmetry
of each proposal's ``genetic fitness'' or likeliness to reproduce.
At each TLC time-step $s$, 
each node announces not only its newly-proposed block,
but also chooses and attaches to its proposal
a random numeric lottery ticket,
which will represent the proposal's ``genetic fitness'' relative to others.
These lottery tickets may be chosen from essentially any distribution,
provided all nodes  correctly choose them at random
from the same distribution:
\eg, real numbers between 0 and 1 will work fine.

By TLC's progress rules,
each node must have collected a threshold number of proposals from step $s$
as its condition to progress to step $s+1$.
Instead of picking an arbitrary one of these proposals
to build on in the next consensus round starting at $s+1$,
each node $i$ must now choose the step $s$ proposal
with the {\em highest-numbered lottery ticket} that $i$ knows about:
\ie, the most ``genetically fit'' or ``attractive'' proposal it sees.
Step $s$ proposals with higher fitness
will be much more likely to ``survive'' and be built upon in subsequent rounds,
while proposals with lower fitness usually disappear immediately
because no one chooses to build on them in subsequent rounds.

\begin{figure}
\begin{center}
\includegraphics[width=1\columnwidth]{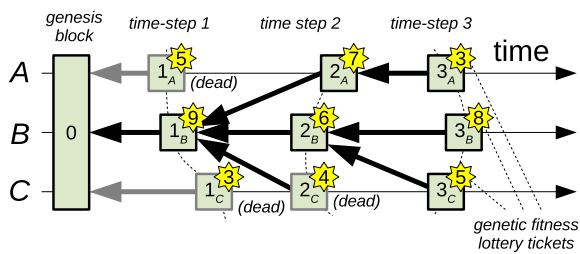}
\end{center}
\caption{Illustration of Strawman 2,
in which each node's proposal at each time step
includes a random genetic fitness.
At the next step, 
each node chooses the most ``fit'' proposal from the previous step
to build on.
}
\label{fig:cons-fitness}
\end{figure}

Figure~\ref{fig:cons-fitness} illustrates this process.
At step 2, all three nodes see and thus choose
node $B$'s ``maximally fit'' proposal
from step 1 to adopt and build on,
thereby achieving instant commitment globally.
At step 3, however, nodes $B$ and $C$ choose
the second-most-fit proposal by $B$,
because $A$'s globally-winning proposal was unfortunately not among those
that $B$ or $C$ collected in progressing to TLC step 3.
With global knowledge, at step 3 we can be certain
that all transactions up through block $1_B$ are committed,
but we remain uncertain whether blocks $2_A$ or $2_B$
will eventually win since both still survive at step 3.

If all nodes correctly follow this process,
then we reduce the number of {\em possible} blockchains
effectively surviving and emerging from any time period
from $n$ down to $f+1$.
This is because when any node $i$ reaches step $s+1$,
there are at most $f$ proposals it might have missed seeing
upon meeting the threshold condition to reach $s+1$,
and hence at most $f$ proposals might have had a better fitness
than the best proposal $i$ saw and picked.
While reducing the possibility space from $n$ possible histories to $f+1$
represents an improvement, it is still far from our goal of course --
but we are moving in the right direction.

\subsection{Strawman 3: a contest of celebrities}
\label{sec:cons:celebrity}

While we have accelerated genetic consensus
and reduced the number of possible histories that can survive at each step,
we still face the problem that no one can be certain
whether consensus has actually been achieved
without seeing {\em all} nodes' choices at each time-step.
If any node, or one of their clients,
tried to collect this information globally,
it might hang waiting to hear from one last long-delayed or failed node,
defeating the high-availability goal of threshold coordination.
It thus appears we can never discern consensus with certainty.

In Figure~\ref{fig:cons-fitness}, for example,
node $B$ may be unable to distinguish between
the ``consensus'' situation at step 2
and the ``lack of consensus'' situation at step 3,
if $B$ has seen only $C$'s step 2 decision and not $A$'s
upon reaching step 3.
$B$ cannot just wait to hear from $A$ as well
without compromising availability,
but $B$ also cannot exclude the risk
that a higher-fitness ``minority opinion'' such as block $2_A$ might exist
and eventually win over those $B$ knew about.

This ``minority report'' problem suggests an appealing solution:
let us restrict the genetic competition at each step only to
{\em celebrity proposals},
or those that a majority of nodes have heard of
by the next time-step when it is time to pick winners.
By having each node choose the most fit only among celebrity proposals,
we hope to prevent an unknown, high-fitness, ``dark horse''
from later ``spoiling'' what might appear to be consensus.
This attempt will fail,
but in a useful way that moves us toward a solution.

TLC's threshold witnessing process in each round
conveniently provides information useful to identify celebrity proposals.
We will say that participant $i$ {\em confirms} proposal $p$
as a celebrity proposal
if $p$ was among the set of threshold-witnessed messages
$i$ used to advance its logical clock to step $s+1$.
Since each participant must collect a threshold number
of threshold-witnessed messages from step $s$ 
in order to transition to step $s+1$,
each node automatically confirms a majority of proposals by $s+1$.

We now require that each participant
choose its best {\em confirmed} proposal,
having the highest-numbered lottery ticket,
as its ``preferred'' step $s$ proposal to build on at step $s+1$.
Step $s$ proposals not in node $i$'s threshold witnessed set --
\ie, the at most $f$ proposals that $i$ did {\em not} wait to be confirmed
before $i$ moved to $s+1$ --
are thus not eligible from $i$'s perspective to build on at $s+1$.

With this added rule,
each proposal from step $s$ that survives to be built on at $s+1$,
is, by protocol construction,
a proposal that {\em most} of the participants (all but at most $f$)
have seen by step $s+1$.
Intuitively, this should increase the chance that ``most'' nodes at $s+1$
will choose and build on the same ``lottery winner'' from step $s$.
This rule still leaves uncertainty, however,
since different participants
might have seen different subsets of confirmed proposals from step $s$,
and not all of them might have seen
the eligible proposal with the globally winning ticket.

\subsection{Strawman 4: seeking universal celebrity}
\label{sec:cons:universal}

To address this lingering challenge,
it would seem useful if we could be certain that
not just a majority of nodes, but {\em all} nodes,
know about any proposal we might see as a candidate for achieving consensus.
Further refining the above celebrity approach, in fact,
we can ensure that celebrity proposals known to a majority of nodes
reach {\em universal celebrity} status --
becoming universally known to {\em all} participants --
simply by ``biding our time'' for a second TLC time-step
during each consensus round.

Recall from Section~\ref{sec:tlc:broadcast}
that with majority thresholds,
any message $m$ that is broadcast at time-step $s$
and is threshold-witnessed by step $s+1$
will have propagated to {\em all} nodes by step $s+2$.
This is because the set $S$ of nodes that witnessed $m$ by step $s+1$
must overlap by at least one node with the set $S'$ of nodes
whose step $s+1$ messages any node must collect in order to reach step $s+2$.

Motivated by this observation,
we now modify the consensus process
so that each round requires two TLC time-steps instead of one.
That is, each consensus round $r$ will start at step $s=2r$,
and will finish at step $s+2$,
the same logical time that consensus round $r+1$ starts.

At step $s$, each node proposes a block as before,
but waits until step $s+2$ to choose a step $s$ proposal
to build on in the next consensus round.
Because the universal broadcast property above holds only for messages
that were witnessed by a majority of nodes by step $s+1$,
we must still restrict each node's choice of proposals at step $s+2$
to those that had achieved majority celebrity status by step $s+1$.
Among these, each node as usual chooses the eligible proposal
from step $s$ with the highest lottery ticket.

\begin{figure}
\begin{center}
\includegraphics[width=1\columnwidth]{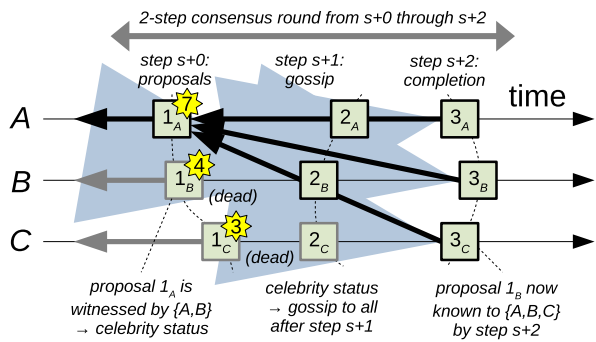}
\end{center}
\caption{Illustration of Strawman 4,
in which knowledge of winning proposal $1_A$ propagates to all nodes
in two steps after being threshold witnessed by step $s+1$.}
\label{fig:cons-celebrity}
\end{figure}

By slowing down consensus,
we ensure the promising property that whichever step $s$ proposal $p$
a node might choose for the next round at $s+2$,
{\em all} nodes know about proposal $p$ by step $s+2$.
Figure~\ref{fig:cons-celebrity} illustrates this process
in a scenario in which $A$'s proposal at step $s+0$
is threshold witnessed by nodes $\{A,B\}$ by step $s+1$
to achieve celebrity status,
then as a result propagates to all nodes $\{A,B,C\}$ by $s+2$.

Are we done?
Unfortunately not.
As discussed earlier in Section~\ref{sec:tlc:broadcast},
the fact that all nodes know the {\em existence} of $p$ by step $s+2$
does not imply that all nodes will know
the crucial fact that $p$ {\em was threshold witnessed},
or thus have {\em confirmed} $p$ as having celebrity status by $s+1$.

Due to message timing, different nodes may reach steps $s+1$ and $s+2$
on the basis of different subsets of threshold-witnessed messages.
For example,
one node $i$ might see that proposal $p$ was threshold-witnessed by step $s+1$,
and eventually choose it as the best eligible proposal by $s+2$.
Another node $j$, in contrast,
might have reached step $s+1$ on the basis of a different
set of witnessed messages than $i$ used.
If proposal $p$ isn't in $j$'s threshold-witnessed set by $s+1$,
$j$ cannot ``wait around'' to see if $p$
eventually becomes fully threshold-witnessed
without compromising $j$'s availability, so $j$ must move on.

\begin{figure}
\begin{center}
\includegraphics[width=1\columnwidth]{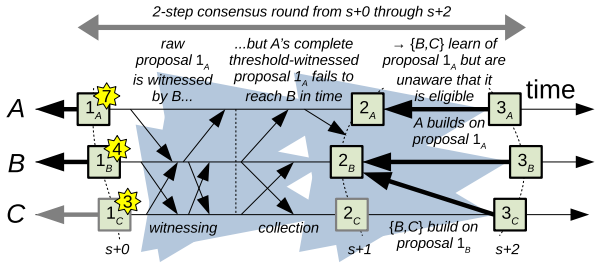}
\end{center}
\caption{Illustration of Strawman 4's failure mode,
where knowledge of proposal $1_A$ propagates to all nodes by $s+2$
but only node $A$ learns that $1_A$ was threshold witnessed.
}
\label{fig:cons-celebrity-fail}
\end{figure}

In this case,
$j$ will definitely learn the {\em existence} of proposal $p$ by step $s+2$,
from at least one of the majority set of nodes that witnessed $p$ by $s+1$.
But this fact tells $j$ only that {\em at least one node} witnessed $p$,
not that a {\em majority} of nodes witnessed $p$ by $s+1$,
as required for $j$ to confirm $p$ as eligible for the next round to build on.
In this situation, nodes $i$ and $j$ may pick different eligible proposals
to build on in the next round,
and neither $i$ nor $j$ has any readily-apparent way to distinguish
this consensus failure situation from one in which all nodes
fortuitously {\em do} choose the same best eligible proposal.
Figure~\ref{fig:cons-celebrity-fail} illustrates such a failure case,
where the globally best proposal $1_A$ is threshold witnessed by $s+1$
but only node $A$ actually learns by then that proposal $1_A$ is eligible.

\subsection{Strawman 5: enter the paparazzi}
\label{sec:cons:paparazzi}

Is there some way a node can tell not only
that a proposal $p$ has reached celebrity status by $s+1$
and thus that $p$'s existence will be known to all nodes by $s+2$,
but additionally that the {\em fact} of $p$'s celebrity status
will also become known to all nodes?
We can, by a second application of the same two-step broadcast principle,
merely shifted one time-step later.
Suppose a node $j$ confirms $p$'s celebrity status at step $s+1$,
then successfully ``gossips'' that fact to a majority of nodes by $s+2$.
Then not only the existence of $p$
but also $j$'s {\em confirmation} of $p$'s celebrity status
will subsequently become known to all nodes by $s+3$.

We therefore extend each consensus round to take three TLC time-steps,
so that round $r$ starts at step $s=3r$ and ends at $s+3$.
In addition, we will try to strengthen the eligibility criteria for proposals
to ensure that both the existence and the celebrity status
of any chosen proposal becomes known to {\em all} nodes by $s+3$.
In particular, for any node $i$ to consider a proposal $p$ broadcast at $s$
to be eligible for the consensus round's genetic lottery,
$i$ must see that:
(a) some node $j$, who we'll call the {\em paparazzi},
observed and reported $p$'s celebrity status in $j$'s broadcast at step $s+1$;
and
(b) that $j$'s broadcast at $s+1$ was in turn threshold witnessed
by a majority of nodes by step $s+2$.

\begin{figure}
\begin{center}
\includegraphics[width=1\columnwidth]{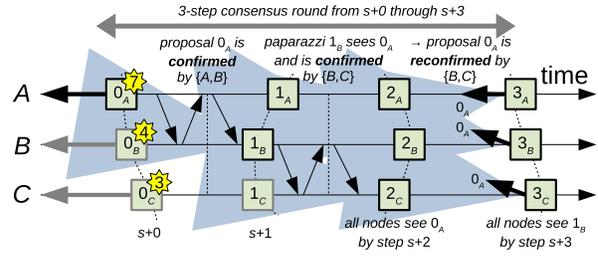}
\end{center}
\caption{Illustration of Strawman 5,
in which paparazzi node $B$ confirms proposal $0_A$ at $s+1$,
then gossips its confirmation to $\{B,C\}$ by $s+2$
and to all nodes by $s+3$.
}
\label{fig:cons-paparazzi}
\end{figure}

For brevity, 
we will say that the paparazzi node $j$ first {\em confirms}
proposal $p$'s celebrity status at step $s+1$,
then $i$ in turn {\em confirms} $j$'s step $s+1$ broadcast in the same way.
When such a ``double-confirmation'' linked by paparazzi node $j$ occurs,
we say that node $i$ {\em reconfirms} proposal $p$.
Node $j$'s confirmation of $p$ at $s+1$ ensures that all nodes
will know the existence of $p$ by $s+2$,
and $i$'s reconfirmation of $p$ at $s+2$ in turn ensures that all nodes
will know of $j$'s confirmation of $p$ by $s+3$.
Figure~\ref{fig:cons-paparazzi} illustrates this process,
with node $B$ acting as paparazzi for $A$'s proposal $0_A$
in an example 3-step consensus round.


Are we done yet?
Unfortunately we've merely kicked the can down the road.
If node $i$ reconfirms $p$ by step $s+2$,
this implies that all nodes will know by $s+3$ that $p$ was confirmed,
but it does not imply that other nodes will have {\em reconfirmed} $p$.
If {\em reconfirmation} and not just {\em confirmation}
is $p$'s new eligibility condition,
then we must account for the fact that we have moved the goalposts.
By the end of the round at $s+3$,
different nodes may still disagree on whether $p$ was {\em reconfirmed}
and hence (still) eligible for the genetic lottery,
once again allowing disagreement about the consensus round's result
in the end.

We could try playing the status gossip and confirmation game yet again,
making triple-confirmation the proposal eligibility condition,
but this approach just leads us in circles.
A proposal's triple-confirmed status will ensure that all nodes
know by $s+4$ that it was double-confirmed,
but will still leave disagreement on whether it was triple-confirmed.
We must therefore try something else:
it is hard to win a game of counting to infinity.

\subsection{Strawman 6: gazing into the crystal ball}
\label{sec:cons:predict}

Since we would appear to need an infinite amount of time
to get ``complete'' information about a consensus round,
let us instead make the best we can of incomplete information.
We will therefore return to using only (single) confirmation
as the eligibility criterion for a proposal to enter the genetic lottery.
We will then use (double) reconfirmation
to give us an unreliable ``crystal ball''
that {\em sometimes} -- when we're lucky --
enables {\em some} nodes to predict
when all other nodes will {\em just happen to}
converge and agree on the same ``best eligible proposal'' during the round.

Our crystal ball will sometimes be clear, allowing a precise prediction,
and sometimes cloudy, offering no useful information.
Further, the crystal ball may appear clear from the perspective of some nodes,
and cloudy to others, even in the same consensus round.
The crystal ball's subjective fickleness may therefore leave
only some nodes aware when consensus succeeds,
while other nodes must wait until future rounds
to learn this fact in retrospect.

\com{
For this reason,
we will give up on the goal of propagating complete information
or making a perfect decision in one consensus round,
and instead find a way to make progress with limited information
only {\em sometimes},
and allow the possibility that some consensus rounds may at other times ``fail''
to make a useful decision at all.

In particular, we now set the more limited goal of identifying some conditions
under which a node can predict during a consensus round
that {\em all} nodes will ``just happen to'' come into perfect agreement
and choose the same time proposal to build on in the next round.
Only some nodes will be able to make this prediction, only some of the time.
There may be some consensus rounds in which {\em no} node
can make such an agreement prediction:
such rounds will produce multiple conflicting proposals,
much like temporary forks in Bitcoin,
which we merely hope and expect
that future consensus rounds will resolve with statistically high probability.
And even in rounds in which some node successfully predicts agreement,
{\em other} nodes might be unaware that this agreement has happened,
and will simply have to assume conservatively that they {\em might} be seeing
only one branch of a fork until they hopefully achieve certainty
during some future round.
}

Since all nodes are again using (single) confirmation 
as their criterion for a proposal $p$'s eligibility,
this implies that no node will choose $p$ in this round
unless at least the {\em existence} of proposal $p$
(though not necessarily its celebrity status)
has become known to all nodes by step $s+2$.
Now suppose that some node $i$ happens to notice that $p$ is not just confirmed
but is in fact reconfirmed (double-confirmed) by step $s+2$.
This does not tell $i$ that other nodes will also reconfirm $p$,
but it {\em does} tell $i$ that all other nodes will have at least
(singly) confirmed $p$ by step $s+3$.
Thus, node $i$ knows -- even if other nodes don't --
that {\em all} nodes will realize by $s+3$ that $p$ is eligible.

Finally, suppose that by step $s+3$, node $i$ is also not aware of
the existence of any other proposal $p'$,
confirmed or not,
having a lottery ticket competitive with that of $p$
(\ie, a value greater than or equal to $p$'s ticket).
Since any such competitive proposal $p'$
cannot become eligible, or be confirmed by {\em any} node,
without at least $p'$'s existence becoming known to {\em all} nodes by $s+2$,
the fact that $i$ has not seen any sign of a competitive proposal
implies that there can be no {\em eligible competitor} to $p$.
There could be proposal $p'$ with a competitive ticket value
that $i$ didn't see,
but to be ``hidden'' from $i$ in this fashion,
 $p'$ must have been seen by only a minority of nodes,
and thus cannot be eligible and cannot have been confirmed by anyone.

Since $i$ now knows from $p$'s reconfirmation
that {\em all} nodes will know and have confirmed $p$ by $s+3$,
and no other eligible proposal competitive with $p$ exists
that {\em any} node could confirm to spoil $p$'s victory,
this means that $i$ has successfully ``gazed into the crystal ball''
and predicted this round's inevitable convergence on $p$.
Node $i$ can predict with certainty that {\em all} nodes will choose $p$
as their best eligible proposal to build in the next round,
even though these other nodes themselves may not be aware of this convergence.
Since all future histories must now build on $p$,
$i$ can consider all transactions in $p$ and all prior blocks that $p$ built on
to be permanently committed.

Since other nodes may not obtain the same information as $i$ in this round,
other nodes may see the crystal ball as cloudy,
and be forced to assume conservatively that consensus may have failed,
and that different nodes might pick different best eligible proposals.
These other nodes will {\em eventually} learn,
in some future round in which they successfully use the crystal ball,
that $p$ has been committed as a prefix to some longer history
that has by then been built atop proposal $p$.
The fact that only some nodes (or even no nodes)
might actually know in this round that all nodes have converged on $p$
does not change the inevitably -- or ``fate'' --
that all future history will build on $p$.

Some consensus rounds may also genuinely fail to converge,
in which case different nodes see and choose different proposals
as the best eligible.
In this case, {\em all} nodes will perceive the crystal ball as cloudy.
Some nodes might fail to discern the eligibility status
of the best (globally) eligible proposal,
instead seeing that proposal as a ``spoiler'' competitive with
some next-best proposal that they {\em do} confirm as eligible.
Other nodes might confirm the best proposal as eligible,
but fail to reconfirm it
because knowledge of the proposal's eligibility failed to propagate
to a majority of nodes,
making the proposal's reconfirmation impossible.
In any case, any consensus round that fails to converge
can still safely yield multiple ``competing'' proposals,
as in earlier cases above,
to be left for resolution by a more-fortunate future consensus round.


\subsection{Something wicked this way routes}
\label{sec:cons:adversary}

Having resigned ourselves to the possibility
that only some consensus rounds may succeed,
and that only some nodes may even realize that a round succeeded,
we would like to know whether and how often 
we can actually anticipate this desirable outcome.
If the network is truly adversarial, however,
choosing message delays and delivery orders intelligently
to prevent consensus from succeeding, for example,
then we still appear to have a problem.

If the adversary can see the lottery ticket for each proposal $p$
as soon as $p$ is broadcast at a consensus round's start time $s+0$,
the adversary can arrange message delivery order
so that no consensus round ever succeeds.
For example, the adversary can first collect all proposals from step $s$
along with their lottery tickets,
then arrange for the proposals with the three highest-valued
lottery tickets each to be witnessed by only a third of the $n$ nodes each,
ensuring that none of these proposals propagate to a majority of nodes 
by step $s+1$ to become eligible.
Any other proposal that any node might confirm as eligible
will always be ``spoiled''
by one of the best three (always-ineligible) proposals,
preventing convergence and keeping all nodes' crystal balls permanently cloudy.

We could just assume that the network schedules message deliveries
arbitrarily but obliviously to the values
computed and used in distributed protocols,
as in {\em oblivious} scheduler
models~\cite{aumann96efficient,aumann05efficient,aspnes03randomized,friedman05simple,aspnes15faster}.
In today's open Internet, however,
the threat of intelligent disruption from network adversaries
is unfortunately all too realistic.

Fortunately, we have a conceptually simple way to ensure
that the adversary cannot interfere with consensus in this fashion.
We simply ensure that the adversary {\em cannot know}
the lottery tickets associated with each proposal until later,
after the consensus round has completed
and the adversary has already ``committed'' its decisions
on network delays and ordering.
In effect, if we force the network adversary to ``play its hand'' first,
by forwarding enough messages to allow the consensus round to complete
{\em before} the adversary can learn any lottery ticket values,
then we can ensure by design that the adversary's decisions
are independent of the lottery tickets --
exactly as if network ordering was arbitrary but oblivious.

How can we ensure that an adversarial network
does not learn the proposals' lottery ticket values
before the consensus round completes?
In the present non-Byzantine case
in which we assume all nodes are correct,
we can rely on them not to leak information to the network adversary directly.
We therefore need only to ensure that ticket values
do not leak to the network adversary while in transit over the network,
which we can accomplish simply by encrypting the lottery ticket values --
or better yet in practice, all inter-node communication --
using a standard pairwise encryption protocol such as TLS~\cite{rfc8446}.
This approach obviously fails as soon as there is even one Byzantine node
that might leak the lottery ticket values to the adversary;
we address this problem later in Section~\ref{sec:tlc:byz}
using Shamir secret
sharing~\cite{shamir79share,stadler96publicly,schoenmakers99simple}.
For now, however, we simply assume that the lottery ticket values
are kept out of the adversary's knowledge ``somehow''
until the consensus round is over,
so that we can assume that they are independent
of network delays and ordering considerations.

\subsection{Calculating the odds of success}
\label{sec:cons:odds}

Given that lottery ticket values
are independent of network scheduling,
we can now analyze the probability that any particular node $i$
will ``get lucky'' and observe a consensus round to succeed.
This occurs only when all nodes converge on the same proposal $p$,
{\em and} node $i$ in particular is able to detect this convergence
by reconfirming (double-confirming) proposal $p$.
We focus this analysis purely on what a particular node $i$ observes,
because we merely want to ensure
that {\em each} node observes success ``often enough''
regardless of any other node's experience.

For simplicity,
we will conservatively focus our analysis on the event
that $i$ observes the {\em globally highest-numbered} proposal $p$ to commit.
This situation is sufficient, but not necessary, for $i$ to observe success.
Network scheduling could cause all nodes to converge on
a proposal other than the global best,
and cause $i$ to witness this as successful commitment,
if any other higher-numbered proposals do not become eligible
and fail to arrive at $i$ to ``spoil'' its view.
But this (likely rare) event can only improve $i$'s observed success rate,
so we ignore it and focus only on commitments of the globally-best proposal.

Recall the two key conditions above for $i$ to see
in its ``crystal ball'' that proposal $p$ has been committed:
(a) that $i$ has reconfirmed $p$, and
(b) that $i$ has seen no other proposal $p'$ from step $s$, confirmed or not,
with a lottery ticket value competitive with $p$'s.
By our assumption that $p$ is the globally-best proposal,
(b) cannot happen since no proposal globally better than $p$ exists.
We also assume here that lottery tickets have enough entropy
that the chance of a tie is negligible,
but accounting for ties of known probability
requires only a minor adjustment to the analysis.

We therefore care only about the probability that $i$ reconfirms $p$:
\ie,
that some paparazzi node $j$ confirms $p$ at step $s+1$
and $i$ subsequently confirms $j$'s step $s+1$ confirmation of $p$.
Recall that $i$ had to collect threshold-witnessed messages
from a majority of nodes to reach step $s+2$.
If any one of these nodes $j$ has confirmed $p$ by $s+1$,
then $i$ will subsequently confirm $j$'s confirmation and hence reconfirm $p$.
The probability that {\em at least one} of these potential paparazzi
confirms $p$ is no less than the probability that {\em any particular one} does,
so we can again focus conservatively on some particular node $j$.

Node $j$, in turn,
had to collect threshold-witnessed proposals
from a majority of nodes in order to reach step $s+1$.
If any one of these proposals
is the proposal $p$ with the globally highest ticket,
then $j$ will certainly confirm $p$ at $s+1$.
Since each of the $n$ nodes' proposals have a $1/n$ chance
of being the globally highest,
and this is unpredictable to the network adversary,
the chance of node $i$ observing any given round to succeed is at least $1/2$.

Although the probabilities that different nodes in the {\em same round}
observe success are interdependent in complex ways,
the probabilities of observing success {\em across successive rounds}
is independent because each round uses fresh lottery tickets.
The success rate any node observes therefore
follows the binomial distribution across multiple rounds.
The probability that a node fails to observe a successful commitment
in $k$ consecutive time steps is less than $1/2^{k}$,
diminishing exponentially as $k$ increases.

\subsection{Summary: whatever will be, will be}
\label{sec:cons:solution}

In summary,
we have defined a simple randomized consensus protocol
atop majority-witnessed TLC.
In each consensus round $r$ starting at TLC time step $s = 3r$,
each node $i$ simply proposes a block with a random lottery ticket,
waits three TLC time-steps,
then uses the communication history that TLC records and gossips
to determine the round's outcome from any node's perspective.

In particular,
each node $i$ always chooses a {\em best confirmed proposal} from round $r$
to build on in the next round $r+1$.
Node $i$ {\em confirms} a proposal $p$ sent in step $s+0$
if $i$ can determine that $p$ was threshold-witnessed
by a majority of nodes by step $s+1$.
A best confirmed proposal for $i$ is any round $r$ proposal $i$ has confirmed
whose lottery ticket is greater than or equal to that of
any other proposal $i$ has confirmed in this round.

In addition,
node $i$ decides that the consensus round
has successfully and permanently committed proposal $p$
if all of the following three conditions hold:

\begin{itemize}
\item	Node $i$ obtains a step $s+1$ message $m$, from some node $j$,
	that $i$ can confirm was threshold-witnessed
	by a majority of nodes by $s+2$;
\item	Node $j$'s message $m$ at $s+1$ recorded that
	proposal $p$ from step $s+0$ was threshold-witnessed
	by a majority of nodes by $s+1$; and
\item	No other step $s+0$ proposal $p' \ne p$
	that $i$ has become aware of by step $s+2$
	has a lottery ticket greater than or equal to that of $p$.
\end{itemize}

Each node $i$ will observe successful consensus in this fashion
with an probability of at least $1/2$ in each round,
independently of other rounds.
Any round that $i$ sees as successful
permanently commits both proposal $p$
and any prior uncommitted blocks that $p$ built on.
Thus, the probability $i$
has not yet finalized a unique proposal for round $r$
by a later round $r+k$ for $k \ge 0$
is at most $1/2^{k}$.


\subsection{Optimizing performance: pipelining}

For simplicity we have described QSC with rounds running sequentially,
each round $r$ starting at TLC time-step $3r$ and ending at step $3r+3$.
A simple optimization, however,
is to pipeline QSC consensus rounds
so that a round starts on {\em every} time-step and overlaps with other rounds.
With pipelining, each consensus round $r$
starts at step $r$ and ends at step $r+3$.
In this way,
we can smooth the communication and computation workload on nodes
at each timestep,
minimize the time clients submitting transactions have to wait
for the start of the next consensus round,
and reduce the average time clients must wait for a transaction to commit,
since commitment occurs with constant probability for each completed round
and pipelining triples the rate at which rounds complete.

One apparent technical challenge with pipelining
is that at the start of round $r+1$
(step $r+1$),
when each node broadcasts its proposal,
we might expect this proposal to include a new block in a blockchain.
To produce a blockchain's tamper-evident
log structure~\cite{schneier99secure,crosby09efficient}, however,
each block must contain a cryptographic hash of the previous block.
But the content of the previous block is not and cannot be known
until the prior consensus round $r$ ends at step $r+3$,
which due to pipelining
is two time-steps after step $r+1$,
when we appear to need it!

The solution to this challenge is to produce complete blocks,
including cryptographic back-links,
not at the start of each round but at the end.
At the start of round $r+1$ (step $r+1$),
each node broadcasts in its proposal only the lottery ticket
and the semantic content to be included in this block,
\eg, a batch of raw transactions that clients have asked to commit.
Only at the {\em end} of round $r+1$, at step $r+4$,
do nodes actually form a complete block based on this proposal.
All nodes, not just the original proposer,
can independently compute the block
produced by round $r+1$'s winning proposer,
deterministically based on the content of its step $r+1$ proposal
and the block it builds on from the previous round $r$,
which we now know because it was fully determined in step $r+3$.

A second-order challenge that this solution creates
is that in transactional systems,
the proposer of a block cannot necessarily know for sure at proposal time
that all of the transactions it is proposing will still be committable
by the completion of the consensus round.
For example, at the start of consensus round $r$,
a node $i$ might propose a batch of transactions
including the payment of a coin from Alice to Bob.
Alice might indeed own the coin to be spent
according to node $i$'s view of the blockchain at step $r$ --
but by the end of round $r$, at step $r+3$,
the coin might have already been spent in a conflicting transaction
appearing in the blocks $i$ is building on
from the rounds completing at steps $r+1$ and $r+2$.
The deterministic block-formation function that all nodes run
at the end of each round
can account for this risk simply by discarding
such transactions that have become uncommittable by the time they were proposed,
leaving them out of the block produced at step $r+3$
without blaming the block's proposer for an event it could not have forseen.

\com{ old likely-obsolete text:

\input{failstop}

\input{fat}

\input{thin}

}

%% file: failstop.tex
\subsection{Fail-stop Consensus}
\label{sec:failstop}

While there are many formulations of the consensus problem,
we adapt the classic one of Paxos~\cite{lamport98parttime,lamport01paxos}.
Any or all $n$ participants may serve as {\em proposers},
who propose {\em values} to agree on (\eg, possible blocks for a blockchain).
Our goal is to ensure that
(a) the protocol eventually {\em chooses} exactly one proposed value, and
(b) all participents eventually learn the chosen value and no other.

As before, we assume for now that nodes are never corrupted ($f_c=0$),
but only fail gracefully or their messages may be indefinitely delayed
($f_d \ge 0$).
We also assume for now that network delays are arbitrary but non-adversarial,
relaxing this assumption later in Section~\ref{sec:adversarial}.

\subsubsection{Leader Election by Random Lottery}

The FLP theorem~\cite{fischer85impossibility}
implies that consensus protocols
must sacrifice one of safety, liveness, asynchrony, or determinism.
Paxos and its leader-based derivatives sacrifice asynchrony
by relying on timeouts to ensure progress.
We instead sacrifice determinism
and use randomness~\cite{rabin83randomized,ben-or83another,bracha85asynchronous,canetti93fast,cachin05random,friedman05simple}.

As in recent blockchain consensus protocols such as
Bitcoin~\cite{nakamoto08bitcoin},
Algorand~\cite{gilad17algorand}, and
DFINITY~\cite{hanke18dfinity,abraham18dfinity},
each proposal will have an associated random ``lottery ticket.''
This lottery
limits the number of proposals in consideration at any time,
serving in lieu of leader election.

While Bitcoin also uses its proof-of-work lottery for pacing
to ensure winning proposals appear not too quickly,
we use TLC for pacing on the basis of network communication.
We build on acknowledged TLC configured with
standard thresholds for fail-stop consensus,
namely \tlc{f+1}{f+1}{n}, where $f_d=f$ and $f_c=0$.

Consensus proceeds in {\em rounds},
each round $r$ taking three logical time-steps,
starting at logical time $3r$ and ending at time $3r+3$.
At start time $\lts=3r$,
each of the $n$ participants includes one proposed value (\eg, block)
in the upper-level payload of its time $\lts$ broadcast.
These messages may be broadcast (and received by other nodes)
at arbitrary and widely-varying moments in real time, of course;
we require only that they occur at the same {\em logical} time $\lts$.
For now, each participant also chooses its proposal's lottery ticket
simply by picking a random number
and including it with the participant's time $\lts$ proposal --
but we will later revise this simplistic design choice.

Once a consensus round has started
and the participants have broadcast proposals at $\lts=3r$,
each participant simply waits until logical time $\lts+3$,
then evaluates the result of the consensus round independently
based on the TLC-driven communication that has transpired in the meantime.
The consensus protocol itself need not produce upper-layer messages
besides the time $\lts$ proposals:
the consensus-level payloads sent at times $\lts+1,\lts+2,\lts+3$ are empty,
while the messages at $\lts+3$ contain the proposals for the next round.
Apart from the proposals,
the nodes learn the information they need for consensus
from the tamper-evident logging and gossip mechanisms
that TLC inherits from the PeerReview framework
(Section~\ref{sec:byz}).

\subsubsection{Invariants and Observations of Consensus}

The consensus process must address two fundamental challenges.
First, depending on the communication patterns and proposal lottery tickets,
a given round may or may not succeed in making any definite choice.
Second, whatever a round's global outcome might be,
{\em any particular node} may finish the round
with an incomplete view of that outcome,
because each node waits to see only a threshold of messages
before proceeding to the next round.
Our goals are to ensure that
(a) each round has a reasonable chance of succeeding,
(b) when a round succeeds,
any given node has a reasonable chance of learning that it succeeded, and
(c) each node's behavior remains {\em safe},
\ie, consistent with whatever {\em might} have been decided,
despite each node's imperfect knowledge.

We define a proposal $p$ broadcast at time $\lts$ as {\em acceptable}
if $p$ has the strictly lowest-numbered lottery ticket
among all time $\lts$ proposals
that are seen by a majority of nodes by time $\lts+1$.
Acceptability implies that proposal $p$ is {\em safe} for this round to choose,
in that no other proposal $p' \ne p$ can also be acceptable in this round.
The fact that $p$ is acceptable means only that $p$ {\em might} be chosen,
and implies neither that $p$ {\em will} be chosen,
nor that any given node will learn during the round
that $p$ is acceptable.

A node $i$ {\em accepts} a time $\lts$ proposal $p$
if by time $\lts+1$,
$i$ can confirm that a majority of nodes have seen $p$,
and $i$ has seen {\em no other} proposal with a lottery ticket
less than or equal to that of $p$.
This rule ensures that node $i$ accepts $p$ only if $p$ is acceptable,
but $i$ might fail to accept an acceptable proposal
if $i$ cannot confirm that $f+1$ nodes have seen $p$,
or if $i$ sees another proposal $p' \ne p$ with a less-or-equal ticket
and cannot rule out the possibility that $f+1$ other nodes may have seen $p'$.
This less-or-equal comparison ensures safety
even if lottery ticket values
are insufficiently large or random to ensure uniqueness.
This provision is merely for defense-in-depth:
in practical systems we expect lottery tickets should always
be large (\eg, 256 bits or more) and cryptographically strong.

We say that a time $\lts$ proposal $p$ is {\em chosen}
if a majority of nodes accept it at time $\lts+1$.  
\xxx{ probably 2 actually...  Not done, to be finished,
and the older text below is inconsistent... }

\subsubsection{Proposal Eligibility and Influential Nodes}

We use the lottery tickets associated with each proposal
to help the nodes converge toward a proposal choice:
in essence, the lowest-numbered lottery ticket in a round ``wins.''
The first problem, however,
is that the globally-winning proposal
might be seen by $f$ or fewer nodes,
leaving other groups of nodes believing
that other proposals with higher-numbered tickets are the winner.

To address this risk and ensure safety, therefore,
we consider a proposal to be {\em eligible}
only if it has been observed by at least $f+1$ nodes by time $\lts+1$.
This way, any two eligible proposals must have been seen by
at least one node in common by $\lts+1$,
and that node will ensure that only one of those proposals can win.
But how does a particular node determine that a proposal is eligible?
This is where TLC acknowledgments become useful.

We say that a node $i$ is {\em influential} at time $\lts$
in the view of node $j$ at time $\lts'>\lts$
if $j$ can determine,
based on the causal history $j$ has collected by time $\lts'$,
that at least $\ta$ nodes acknowledged $i$'s time $\lts$ message.
If $j$ itself saw and collected at least $\ta$ acknowledgments
for $i$'s time $\lts$ message,
\eg, as part of the evidence $j$ collected to progress to time $\lts+1$,
then this condition is satisfied.
Since $j$ may also have received indirectly via gossip
message and acknowledgment sets
that {\em other} nodes used to progress from $\lts$ to $\lts+1$,
however,
all unique acknowledgments of $i$'s message at $\lts$
that $j$ collected by $\lts'$ count towards the threshold.

Now we have a criterion by which nodes
can conservatively determine a proposal to be eligible.
If $p$ is a proposal broadcast by node $i$ at time $\lts$,
and $i$ is influential at time $\lts$
in the view of another node $j$ at time $\lts+1$,
then this is sufficient for $j$ to confirm that $p$ is eligible.
This means, in effect,
that $j$ has collected by $\lts+1$ a threshold of ``evidence'' --
in the form of TLC acknowledgments --
to confirm that $p$ was seen by at least $\ta>f$ nodes by $\lts+1$.
Thus, $j$ at least can determine that $p$ appears safe to commit
if $p$ is also the proposal with the lowest-numbered ticket $j$ has seen.

\subsubsection{Proposal Commitment}

If only one node knows that a proposal is eligible
and tries to ``commit'' it unilaterally, however,
then that node might fail or its messages be indefinitely delayed,
causing its knowledge of the commitment to be lost
as the majority proceeds without it.
As with other ``two-phase'' consensus protocols
like PBFT~\cite{castro99practical}, therefore,
after a threshold of nodes has acknowledged a proposal
to confirm it to be safe to commit,
a threshold of nodes must subsequently
{\em learn that it is safe to commit}
and promise to record that fact permanently.
We do not do anything special to make this desirable event occur,
but instead merely enable nodes to identify when it has fortuitously happened.

We again use the causal histories TLC collects,
and the notion of influential nodes,
for this purpose.
Let $i$ be any node that has reached time $\lts+3$,
let $p$ be the time $\lts$ proposal with the lowest-numbered ticket
appearing in $i$'s causal history by $\lts+3$,
and let $i$ be the node that sent $p$ at $\lts$.
A node $j$ is a {\em key node} for $i$ 
if $i$'s causal history by $\lts+3$ if $i$ can confirm that:
(1) $i$ was influential at time $\lts$ in the view of $j$ at time $\lts+1$, and
(2) $j$ was influential at time $\lts+1$ in the view of $i$ at time $\lts+2$.
The first condition implies that key node $j$
was able to determine at $\lts+1$ that $p$ was eligible.
The second condition implies that $i$ can confirm
that $j$'s knowledge of $p$'s eligibility
propagated to a majority $\ta$ of nodes by time $\lts+2$.

If $i$ can identify any key node in its causal history by $\lts+3$,
then $i$ knows not only that the winning proposal is safe to commit,
but also that a majority set of nodes $S$ learned that fact by $\lts+2$.
Since any node must obtain a time $\lts+2$ message from at least one node in $S$
as a condition for reaching time $\lts+3$,
this further implies that {\em all} nodes will know about $p$'s commitment
by the time they reach logical time $\lts+3$.
Thus, if $i$ sees these conditions to be satisfied,
$i$ knows that $p$ is committed --
and that all {\em other} nodes will also know this fact by $\lts+3$.

\subsubsection{Probability of Successful Commitment}

Randomized consensus would not be useful in practice
if each consensus round has a vanishing chance of succeeding,
a problem typical of ``private-coin'' consensus
protocols~\cite{ben-or83another,bracha84asynchronous,bracha85asynchronous}.
The above protocol based on TLC, fortunately,
ensures that each consensus round succeeds with probability at least $1/2$ --
given the temporary assumption stated above
that the network schedules message deliveries
arbitrarily but non-adversarially,
by which we specifically mean {\em independently}
of the lottery ticket values the nodes assign to their proposals at time $\lts$.

Let us focus on any particular node $i$
and the probability that $i$ observes
a particular consensus round starting at $\lts$ and ending at $\lts+3$ to succeed.
By time $\lts+2$,
$i$ knows a set $S$ of at least $\tm$ nodes that are influential at $\lts+1$
in the view of $i$ at $\lts+2$:
namely, the $\tm$ nodes for whose time $\lts+1$ messages
$i$ collected $\ta$ acknowledgments each in order to advance to $\lts+2$.
(If $i$ reached $\lts+2$ virally based on a message from another node $j$
that had already reached $\lts+2$,
then the message-and-achknowledgment matrix $i$ receives from $j$
serves in lieu of the matrix $i$ would have built itself.)
Each of these $\tm$ nodes in $S$ is a ``candidate'' to become a key node.

The chance that {\em at least one} of these $\tm$ nodes in $S$
actually becomes a key node is at least as great
as the chance that {\em any particular one} of these nodes becomes a key node,
regardless of these events' independence or lack thereof.
We therefore restrict focus conservatively on
any particular one of these nodes $j \in S$.
We next examine the probability that $j$ satisfies
the other condition on becoming a key node:
namely, that whichever node $i$ broadcasts the winning proposal at time $\lts$
is influential in the view of $j$ at time $\lts+1$.
By the same counting logic as above,
there is a set $S'$ of at least $\tm$ nodes that are influential at $\lts$
in the view of $j$ at $\lts+1$.
Because of our simplifying assumption above that the network adversary
schedules message delivery independently of proposals' lottery tickets,
as well as the assumption that all nodes behave honestly ($f_c=0$)
and pick their tickets from the same distribution,
each of the influential proposers in $S'$
has a uniform probability of $1/n$ of receiving the globally winning ticket.

Since $\tm > n/2$, therefore,
there is at least a $1/2$ probability that some node $i \in S'$ wins,
ensuring in turn that $j$ becomes a key node
in $i$'s view at time $\lts+2$,
and hence that $i$ observes the consensus round to succeed.
And if $i$ observes the consensus round to succeed by $\lts+3$,
then by extension {\em all} nodes observe this by $\lts+3$,
because of the guaranteed propagation of $j$'s knowledge at $\lts+1$ to
at least $f+1$ nodes at $\lts+2$ and in turn to all nodes at $\lts+3$.

\subsubsection{Handling Adversarial Network Scheduling}
\label{sec:adversarial}

We now revisit the simplifying assumption
that the scheduling of message delivery is non-adversarial.
In particular, what if the network adversary can and does
``peek into'' the proposals broadcast at time $\lts$
and use that information to schedule messages from them until $\lts+3$?
This is a realistic attack in practice
given the pervasive reality of ``deep packet inspection'' (DPI) technologies.

Unfortunately this attack is immediately devastating
to the probabilistic liveness of the above consensus protocol,
allowing the adversary to reduce the chance of each round's success
not just to ``vanishingly small'' but literally to zero.
The adversary merely needs to observe, for example,
which three time $\lts$ proposals $p_1,p_2,p_3$
contain the three lowest-valued lottery tickets,
and schedule message delivery so that each of them
is seen by disjoint sets of $f$ or fewer nodes at $\lts+1$,
thereby ensuring 100\% reliably that no winning proposal will ever be eligible.

While we retain our assumption that all nodes are honest ($f_c=0$),
there is an easy defense to this attack.
We simply ensure that all communication between nodes is encrypted
(\eg, via TLS~\cite{rfc8446}),
thereby denying the network adversary
any information about the lottery ticket values
before the consensus round completes at time $\lts+3$.
Simply encrypting the network links
will obviously be an inadequate defense
as soon as there is even one corrupt node ($f_c>0$),
which can leak all the ticket values it observes to the adversary,
necessitating the use of threshold randomness
in the Byzantine consensus variants discussed below.
Nevertheless, the fundamental gist of the solution --
denying the adversary any {\em information} about the lottery tickets
associated with the proposals
until it is ``too late'' for the adversary
to utilize that information in scheduling attacks --
will remain the same throughout.

\com{	XXX old description: delete after verifying nothing iimportant
	failed to make it into the write-up above.

Use T=f+1 with n=2f+1 nodes total, simple-majority threshold.  Use (T,T)-TLC,
where at each time-step each node waits to receive a simple-majority T of
messages each with at least T acknowledgments from distinct nodes, before
moving to the next time-step.

At $\lts=0$, each node just picks a ticket and sends it in proposal.  Then it waits
for the next time-step $\lts=1$.  At this point, it broadcasts the proposal it saw
in the first time-step with the numerically lowest ticket. Each node then waits
for time-step 2, at which point it has collected T messages from time-step 1
indicating the lowest-valued tickets observed by T different nodes. 

At the point a given node reaches time-step 2, it now determines whether or not
the protocol successfully committed. If the node sees that the absolute
lowest-valued ticket observed by any of these nodes was seen by at least T of
them, then the protocol successfully committed. Otherwise the did not commit
from this node’s perspective (although it is possible that other nodes might
observe it as committed at time 2).

Because of the (T,T)-TLC, by the time any node reaches time $\lts$=1, it has
received T proposals from time-step 0, each of which has been seen and
acknowledged by at least T nodes including itself.  

We don’t expect the protocol to succeed every run; just “often enough.”
Reasoning:

Safety: Suppose two nodes i and j believe the protocol has committed with two
values vi and vj.  They must both have observed a majority of nodes declare
these respective proposals to have the lowest ticket and with each known to a
majority of nodes. These two majority sets have at least one node overlap, and
(because all nodes are honest) would have declared this for only one value, so
v1 and v2 must be equal. 

Probability of success: Assuming simplistically that the non-adversarial
network delivers messages sent in a given time-step in uniformly random order,
each node will get a given proposal in its T threshold of T-acknowledged
proposals with probability slightly over 1/2.  The proposal with the lowest
ticket will thus be seen, in expectation, by slightly over half the nodes, give
or take based on luck. In at least about half of the runs, the number of nodes
with the lowest ticket in their set will be no less than the expectation (XXX
expand logic), and hence produce a success round. 

Liveness: since the network eventually delivers all messages, TLC guarantees
that time-steps advance as quickly as the network allows. While the consensus
protocol doesn’t guarantee success in each round, it guarantees at least about
a 1/2 probability (?) of success each round, and thus w.h.p. success in O(log
epsilon) rounds...

spin-the-bottle analogy: sometimes the bottle stops pointing close to the
boundary between two players, in which case people may disagree on which player
it is pointing to. In this case, if a majority can’t agree, the solution is
simply to spin again.

Step 2: adversarial network but all nodes honest.  

Each just picks a random ticket and encrypts it so all the other nodes (but not
the network adversary) can see it.
}

%% file: fat.tex
\subsection{Fat Byzantine Consensus}
\label{sec:fat}

\xxx{ this section will be dropped for now due to complexity
	and lack of certainty that it actually works }

Fat Consensus

Requires threshold logical clocks with two-thirds supermajority threshold
messages and acknowledgements: i.e., parameterized with $\tm >= 2f+1$ and $\ta >=
2f+1$ where $n = 3f+1$.

Two phases: $\lts+0$: Propose.  All nodes send a value.  
$\lts+1$: Commit.  $\lts+2$: Reveal.
End of consensus round (and perhaps beginning of next round).

Each node, upon reaching logical time $\lts+0$, broadcasts a value/block proposal,
including its “view of history.”  (Maybe the $\lts+0$ and the view of history can be
simplified out for initial exposition.)

As per the $(T,T)$-TLC protocol, each node then waits to receive at least $T=2f+1$
distinct proposals, each with at least $T$ acknowledgments from distinct nodes.
While waiting to meet this threshold and collect the evidence required to move
to time $\lts+1$, each node broadcasts acknowledgments to proposals it receives
other than its own.  (Each node implicitly acknowledges its own proposal as
part of sending the proposal itself, so each proposal already has one
acknowledgment immediately upon being sent.)

Upon reaching $\lts+1$, each good node broadcasts a message stating what specific
specific set of proposals it received from time $\lts+0$, and the number of
acknowledgments it received for each before moving to time $\lts+1$.  (The latter
number obviously must be at least $T$ for at least $T$ of the proposed messages,
but may be smaller for some proposers that the node received but did not reach
the acknowledgment threshold by the time the node moved to time $\lts+1$.)

Upon reaching time $\lts+2$, each node determines whether it can conclude that the
consensus round completed successfully.  Each node has collected $T$ messages
from time $\lts+1$, each representing a collection of at least $T$ proposals from time
$\lts+0$l.  

As soon as at least $2f+1$ nodes have reached time $\lts+2$ (of which a stable core of
at least $f+1$ are good), any node receiving a set of $2f+1$ shares of the beacon
for time $\lts+2$ can produce the randomness that determines the ticket values
associated with the proposals broadcast earlier at $\lts+0$.

Upon calculating the ticket values for the known proposals, each node $i$
identifies the numerically lowest ticket value among all the proposals at time
$\lts+0$, and checks whether this apparently-winning proposal also received a
threshold $T$ of acknowledgments by time $\lts+1$.  If so, node $i$ considers the
consensus round definitely successful; otherwise, node $i$ considers the
consensus round as potentially unsuccessful.  (Other nodes with different views
might potentially be able to determine the round to be successful, however.)

Safety reasoning:  Suppose two honest nodes $i$ and $j$ at time $\lts+2$ believe that
the consensus round has successfully committed two proposals $p_1$ and $p_2$.  Each
of these honest nodes has verified that their respective proposals has been
acknowledged by at least $2f+1$ nodes total by time $\lts+1$, of which at least $f+1$ of
these nodes must be good, and of which at least one must be a good node that
observed both proposals.  If proposals $p_1$ and $p_2$ were different, then this
overlapping good node at time $\lts+1$ would have seen both proposals and
communicated their existence to the good nodes $i$ and $j$ in question.  But
one of those proposals must have a ticket value higher than the other, which
means neither honest nodes $i$ nor $j$ could have considered the
higher-ticket-valued proposal to be committed, leading to a contradiction.
This, for nodes $i$ and $j$ to consider $p_1$ and $p_2$ to be comitted, $p_1$ must equal $p_2$,
ensuring safety.

Liveness reasoning: …

%% file: thin.tex
\subsection{Thin Byzantine Consensus}
\label{sec:thin}

Thin Consensus

Five phases:
$\lts=0$: Propose.
$\lts=1$: Collect.
$\lts=2$: Select.
$\lts=3$: Commit.
$\lts=4$: Reveal.

%% file: byz.tex
\section{Tolerating Byzantine Nodes}
\label{sec:byz}
\label{sec:tlc:byz}

For simplicity we have assumed so far that only the network,
and not the participating nodes themselves,
might exhibit adversarial behavior.
Both TLC and QSC may be extended to tolerate Byzantine behavior
using well-known existing techniques, however,
as we outline in this section.
We address this challenge in three main steps,
roughly corresponding to three key layers of functionality from bottom to top:
first, enforcing the causal ordering that TLC depends on;
second, ensuring TLC's correct time progress
in the presence of Byzantine nodes;
and third, protecting QSC consensus from adversarial nodes.

\com{
While we have assumed above that $f_c=0$, 
we now relax this constraint and address the corruption
of nodes and their messages.
Corrupt nodes might violate the above properties
by advancing their logical clocks too quickly,
not advancing their clocks when they should,
or confusing other nodes in various ways,
for example.
}

\com{
\subsubsection{Byzantine Protection: Architectural Perspective}

\begin{figure}[t]
\begin{center}
\begin{footnotesize}
\begin{tabular}{|c|l|}
\multicolumn{1}{c}{\bf Layer}	& \multicolumn{1}{l}{\bf Relevant Byzantine protection mechanism(s)} \\
\hline
consensus	& asynchronous Byzantine consensus~\cite{XXX} \\
\hline
keysensus	& threshold distributed key generation~\cite{XXX} \\
\hline
randomness	& threshold randomness protocols~\cite{XXX} \\
\hline
time release	& verifiable secret sharing~\cite{XXX}, threshold encryption~\cite{XXX} \\
\hline
threshold time	& accountable state machines, peer review~\cite{XXX} \\
\hline
witnessing	& threshold signing~\cite{XXX}, witness cosigning~\cite{XXX} \\
\hline
gossip		& transferable authentication, digital signatures~\cite{XXX} \\
\hline
vector time	& cryptographic hashes, timeline entanglement~\cite{XXX} \\
\hline		
logging		& tamper-evident logging~\cite{XXX} \\
\hline
timestamping	& delay imposition for timestamp consistency \\
\hline
link		& encryption, transport-layer security (TLS)~\cite{XXX} \\
\hline
\end{tabular}
\end{footnotesize}
\end{center}
\caption{Overview of Byzantine protection machanisms by architectural layer}
\label{fig:byz-arch}
\end{figure}
}

\subsection{Causal Logging and Accountability}
\label{sec:byz:log}

While TLC's goal is to create a ``lock-step'' notion of logical time,
to build TLC and secure it against Byzantine nodes,
it is useful to leverage the classic notion of
{\em vector time}~\cite{fischer82sacrificing,liskov86highly,mattern89virtual,fidge91logical}
and associated techniques such as
tamper-evident logging~\cite{schneier99secure,crosby09efficient},
timeline entanglement~\cite{maniatis02secure},
and accountable state
machines~\cite{haeberlen07peerreview,haeberlen10accountable}.

\subsubsection{Logging and Vector Time}

Our approach hinges on transparency and accountability
through logging and verification of all nodes' state and actions.
Each node maintains a sequential log
of every significant action it takes,
such as broadcasting a message.
Each node's log also documents the nondeterministic inputs,
such as messages it received,
that led it to take that action.
Each node assigns consecutive {\em sequence numbers} to its log entries.
A node's sequence numbers effectively serve as a node-local logical clock
that simply counts all the events the node records,
independent of both wall-clock time and other nodes' sequence numbers.

In addition to logging its own events,
each node $i$ also maintains a mirror copy
of all {\em other} participating nodes' logs,
and continuously tracks their progress by maintaining a vector containing
the highest sequence number $i$ has seen so far from each other node $j$.
This is the essence of the classic concept of
vector clocks~\cite{mattern89virtual,fidge91logical}.

Because a vector clock indicates only that some node $i$
has seen all the logged events of some other node $j$
up to a particular sequence number in $j$'s local sequence space,
node $i$ must process messages from $j$,
and update its vector clock accordingly,
strictly in the order of $j$'s sequence numbers.
Suppose $i$ has seen all messages from $j$ up to sequence number $3$,
for example,
then receives a message containing $j$'s event $5$ out of order.
In this case,
$i$ must hold this out-of-order message in a reordering buffer
and delay its actual processing and internal delivery
until $i$ receives a message from $j$
filling in the missing sequence number 4.
This reordering process is no different from
classic in-order delivery protocols such as TCP~\cite{rfc793}.

Whenever node $i$ records a new entry in its log,
it includes in the new entry a {\em vector timestamp},
which documents, for the benefit and validation of other nodes,
which messages from {\em all} nodes $i$ had seen when it wrote this entry.
This vector timestamp precisely documents all the nondeterministic information
that led $i$ to take the action this log entry describes.
This is also precisely the information that {\em other} nodes need
to ``replay'' $i$'s decision logic and verify that $i$'s resulting action
is consistent with the protocol that all nodes are supposed to follow,
the essential idea underlying
accountable state machines~\cite{haeberlen07peerreview,haeberlen10accountable}.

\xxx{ illustrate}

\xxx{ optimization to discuss later: no need to log/order message receives,
just record the vector of messages seen before each message send,
which works as long as our handling of causally-independent messages
received from different nodes is order-independent.
That is the case in practice:
while we're tracking messages we received from other nodes,
we only need to verify that what that particular node is saying and doing
seems to be valid and self-consistent or not based on its causal history.
}

\subsubsection{Exposing Node Misbehavior}
\label{sec:byz:expose}

To hold nodes accountable,
we require each node to make its log cryptographically tamper-evident
according to standard practices~\cite{schneier99secure,crosby09efficient}.
In particular,
each node chains successive log entries together
using cryptographic hashes as back-links,
and digitally signs each complete log entry
including its back-link and vector timestamp.
This construction ensures that nothing in a log's history can be modified
without changing the back-links in all subsequent log entries,
making the modification evident.

If a node ever misbehaves in a way
that is manifestly identifiable from the contents of its log --
\eg, producing a log entry representing an action
inconsistent with the prescribed protocol
applied to the node's documented history leading up to that log entry --
then the digital signature on the invalid log entry
represents transferable, non-repudiable ``evidence'' of the node's misbehavior.
Correct nodes can gossip this transferable evidence
to ensure that all correct nodes eventually know about the misbehavior
and can respond appropriately,
\eg, by alerting operators and excluding the misbehaving node from the group.


\subsubsection{Exposing Equivocation and Forked Histories}
\label{sec:byz:equiv}

Besides producing invalid histories,
another way a node can misbehave is
by producing multiple conflicting histories,
each of which might individually appear valid.
For example, a malicious node might produce
only one version of history up to some particular event,
then {\em fork} its log and produce two histories building on that event,
presenting one version of its history to some nodes
and the other version of its history to others.

To fork its history,
a malicious node must by definition {\em equivocate} at some point
by digitally signing two or more different messages
claiming to have the same node-local sequence number.
If the malicious node is colluding with a powerful network adversary,
we cannot guarantee that correct nodes will immediately -- or even ``soon'' --
learn of this equivocation.
The adversarial network could schedule messages carefully
to keep different correct nodes in disjoint ``regions of ignorance''
for an arbitrarily long time,
each region unaware that the other is seeing a different face
of the equivocating node.

Nevertheless, provided the network adversary
cannot partition correct nodes from each other indefinitely,
the correct nodes will {\em eventually} obtain evidence of the equivocation,
by obtaining two different messages signed by the same node
with the same sequence number.
These correctly-signed but conflicting messages again
serve as transferable, non-repudiable evidence of the node's misbehavior,
which the correct nodes can gossip and respond to accordingly.
In a general asynchronous setting with no added assumptions,
this eventual detection of equivocation is the best we can do.

\subsubsection{Causal Ordering in Verification Replay}

In order for any node $i$ to validate the logged actions of another node $j$,
$i$ must replay the deterministic logic of $j$'s protocol state machine,
and compare it against the resulting event that $j$ signed and logged.
Since this action by $j$ likely depended
on the messages $j$ had received from other nodes up to that point,
this means that $i$ must use {\em exactly the same} views
of all other nodes' histories as $j$ used at the time of the event,
in order to ensure that $i$ is ``fairly'' judging $j$'s actions.
If $i$ judges $j$'s logged event from even a slightly different ``perspective''
than that in which $j$ produced the log entry,
then $i$ might might incorrectly come to believe that $j$ is misbehaving
when it is not.

Because the verifier node $i$'s perspective
must line up exactly with that of the verified node $j$'s perspective
as of the logged event,
this implies first that $i$ must have received and saved
all the causally prior messages --
from {\em all} nodes --
that $j$ had seen upon recording its event.
This means that $i$ must process $j$'s messages, and replay its state machine,
not just in sequential order with respect to $j$'s local log,
but also in {\em causally consistent} order
with respect to the vector timestamps in each of $j$'s log entries.
If one of $j$'s log entries that $i$ wishes to validate
indicates that $j$ had seen message 3 from another node $k$, for example,
but $i$ has not yet received message 3 from node $k$,
then $i$ must defer its processing and validation of $j$'s log entry
until $i$'s own vector clock ``catches up'' to $j$'s logged vector timestamp.
Only at this point can $i$ then be sure that it has $k$'s message 3
and all others that $j$'s logged decision might have been based on.

\subsubsection{Handling Equivocation in Log Verification}

Equivocation presents a second-order challenge in log verification,
because correct nodes can expect to detect equivocation
only eventually and not immediately.
Suppose that correct node $i$ is validating a log entry
of another correct node $j$,
which indicates that $j$ had seen message 3 from a third node $k$.
If $k$ is an equivocating node that forked its log,
then $i$ might have seen a {\em different} message 3 from $k$
than the message 3 from $k$ that $j$ saw in recording its logged action.
In this way, $k$ might try to ``trick'' $i$
into thinking that $j$ misbehaved,
when in fact the true misbehior was not-yet-detected equivocation by $k$.

Node $i$ must therefore ensure that when validating another node $j$'s log,
$i$ is referring to {\em exactly the same messages} that $j$ had seen,
even if these might include equivocating messages from other nodes like $k$
that have not yet been exposed as misbehaving.
One way to ensure this property
is for $j$ to include in its logged vector timestamps
not just the sequence numbers of the last message
it received from each other nodes,
but also a crypographic hash of that last message $j$ received from each node.
Thus, a vector timestamp is in general a vector
of both sequence numbers and cryptographic hashes of ``log heads''.

If a correct node $i$ obtains such a generalized vector timestamp from $j$,
representing a set of messages of other nodes that $i$ ``should'' have already
according to their sequence numbers,
but the cryptographic hash of $k$'s last message
according to $j$'s vector timestamp
does not match the message $i$ already has from $k$ with that sequence number,
then $i$ knows that it must defer judgment of whether $j$ or $k$ is misbehaving.
Node $i$ asks $j$ for copies of the signed messages from $k$
that $j$ had received and logged.
If any of these are correctly-signed by $k$
but inconsistent from those $i$ had seen,
then $i$ now has evidence of $k$'s misbehavior.
In addition, $i$ uses the version of $k$'s history that $j$ documented,
instead of $i$'s own version of $k$'s history,
to replay and validate $j$'s logged actions,
to avoid the risk of incorrectly judging $j$ as misbehaving.

\subsection{Byzantine Hardening TLC}

None of the above methods above for holding nodes accountable are new,
but rather a combination of existing techniques.
These techniques provide all the foundations we need
to make TLC resistant to Byzantine node misbehavior,
as we explore in more detail now.

\subsubsection{Enforcing correct logical time progress}
\label{sec:byz:time-progress}

To protect TLC against Byzantine node behavior,
correct nodes must prevent Byzantine nodes both
both advancing their logical clocks incorrectly,
and from tricking other correct nodes into incorrect behavior.
For example, a Byzantine node might improperly attempt to:
advance its clock faster than it should,
before it has received the threshold of messages required for it to advance;
claim that a message has been threshold witnessed when it has not;
fail to advance its clock when it {\em has} received and logged
a threshold of messages;
or violate logical clock monotonicity by ``turning back'' its clock.
This is merely a sample, not an exhaustive list of potential misbehaviors.

By encoding TLC's rules into the accountable state machine logic
by which all nodes verify each others' logs, however,
we can detect most of these misbehaviors automatically.
Haeberlen's PeerReview framework for
accountable state machines~\cite{haeberlen07peerreview,haeberlen10accountable}
lays out all the necessary principles in general,
though simplifications and optimizations are of course possible
in specializing this framework
to particular protocols such as TLC and QSC.

In order to advance its clock, for example,
any node must not just {\em claim} to have received a threshold of messages,
but must actually {\em exhibit evidence} of its claim.
This evidence consists of an appropriate collection of
TLC proposal messages from the appropriate time-step,
each embedded in the valid and properly-signed logs of
a suitable threshold of distinct participating nodes,
all with sequence numbers causally prior to (``covered by'')
the vector clock with which the node announces its time advancement.
Since all of this evidence lies in messages causally prior to
the time advancement in question,
correct nodes will automatically obtain and verify this body of evidence
prior to processing or verifying the time advancement message itself.
As long as the message threshold $t_m$
is larger than the number of maliciously colluding nodes,
therefore,
the colluding nodes cannot advance time without the cooperation
of at least one correct node.

The same verification mechanism precludes nodes from
incorrectly claiming a message has been threshold witnessed,
since no correct node will believe such a claim
without seeing the digitally-signed evidence
that a threshold of nodes have indeed witnessed the message.
Similarly, a malicious node cannot turn back its logical clock
without either equivocating and forking its log,
which correct nodes will eventually detect as discussed above,
or producing a log that self-evidently breaks the monotonicity rule
that logical clocks only ever increase,
a violation that correct nodes will immediately detect.

A malicious node can, of course, fail to advance time when it should
by delaying the processing and logging of messages it in fact received.
This behavior is merely a particular variant
of a node maliciously running slowly,
which we fundamentally have no way of distinguishing
from a node innocently running slowly or failing to receive messages
for reasons outside of its control,
such as network delays or DoS attacks attacks.
Nevertheless, if a malicious node does receive {\em and acknowledge in its log}
a threshold of suitably-witnessed messages from a given time step,
then it {\em must} advance its logical clock in the next action it logs,
otherwise correct nodes will detect its misbehavior.
Similarly, if a malicious node is at step $s$
and acknowledges in its log any broadcast received from some node
at a later time step $s+\delta$,
then the malicious node {\em must} catch up
by advancing its clock immediately to step $s+\delta$
or being caught in non-compliance by correct nodes' verification logic.
In effect, in order to ``drag its heels'' and avoid advancing its clock
without being caught,
a malicious node must entirely stop acknowledging any new messages
from other nodes that would force it to advance its clock,
thereby eventually becoming behaviorally indistinguishable from a node
that is merely offline or partitioned from the network for an extended period.

\subsubsection{Majoritarian Reasoning in Byzantine TLC}
\label{sec:byz:maj}

To adapt the majoritarian reasoning tools
described earlier in Section~\ref{sec:tlc:maj}
to a Byzantine environment,
we must adjust the thresholds in much the same way as in
existing Byzantine consensus algorithms~\cite{castro99practical,kotla09zyzzyva,clement09making,clement09upright,bessani14state,aublin15bft}.
In particular,
we must set the thresholds to ensure that they cover
a majority of {\em correct} nodes
{\em after} accounting for the potentially arbitrary behavior
of Byzantine nodes.

To define these threshold constraints precisely
while maintaining maximum configuration flexibility,
we distinguish between {\em availability failures},
in which a node follows the prescribed protocol correctly
but may go offline or be unable to communicate due to DoS attacks,
and {\em correctness failures},
in which a node may be compromised by an adversary,
leaking its secrets and
sending arbitrary messages to other nodes (including by equivocation)
in collusion with other compromised nodes and the network.
We assume any TLC configuration is motivated by some threat model
in which there is a particular assumed limit $f_a \ge 0$
on the maximum number of availability (fail-stop) failures,
and another assumed limit $f_c \ge 0$
on the maximum number of correctness (Byzantine) failures.
This decoupling of availability from correctness failures
is closely analogous to that done in UpRight~\cite{clement09upright}.

To apply the majoritarian reasoning from Section~\ref{sec:tlc:maj}
in such a Byzantine threat model,
the message and witness thresholds must satisfy the following two constraints:

\begin{enumerate}
\item	$t \le n - f_a$:
	This constraint ensures that TLC time can advance,
	ensuring the system remains live,
	in the absence of any communication from up to $f_a$ nodes.

\item	$t > f_c+\frac{n-f_c}{2}$ (or $t>\frac{n+f_c}{2}$):
	This constraint ensures that the threshold $t$
	is large enough to include
	{\em all} of the $f_c$ potentially Byzantine nodes,
	plus a majority (strictly greater than half)
	of the $n-f_c$ correct nodes.
\end{enumerate}

Here we use a single threshold $t$ to represent either $t_m$ or $t_w$,
which will typically be the same in practice,
except when $t_w=0$ in the case of unwitnessed TLC.

While we leave the formal definitions and details
for later in Section~\ref{sec:analys},
this majoritarian reasoning works in TLC (and QSC)
for {\em arbitrary} nonnegative combinations of $f_a$ and $f_c$.
These parameters can in principle represent separate and independent sets of
unavailable and Byzantine nodes, respectively,
which may or may not overlap.
That is, TLC can tolerate $f_a$ correct but unreachable nodes
and an {\em additional} $f_c$ responsive but malicious nodes.

If we assume just one set of $f$ generic ``faulty'' nodes,
each of which might be unresponsive and/or malicious
(\ie, $f=f_a=f_c$),
and we set $t=n-f$,
then the above constraints reduce to the classic $n>3f$
(or equivalently $n \ge 3f+1$)
commonly assumed by Byzantine consensus algorithms.
But this represents only one possible and sensible configuration
of TLC's thresholds.

If we set $f_c=0$,
then the above constraints reduce to the basic fail-stop model
as we assumed in Section~\ref{sec:tlc:maj},
where a ``simple majority'' threshold $t > \frac{n}{2}$ is adequate.

But arbitrary intermediate values of $f_c$ are possible and interesting as well.
Suppose, for example, we set up a classic BFT-style group
of $n$ nodes where initially $f_a=f_c=f$ and $n > 3f = 2f_a+f_c$.
If during the group's operation,
a malicious node is actually {\em exposed} as malicious
by triggering the accountability mechanisms discussed above,
then one reasonable automated response may be to expell it from the group.
Doing so reduces both $n$, $f_c$, and $t$ by 1,
while leaving $f_a$ unaffected since correct-but-slow nodes aren't expelled.
In the limit case where all $f_c$ malicious nodes eventually expose themselves,
the group gradually reconfigures itself from a classic BFT configuration
($n > 3f = 2f_a+f_c$)
into a classic Paxos-like fail-stop configuration
($n > 2f = 2f_a$).

TLC also does not inherently assume or require
that correct nodes outnumber Byzantine nodes:
that is, $n-f_c$ may potentially be less than $f_c$.\footnote{
	Specific distributed protocols built atop TLC may, of course,
	require that correct nodes outnumber malicious nodes.
	One such example is the AVSS-based 
	asynchronous distributed key generation protocol
	we develop later in Section~\ref{sec:dkg:qsdkg}.
}
In the limit case where $f_a=0$,
the above constraints reduce to $n > f_c$,
the {\em anytrust} model~\cite{wolinsky12scalable}.
In such a configuration,
liveness and progress require all $n$ nodes to participate,
tolerating no unavailability or unreachability,
but there need be only one correct node.
All other nodes may collude,
and no one needs to know or guess which node is correct.

\xxx{ clarify threat model:
	\ie, briefly and intuitively,
	what the adversary can do to the $f_c$ corrupt nodes
	and what it can do to the $f_a$ unavailable nodes.}

\subsubsection{Proactive anti-equivocation via witnessing}
\label{sec:byz:wit}

Although the accountability mechanisms above ensure that correct nodes
will {\em eventually} expose equivocation by malicious nodes,
protocols built atop TLC might still be subverted in the short term
by equivocation attacks before the equivocation is detected.
In witnessed TLC with the witness threshold $t_w$
satisfying the majoritarian constraints above, however,
the threshold witnessing process built into each TLC time-step
offers a natural {\em proactive} form equivocation protection,
a consequence of the proactive accountability
offered by witness cosigning~\cite{syta16keeping}.

In particular, protocols built atop TLC
with a majoritarian witness threshold
can rely on never seeing two equivocating {\em threshold witnessed} messages.
This is because for any malicious node to get two equivocating messages
for the same time step threshold witnessed,
it would have to obtain a transferable body of evidence
including a witness threshold $t_w$
of valid, properly-signed witness acknowledgment messages for each.
This threshold would require a majority of correct nodes
to witness and cosign each equivocating message,
implying that at least one correct node would have to sign both messages.
But if a correct node ever sees a second messages with the same sequence number
from the same node,
it does not accept or witness the second,
but instead uses it as evidence
to expose the equivocating node's self-evident misbehavior.

\subsubsection{Majoritarian time period delineation}
\label{sec:byz:periods}

\begin{figure}
\begin{center}
\includegraphics[width=1\columnwidth]{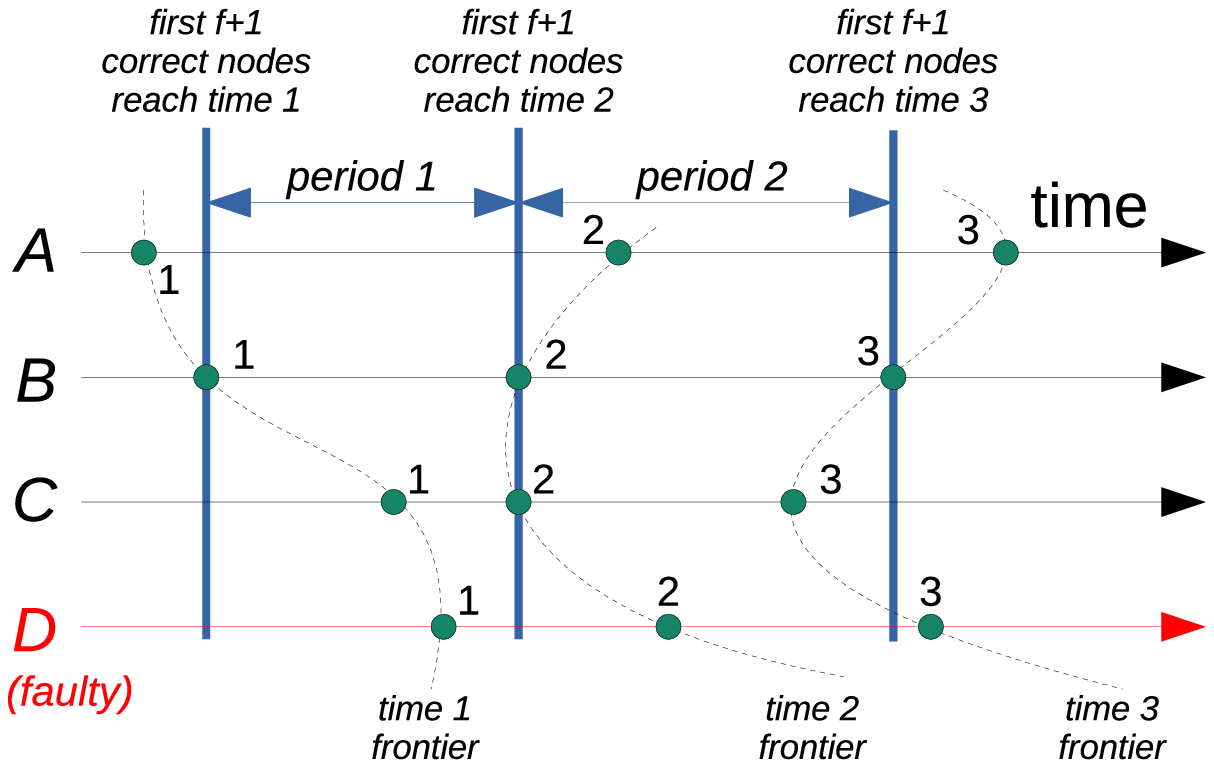}
\end{center}
\caption{Global time periods demarked by the moments a majority of
	correct nodes reach a threshold time $t$.}
\label{fig:time-periods-byz}
\end{figure}

With the above adjustments to the thresholds,
the time period delineation described earlier in Section~\ref{sec:tlc:periods}
extends naturally to the Byzantine environment.
In particular, the ``critical moment''
that determines when a new time period $s$ begins
is the moment when {\em a majority of correct nodes} reach step $s$.
When any the $f_c$ Byzantine nodes advance their clocks
is thus irrelevant to the conceptual delineation of time periods.
Figure~\ref{fig:time-periods-byz} illustrates this process.

Even though the correct nodes have no realistic means of determining
either which other nodes are correct
or precisely when each time period begins and ends,
nevertheless this conceptual delineation imposes strict bounds
on when any valid message labeled for a given time step
may have been formulated.

\begin{itemize}
\item
First,
as long as the message threshold $t_m$ satisfies the above constraints,
no one can reach or produce a valid message for time step $s+1$ or later
before time period $s$ has started.
Reaching step $s+1$ requires exhibiting a ``body of evidence'' that includes
valid, properly-signed messages from a threshold $t$ of messages from step $s$.
This threshold set
must include a majority of the correct nodes even {\em after} being
``maximally corrupted'' by up to $f_c$ Byzantine nodes.
Since such a majority of correct nodes is unavailable
until a majority of correct nodes reach step $s$
and thus collectively enter time period $s$,
no malicious node can produce a valid message labeled $s+1$ or later
before period $s$ starts, 
without being exposed as corrupt.

\item
Second, in witnessed TLC where $t_w$ satisfies the above constraints,
no one can formulate and produce
any new threshold witnessed message for step $s$
after time period $s$ ends and period $s+1$ begins.
This is because such a message would have to be verifiably witnessed
by a threshold $t_w$ that includes a majority of correct nodes
even after being maximally corrupted by up to $f_c$ Byzantine nodes.
Such a majority of correct nodes is unavailable after period $s$ ends,
because correct nodes refuse to witness messages for step $s$
after having advanced to step $s+1$.
A node that formulates a message $m$ and gets {\em at least one}
witness cosignature for it before period $s$ ends
might still be able to {\em finish} getting $m$ threshold witnessed
after period $s+1$ starts,
but this does not violate the time bounds
because $m$ was {\em formulated} during step $s$.
\end{itemize}

\subsubsection{Two-step broadcast}
\label{sec:byz:bcast}

Byzantine-protected majoritarian witnessed TLC
similarly enforces the two-step broadcast property
described earlier in Section~\ref{sec:tlc:broadcast}.
Any message $m$ a node broadcasts at some step $s$
that is threshold witnessed and used in advancing to step $s+1$
is guaranteed to have been witnessed by a majority of correct nodes
by the time they reach $s+1$.
This majority must overlap by at least one correct node
with the majority of correct nodes from which
any node must gather step $s+1$ messages to advance to step $s+2$.
This overlapping correct node will always reliably propagate knowledge of $m$,
even if other malicious nodes might ``forget'' or equivocate about it.
Thus, the majorities of correct nodes alone ensure that
knowledge of each message threshold witnessed at $s+0$
propagates to {\em all} nodes by the time they reach $s+2$.

Even a malicious node cannot pretend not to have seen $m$ by $s+2$,
because the malicious node must exhibit the appropriate body of evidence
to convince correct nodes it has reached $s+2$ in the first place.
That body of evidence must have a threshold of signed messages from other nodes
including at least one from a correct node
that conveys knowledge of $m$,
and the malicious node can neither omit nor forge this message referring to $m$
without producing an invalid log and being exposed as corrupt.

...

\xxx{ diagram: epochs separated by events of majority of {\em honest} nodes
	reaching next logical time-step }

\xxx{ rewrite to use witness cosigning~\cite{syta16keeping} }

\begin{figure}
\begin{center}
\includegraphics[width=1\columnwidth]{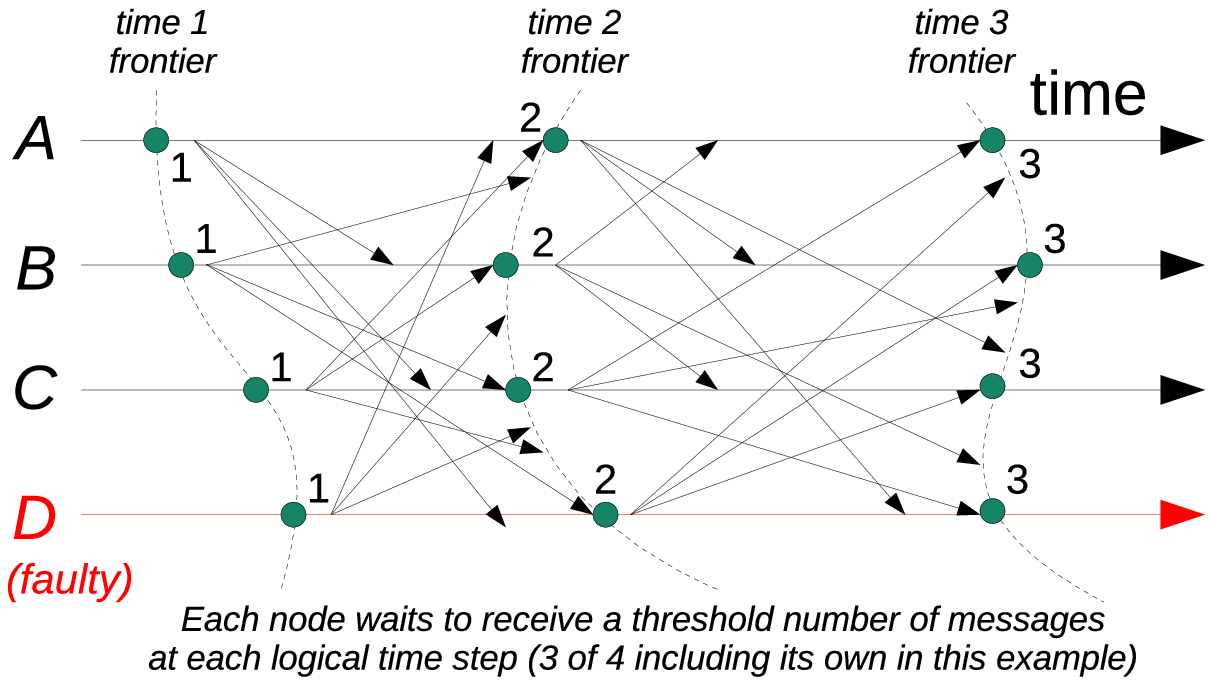}
\end{center}
\caption{Illustration of basic threshold logical clocks
	without threshold witness certification.}
\label{fig:tlc-basic-byz}
\end{figure}

\begin{figure}
\begin{center}
\includegraphics[width=1\columnwidth]{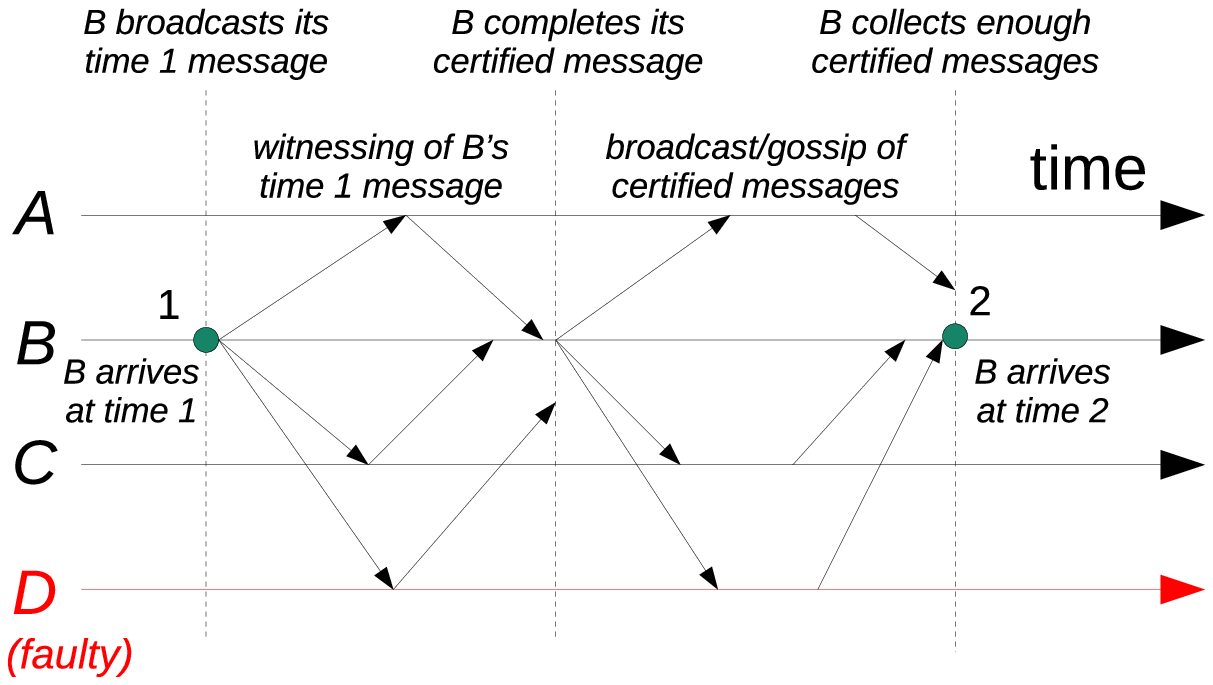}
\end{center}
\caption{Illustration of one witnessed TLC time-step
	from the perspective of one particular node $B$.}
\label{fig:tlc-witnessed-byz}
\end{figure}


\xxx{	this probably belongs in the consensus section.

\subsection{Equivocation Protection}

We’ll now assume T = 2f+1, since at least this threshold is fundamentally
required under the full asynchronous network model to offer any protection
against an equivocating adversary.  [summarize briefly why.]

Stable core concept: once the first f+1 honest nodes have reached a given time
t, and are in agreement in some fashion (either in terms of consensus or any
other shared viewpoint fact), no adversarial behavior can successfully
contradict their agreement in a way that will be accepted by any honest node in
any subsequent tilmestep t+1 or later.  This is because any honest node
reaching time t+1 or later will have heard from at least 2f+1 nodes at time
step t, of which at least f+1 are good, and of which at least 1 good node
overlaps with the stable core (the first f+1 good nodes to reach time t+1).  If
any message from any malicious node attempts to contract the knowledge of the
stable core through equivocation, then this 1-good-node overlap at time t will
ensure that the potential good victim node at time t+1 learns of the
equivocation (both the stable core version and at least one alternate attack
version), exposes the equivocation, and ensures that the good victim node
permanently eliminates and ignores messages from the equivocating node(s) going
forward, effectively turning the equivocating nodes into virtual “fail-stop”
nodes.

Discuss how this affects (T,T)-TLC in particular.

Discuss how this affects (T,0)-TLC in particular.
}

\subsection{Byzantine Consensus with QSC}
\label{sec:byz:cons}

Byzantine-protecting the QSC3 consensus protocol
described in Section~\ref{sec:cons}
involves two challenges:
first, protecting the basic consensus logic and state machine
from Byzantine manipulation,
and second,
ensuring that Byzantine nodes cannot leak the proposals'
genetic fitness lottery tickets to the network adversary
before the 3-step consensus round is complete.

\subsubsection{Protecting the QSC consensus logic}
\label{sec:byz:cons-logic}

Each node's QSC consensus logic must make two key decisions
in a consensus round starting at a given TLC step $s$.
First, the node must choose the best eligible (confirmed) proposal
the node is aware of by step $s+3$,
as the proposal to build on in the next round.
Second, the node must determine whether it may consider
this best eligible proposal to be {\em committed},
according to whether the proposal is reconfirmed (doubly confirmed)
and not ``spoiled'' by another proposal with a competitive lottery ticket.

The standard state machine accountability and verification mechanisms above
are sufficient to force even a malicious node $i$
to make these decisions correctly,
or else be exposed as misbehaving before their incorrect decisions
can affect any correct node.
This is because both $i$'s best eligible proposal decision
and its commitment decision
are well-known, deterministic functions
of the precise set of causally prior messages
$i$ had seen by step $s+3$ as documented in $i$'s log.
Upon receiving and processing $i$'s broadcast at step $s+3$,
each correct node simply replays $i$'s QSC consensus decisions independently
based on the same set of causally prior messages that $i$ used,
and expose $i$ as misbehaving if its logged decisions are incorrect.

\subsubsection{Protecting the lottery ticket values}
\label{sec:byz:tickets}

As discussed in Section~\ref{sec:cons:odds},
QSC's guarantee that each round enjoys a reasonable (at least $1/2$)
independent probability of success
holds only if the adversary cannot learn the lottery ticket values early
and use them to schedule message delivery maliciously based on them.
In addition, QSC's success probability also depends on all nodes
choosing their proposals' lottery tickets fairly
from the same random distribution.
As soon as we admit even a single Byzantine node ($f_c > 1$),
that node might break the consensus protocol
either by leaking all proposals' lottery ticket values
to the network adversary during step $s+0$,
or by choosing its own proposals' lottery tickets unfairly,
\eg, so that the Byzantine node always wins.

Since we must assume a Byzantine node
will leak anything it knows to the adversary,
we must force all nodes to choose their proposals' lottery tickets
such that even {\em they} cannot learn, predict, or bias their choice of ticket
before it is revealed to all at step $s+3$.
Fortunately, a protocol such as RandHound~\cite{syta17scalable},
which implements public randomness through leader-based
{\em verifiable secret sharing} (VSS)~\cite{shamir79share,stadler96publicly,schoenmakers99simple}
provides the required functionality.
We can readily incorporate such a protocol into QSC
for unpredictable and unbiasable lottery ticket selection.

\subsubsection{QSC4: protecting the lottery tickets with PVSS}
\label{sec:byz:pvss}

One  solution is to use a
Publicly Verifiable Secret Sharing (PVSS) scheme
that permits the homomorphic addition of multiple shared secrets
generated by independent
dealers~\cite{stadler96publicly,schoenmakers99simple,cascudo17scrape}.
We extend QSC by one additional TLC time-step,
so that a full consensus round requires four steps total (QSC4).

Initially at $s+0$, 
each node $i$ chooses a random secret $s_i$
and a secret-sharing polynomial $p_i(x)$
such that $p_i(0) = s_i$.
Node $i$'s polynomial $p_i(x)$ is of degree $t_m-f_{ck}-1$,
where $f_{ck}$ is the number of {\em known} corrupt nodes
that have been exposed so far in $i$'s view.
Node $i$ then deals PVSS secret shares
only to the $n-f_{ck}$ nodes {\em not} known to be corrupt.
Node $i$ includes its commitments to $p(x)$
and its publicly-verifiable encrypted shares
in its step $s+0$ proposal.
Because these deals are publicly verifiable,
all correct nodes can immediately ensure that each node's deal is valid
and immediately expose any node that deals an incorrect secret
(\eg, containing secret shares that the intended recipients cannot decrypt).

We then embed the QSC3 consensus protocol into steps 1--4 of PVSS-based QSC4.
At step $s+1$,
each node has received a threshold $t_m$ of PVSS secrets
dealt at $s+0$.
Each node $i$ chooses at least $f_{cu}+1$ such valid secrets
dealt by nodes not yet exposed in misbehavior from $i$'s viewpoint,
where $f_{cu}$ is the maximum number of {\em unknown}
potentially-corrupt nodes not yet exposed
($f_{ck} + f_{cu} = f_c$).
Because the set of $f_{cu}+1$ deals that $i$ chooses
must include at least one by a correct node not colluding with the adversary,
this correct node's deal both uniformly randomizes $i$'s lottery ticket
and ensures that it remains unknown and unpredictable to $i$ or anyone else
until $s+4$.
Nevertheless, $i$'s choice of a {\em specific} set of $f_{cu}+1$ deals at $s+1$
represents a commitment to one and only one lottery ticket,
which thereafter cannot be changed or biased by anyone including $i$.

\xxx{ introducing this unknown potentially-corrupt node metric here
	is probably unnecessary complexity at this point: simplify. }

At the end of the consensus round,
each node includes in its step $s+4$ message
its decrypted shares from all the deals it saw from step $s+0$.
To determine the lottery ticket
for a given node $i$'s proposal $p_i$ from $s+1$,
each node $j$ must obtain and linearly interpolate
$t_m$ shares,
minus any missing shares for nodes known corrupt by $s+1$,
from each of the $f_{cu}+1$ deals that $p_i$ committed to,
and combine them to reconstruct the joint secret $S_i$
represented by $i$'s chosen set of deals.
While unpredictable and unbiasable,
$S_i$ will be the same for all nodes
that used the same set of deals in their proposals,
whereas we need each lottery ticket $T_i$ to be unique and independent
for each node $i$.
We therefore use $S_i$ not directly as proposal $i$'s lottery ticket,
but as a key for a hash of some unique consensus group ID $G$,
consensus round number $r$, and node number:
$T_i = \mathrm{H}_{S_i}(G,r,i)$.

Because decrypted shares from a majority of correct nodes are required
to reconstruct the secret dealt by any correct node,
anyone including the adversary can reconstruct these secrets
only after a majority of correct nodes have reached $s+4$,
and have thereby collectively entered the next global time period
following the completion of the consensus round.
At least for this majority of correct nodes, therefore,
the adversarial network's schedule of message deliveries for this round
is complete, fixed, and ``in the past'' at this point,
ensuring that this majority of correct nodes observes
the correct probabilities of success discussed in Section~\ref{sec:cons:odds}.
The network adversary might affect the delivery schedules
seen by the minority of correct nodes that reaches $s+4$ later,
and hence the odds of success that these nodes directly observe.
But even such a ``latecomer'' node $j$ will have by $s+5$ heard from
at least one member $i$ of the majority that was first to reach $s+4$,
and hence if $i$ observed the round to succeed at $s+4$
then $j$ will know that fact as well by $s+5$.

It is possible that some PVSS deals used in proposals at step $s+1$
may not become known to all nodes by $s+4$,
making their dealt secrets unrecoverable
by nodes that obtain fewer than necessary shares at $s+4$.
The only proposals {\em eligible} for consensus, however,
are those that were threshold witnessed during step $s+1$.
As described in Section~\ref{sec:cons:universal},
this guarantees that {\em all} nodes will have seen any eligible proposal,
and hence the deals it relies on, by $s+3$.
If a node cannot reconstruct some proposal's lottery ticket at $s+4$,
therefore,
the node may assume this means the proposal is ineligible
and simply discard it.

An advantage QSC4 has
over most randomized asynchronous Byzantine consensus
protocols~\cite{rabin83randomized,ben-or85fast,canetti93fast,cachin01secure,cachin05random,friedman05simple,correia06consensus,correia11byzantine,mostefaoui14signature,miller16honey,duan18beat,abraham19vaba}
is that it needs no ``common coins''
or the distributed key generation (DKG) processes
needed to establish them in practice without a
trusted dealer~\cite{gennaro07secure,kate12distributed}.
Each proposer $i$ effectively {\em chooses its own} lottery ticket at $s+1$,
through its particular choice of $f_{cu}+1$ deals to use,
although $i$ cannot tell in advance what ticket {\em value}
it chose and committed to.
A disadvantage of QSC4 is that
because the readily-available PVSS schemes may be used only once,
all nodes must incur the computational and communication costs
of dealing, verifying, and reconstructing fresh PVSS secrets
for every consensus round.
We will explore later in Section~\ref{sec:dkg}
how we can build on QSC to implement
an asynchronous distributed key generation (DKG) protocol
that produces {\em reusable} shared secrets,
amortizing the costs of this bootstrap over many subsequent consensus rounds
that can generate public randomness much more efficiently
as in RandHound~\cite{syta17scalable} or drand~\cite{kozlov19league}.

%% file: dkg.tex
\section{Distributed Key Generation}
\label{sec:dkg}

A large variety of security applications and services
require, assume, or can benefit from
a {\em distributed key generation} (DKG) protocol.
DKG enables a group of nodes to generate
a threshold secret-shared public/private key pair cooperatively,
so that {\em none} of the members ever know or learn the composite private key.
Each member knows a share of the private key, however,
so that any threshold number of members can work together to use it
according to an agreed-upon policy.
Example applications that typically depend on DKG include
threshold schemes for
encryption and decryption~\cite{desmedt89threshold,shoup98securing},
digital signing~\cite{shoup00practical,boldyreva03threshold},
identity-based encryption~\cite{boneh01identity,baek04identity,waters05efficient,kate10distributed},
public randomness beacons~\cite{cachin05random,syta17scalable,kozlov19league},
secure data deletion~\cite{geambasu09vanish},
accountable data access control~\cite{kokoris18calypso},
credential issuance and disclosure~\cite{sonnino18coconut},
electronic voting~\cite{schoenmakers99simple},
and general multiparty
computation~\cite{gennaro98simplified,cramer00general,bogetoft09secure}.

\subsection{The Challenges of DKG}
\label{sec:dkg:challenges}

Distributed key generation in general is not easy, however.
We could rely on a trusted dealer to deal shares of a public/private keypair
via verifiable secret sharing
(VSS)~\cite{chor85verifiable,herzberg95proactive,cachin02asynchronous},
but the trusted dealer is a single point of compromise.
We could require {\em all} $n$ participating nodes to deal secrets using VSS
and combine all $n$ deals homomorphically to produce a joint secret
that no one can know or bias as long as at least one node is correct
(the anytrust model~\cite{wolinsky12scalable}),
but this DKG process can tolerate no unavailability or unreachability
and hence is highly vulnerable to denial-of-service.

Combining only $f_c+1$ VSS deals is sufficient in principle
to ensure that it includes at least one contribution by a correct node.
There are $\binom{n}{f_c+1}$ possible choices of such subsets, however,
and the group must agree on one and only one {\em particular} subset,
an agreement task that requires consensus.
Most of the practical and efficient asynchronous algorithms
rely on common coins~\cite{rabin83randomized,ben-or85fast,canetti93fast,cachin05random,friedman05simple,correia06consensus,correia11byzantine,mostefaoui14signature,miller16honey,duan18beat,abraham19vaba},
yielding a chicken-and-egg problem.
We need common coins to enable asynchronous consensus
to agree on a particular set of VSS deals to generate a distributed keypair
to produce common coins.

One way around this challenge is to drive the DKG process
using traditional leader-based consensus,
which introduces partial synchrony assumptions
to ensure liveness~\cite{gennaro07secure,kate12distributed}.
Another circumvention is to assume the group evolves gradually via
a series of occasional group reconfiguration and DKG events.
The first DKG runs manually or synchronously.
For each subsequent DKG event,
an existing asynchronous consensus group
using common coins from the {\em previous} DKG configuration
agrees asynchronously on a set of VSS deals
representing the {\em next} configuration.
While supporting full asynchrony after initial launch,
this approach unfortunately makes the security of every DKG configuration
critically dependent on that of {\em all} past configurations.
If any one configuration is ever threshold compromised,
then the adversary can retain control forver and security is never recoverable.

\subsection{Que Sera Distributed Key Generation}
\label{sec:dkg:qsdkg}

Because QSC requires no already-agreed-upon common coins,
we can adapt it for asynchronous DKG
without partial synchrony or secure history assumptions.
We call the result {\em que sera distributed key generation} or QSDKG.

The main remaining technical challenge is that to give all nodes
reusable long-term key shares,
we cannot use PVSS schemes that encrypt the shares into exponents
to make them amenable to zero-knowledge proofs.
We must therefore make do with
(non-publicly-)verifiable secret sharing (VSS) schemes,
in which an encrypted share is verifiable only by the share's recipient.

To protect liveness,
the DKG protocol will have to wait until only a threshold $t_w$
of nodes have had a chance  to verify their shares of any given deal
before moving on.
This leaves a risk, however, that a misbehaving node may
deal incorrect shares to up to $f_a$ correct nodes
undetectably during the DKG.
Since we cannot detect this situation before the DKG completes,
the network adversary could compromise liveness later
if any correct nodes are missing shares dealt for a full $t_m$ threshold.
We must therefore ensure
that {\em all} correct nodes obtain correct shares,
including the $f_a$ that couldn't verify their shares during DKG itself.
For this purpose we adapt techniques from
asynchronous verifiable secret sharing (AVSS)~\cite{cachin02asynchronous}.

In addition to the majoritarian message and witness thresholds $t_m$ and $t_w$
each satisfying $\frac{n+f_c}{2} < t \le n - f_a$
as discussed in Section~\ref{sec:byz:maj},
QSDKG also relies on a {\em share recovery} threshold $t_r$
satisfying the constraints $f_c < t_r \le n - f_a - f_c$.
In a classic Byzantine consensus configuration in which
$f_a = f_c = f$ and $n > 3f$, for example,
we set $t_m = t_w = n-f$ and $t_r = n-2f$,
so $t_r > f$.

To generate a new distributed keypair starting at TLC time step $s+0$,
each node deals a fresh secret
by choosing a random bivariate polynomial $p(x,y)$
of degree $t_m-1$ in $x$ and
of degree $t_r-1$ in $y$.
The dealer's secret is $p(0,0)$.
The dealer includes in its TLC step $s+0$ broadcast
a $t_m \times t_r$ matrix of commitments to the polynomial,
and an $n \times n$ matrix of secret shares,
each share $S_{ij}$ encrypted with a random blinding exponent $r_{ij}$
such that $S_{ij} = g^{r_{ij}} p(i,j)$.
Finally, for each $i$ and $j$,
the dealer includes in its step $s+0$ broadcast
ElGamal encryptions of $g^{r_{ij}}$
to each of node $i$'s and node $j$'s public keys,
along with a zero-knowledge proof that the dealer knows $r_{ij}$
and that these ElGamal encryptions to nodes $i$ and $j$ are consistent.
A misbehaving dealer can still incorrectly encrypt the share $S_{ij}$,
but if it does so,
{\em both} nodes $i$ and $j$ will be able to detect and expose this misbehavior
by revealing the blinding factor $r_{ij}$ along with a zero-knowledge proof
of either ElGamal ciphertext's correct opening.
For this deal to be eligible for subsequent use in the DKG,
the dealer must obtain a witness threshold $t_w$ of cosignatures
on it its $s+0$ broadcast.
Correct witnesses provide these signatures
only if both their rows and columns of the dealer's share matrix check out;
otherwise they expose the dealer's misbehavior by opening an incorrect share.

At step $s+1$,
each node $i$ then chooses and proposes a particular set
of $f_c+1$ threshold witnessed deals from step $s+0$,
then commences a series of 3-step consensus rounds
at least until all nodes have observed commitment.
Each node $i$'s particular choice of $f_c+1$ deals at $s+1$
determines the lottery ticket associated with $i$'s proposal
in the first consensus round starting at $s+1$.
In subsequent rounds, each proposal's lottery ticket is determined
by the set of deals from the first proposal
{\em in the history the proposer builds on}.
If in the first consensus round
node $i$ chooses node $j$'s step $s+1$ proposal as the best eligible,
then the lottery ticket for node $i$'s proposal
in the second round starting at step $s+4$
is determined by the deal node $j$ chose at $s+1$,
since that is the deal at the ``base'' of the history $i$ adopted.
In this way, as soon as all nodes have observed commitment at least once,
they will all have agreed on a common prefix history
including a common choice of $f_c+1$ deails to serve as the DKG result.
The participating nodes can then cease consensus rounds
if only the DKG result was needed,
or continue them if regular consensus is still needed for other purposes.

Accounting for the $f_a$ correct nodes that may not have a chance
to verify their shares in a given deal,
plus the $f_c$ nodes that might dishonestly verify their shares,
we can be sure that at least $n-f_a-f_c$ full rows and columns
of the encrypted share matrix were properly verified by correct nodes.
Since $t_r \le n-f_a-f_c$,
this ensures that {\em every} node $i$ will obtain enough correct shares
of its re-sharing polynomial,
represented by $p(i,y)$ with threshold $t_r$,
to reconstruct its share
of the main secret sharing polynomial,
represented by $p(x,0)$ with threshold $t_m$.
Since $t_r > f_c$, however,
the $f_c$ misbehaved nodes cannot alone reconstruct and learn 
the secret shares of correct nodes.

\xxx{ illustrate appropriately}

\xxx{

Also relevant: will the above approach provide a simpler way
to perform asynchronous DKG for IBE?
Kate/Goldberg "Asynchronous Distributed Private-Key Generators for Identity-Based Cryptography"
\url{http://citeseerx.ist.psu.edu/viewdoc/download?doi=10.1.1.433.3027&rep=rep1&type=pdf}
}

\xxx{ talk about random beacons, e.g., RandHerd and drand
	and an asynchronous version, here or elsewhere? }

%% file: time.tex
\section{Logical Time Meets Real Time}
\label{sec:time}

As discussed earlier in Section~\ref{sec:app:time},
correctly observing and interacting with real wall-clock time is often important
even in distributed protocols and services
we would like to {\em pace} asynchronously
as fast as network connectivity permits.
Beyond basic time-centric services such as clock synchronization,
application-logic and policies often depend on real time.
In trusted time stamping or blockchain-based content notarization, for example,
we would like to produce proof that content existed at a particular real time.
In smart contract systems such as Ethereum~\cite{wood14ethereum},
we often want a smart contract to allow or trigger some action
at a particular future time,
or allow an action only until a deadline.
In games and markets systems,
users would like to encrypt their bids for release
only at a synchronized closing time,
to guard against
front running~\cite{czernik18frontrunning,eskandari19transparent}.
In all of these situations,
even if we might like the consensus and application logic
to run as quickly as network conditions permit,
time-dependent applications typically would like to refer to
real wall-clock times rather than logical times or block numbers.
This section explores methods for ensuring that logical time
can observe and interact with real time securely.

\subsection{Securing timestamps in blockchains}

We first address the problem of merely {\em observing} real time accurately
in asynchronous systems driven by TLC.
In either a basic distributed timestamping or beacon service
where each node produces its own log (Section~\ref{sec:app:time:sync}),
or a consensus-based service in which nodes use consensus
to agree on a common blockchain,
we would like each new log entry or block a node produces
to have an accurate wall-clock timestamp.
But how can we ensure these timestamps are accurate,
given that different nodes' clocks may lose synchronization for many reason,
and corrupt nodes might even deliberately set their clocks
arbitrarily forward or back with respect to reality?

Witness cosigning~\cite{syta16keeping,ford16apple,nikitin17chainiac}
offers a partial solution:
the proposer of a new log entry or block simply includes in the block
a wall-clock timestamp based on the proposer's notion of the current time.
The proposer must then obtain cosignatures
from a threshold number of group members serving as witnesses.
An obvious idea is for witnesses to sanity-check the proposer's time stamp
against their own clocks,
rejecting and refusing to witness the proposal for example
if the proposed time stamp is outside a tolerance winder
either before or after the witness's real-time clock.
This way, the fact that a proposal has been threshold witnessed
should indicate that the block's time stamp is ``reasonably'' accurate
according to a number of nodes, to within some fixed tolerance:
a malicious proposer cannot maliciously time stamp a block
either way in the past or way in the future.

The need to pick an arbitrary before-and-after tolerance window, however,
seems akin to a timeout,
inconsistent at least in spirit with fully-asynchronous systems,
and works against the principle that they should be self-timed.
Too large a tolerance window gives malicious nodes
greater leeway to manipulate time stamps they produce,
while too small a tolerance window may trigger false positive 
in which witnesses refuse to cosign a time stamp that is out-of-window
merely because of exceptional network delays or DoS attacks.
We would prefer a way for witnesses to ``keep proposers honest''
in their time stamps without imposing arbitrary thresholds.

In a group that uses TLC and QSC to produce
a collective time stamped blockchain,
we can leverage TLC's delay-tolerance to constrain the inaccuracy
of generated time stamps without imposing artificial tolerance windows.
At the beginning of each QSC consensus round,
each node proposing a potential block includes a time stamp
containing the current time according to the proposer's internal clock.
When another node receives this time stamped proposal,
it first verifies that the proposal's time stamp does not violate monotonicity
by ``turning back the clock'' or failing to increase it
with respect to whichever previous block the proposal builds on.
A monotonicity violation is an immediately-detectable correctness failure,
which the receiver can expose simply by gossiping
the signed but invalid proposal.
Since QSC guarantees that the correct nodes in a group
win a significant percentage of the proposed blocks,
and we assume that correct nodes have reasonably correctly-synchronized clocks,
the most a badly-synchronized or malicious node can date a proposal in the past
is back to just after the time stamp of the most recent block
proposed by a correct node.
Since the block consensus rate depends on network conditions,
the faster the rate at which the network permits TLC to pace the group,
the more tightly-constrained a slow or malicious node's time stamps will be
against accidental or deliberate proposal back-dating.

After verifying monotonicity,
the receiver of a proposal also checks
if the proposed time stamp is in the future
with respect to its own real-time clock.
If so, the receiver does not reject the proposal,
but instead merely delays its processing internally
until the indicated time has passed according to the receiver's clock.
If the proposer's clock is ahead of the receiver's,
the arrival of a future-dated time stamp at the receiver
will thus simply cause the receiver to add a corresponding delay.
If the proposer's clock is significantly fast with respect to correct nodes,
then {\em all} correct nodes will similarly delay the future-dated proposal.
If the future-dated proposal eventually wins the QSC consensus lottery,
then by the time it commits it will no longer be in the future
from the perspective of a majority of correct nodes,
and thus will ``no longer'' be violation of time stamp correctness.
If proposer is future-dated enough, however,
then the delays that all correct nodes impose on its processing
will decrease and potentially eliminate the chance the proposal has
of being threshold witnessed or chosen by QSC consensus,
just as if the proposer was actually just a too-slow or unavailable node
that the rest of the group cannot ``wait around for''
without violating its threshold liveness.

Delaying the processing of forward-dated proposals at correct nodes
serves simultaneously both to ``correct'' the time stamp
by ensuring the proposal cannot be agreed on by consensus 
until a majority of correct nodes agree that its timestamp has passed,
and also serves to ``punish'' the proposer gracefully
by making the forward-dated proposal less likely to win consensus
to whatever extent the added delays disadvantage the forward-dated proposal
with respect to those of correct nodes.
Because chance of a forward-dating node's proposals getting picked by QSC
will disappear as soon as the time stamps in its proposals for a given round
are higher than those of most correct and responsive nodes in the group,
this provision effectively constrains the amount by which a proposal
may be forward-dated and still get in the blockchain,
according to the distribution and variants of other clocks in the group.
Between the enforcement of time stamp monotonicity
and the delay of received messages with future time stamps,
therefore,
the range in which poorly-synchronized or malicious nodes
can produce inaccurate block time stamps
is naturally constrained to an effectively self-timed tolerance
that becomes tighter as network conditions allow the group to proceed faster.

\subsection{Asynchronous encrypted time vaults}
\label{sec:time:vaults}

As discussed earlier in Section~\ref{sec:app:time:vault},
threshold identity-based encryption
(IBE)~\cite{boneh01identity,waters05efficient},
together with the asynchronous distributed key generation
needed to set it up (Section~\ref{sec:dkg}),
suggest an attractive approach to encrypted time vaults
allowing a ciphertext to be decrypted at a designated future time.
The sender simply encrypts a message to an IBE ``identity''
representing some future time.
The threshold group collectively holding the IBE master key
then simply generates and publicly releases the ``private key''
for each time ``identity'' once that time has arrived.
Anyone can then use the released IBE private key to decrypt
any ciphertexts that were encrypted for that time.

If the threshold group generates and releases IBE private keys
for ``time identities'' representing TLC logical clocks or block numbers,
or wall-clock times in a fixed-period schedule in which the group
promises to release exactly one private key per minute on the minute,
for example,
then this works fine.
Users of most applications will probably not want to time-lock their messages
for logical clocks or block numbers with no predictable correspondence
to wall clock time, however,
and using a fixed-period release schedule again defeats
the potential benefits of asynchronous operation.
If the fixed period is too short,
the group's TLC coordination may not keep up with it,
requiring the group sometimes to release multiple keys per TLC step
to ensure that messages encrypted to certain times aren't left un-decryptable
because of missing IBE private key releases.
If the fixed period is too long,
users (or smart contracts) are unnecessarily limited in the precision
with which they can schedule future information releases.

We can address this problem, however,
by agreeing on a convention between message encryptors
and the threshold group
that accounts for uncertainty in the future rate and schedule
of IBE private key releases.

First, message encryptors produce ciphertexts encrypted for not just one
but a logarithmic number of future wall-clock time ``identities''.
This is typically straightforward and efficient,
since messages are typically symmetric-key encrypted
with a random ephemeral key,
and that ephemeral key in turn public-key encrypted.
Encrypting to multiple future time identities
simply requires IBE-encrypting the ephemeral key several times,
increasing the message size only slightly and not multiplicatively.

In particular,
if the message sender's ideal desired time-release point is $t$,
then the sender first encrypts to the time identity
for the exact binary representation of $t$.
Then the sender performs IEEE floating-point-style round-to-larger
to round $t$ to an approximation $t'>t$
having at least one fewer significant bits than $t$ does,
and encrypts the message to the time identity corresponding to $t'$ as well.
The sender repeats this successive round-to-higher and encryption process
until the target time representation has only one significant bit.
In this way, the ciphertext will be decryptable by any of
a logarithmic-size set of approximations to the target time,
each less-precise approximation being more conservative (\ie, later).

When the threshold time vault beacon 
is operating asynchronously and periodically releasing IBE private keys,
it similarly releases not just one but a small (logarithmic) set of keys
at each TLC time step.
Suppose the previous block in the beacon's blockchain
was time stamped $t$ using the secure time stamping approach above,
and the next committed block built on it has time stamp $t' > t$.
The precise wall-clock time stamp delta from one block to the next, of course,
depends on asynchronous network communication progress,
unpredictable delays and jitter in the network and nodes,
and the QSC-randomized selection of the winning proposal each consensus round.

But regardless of the time stamp delta,
the threshold group releases IBE private keys for time identities
representing not just the new time stamp $t'$,
but also to the time identities resulting from rounding $t$ up
to larger binary numbers with progressively fewer significant bits,
and also to the time identities resulting from rounding $t'$ {\em down}
to smaller binary numbers with progressively fewer significant bits,
until these approximation processes ``meet in the middle''
at some $t_m$ such that $t \le t_m \le t'$.

This process ensures that the time vault beacon's release of IBE private keys
effectively traverses a binary tree of all possible time stamps,
producing a private key for a more-approximate time stamp
with fewer significant bits whenever the real time
representing that position in the conceptual binary time stamp tree passes.
Since message senders encrypt their messages to each possible precision,
corresponding to interior nodes in this binary time stamp tree,
the set of IBE private keys the time vault beacon generates
is guaranteed to ``hit'' one of the time identities 
the message sender encrypted the message for, eventually --
and with a maximum error approximately proportional to 
the time stamp delta between the TLC consensus rounds stamped
immediately before ($t$)
and immediately after ($t'$)
the sender's ideal target time step.

In this way,
senders can time-lock their messages (or schedule events using them)
for any desired time stamp at any precision,
without having to predict or guess the rate at which
the asynchronous IBE key-holder group will progress and release keys
at that future time.
The time vault beacon will release some key allowing decryption of the message,
with a varying time precision depending on how quickly or slowly
the group is actually progressing at that time due to network conditions.

%% file: causal.tex
\section{Robust, Efficient Causal Ordering}
\label{sec:causal}

{\em In preparation.}

\xxx{
Various approaches:

- Strawman 0: include cumulative histories in every message
	as discussed earlier. works, but completely impractical.

- Strawman 1: use vector clocks,
	and delay received messages until dependent messages received.
	Works, but only if nodes never crash and the network
	eventually delivers all messages.
	Otherwise, a missing dependency can deadlock a receiver.

- Strawman 2: use a classic reliable broadcast protocol.
	Works, reliably, but causes each node to send and receive
	$O(N)$ copies of each message.

- Solution: use an efficient causal gossip protocol.
	Each node maintains vector clocks,
	and in pairwise communication, uses IHAVE/SENDME type protocol
	to fill in all dependencies before an interaction
	is considered complete.

	optimization: compressed vector timestamps while no equivocation.

	optimization: subscriptions for eager transmission,
	approximating the efficiency of tree-based broadcast in steady state.
}

%% file: arch.tex
\section{A Coordination Architecture}
\label{sec:arch}

In the above expositions of TLC and QSC
we have made many simplifying assumptions for clarity,
such as assuming no Byzantine nodes and causally-ordered message propagation.
We also ignored many additional requirements,
and optional but ``nice-to-have'' features,
that we often want in a practical distributed coordination system.

We now address the challenge
of building practical asynchronous distributed systems
logically clocked by TLC.
Following the tradition of layering
in network architectures~\cite{zimmermann80osi,clark90architectural},
we will adopt a layered approach to build progressively
the functionality and abstractions we need to implement TLC and QSC
in a conceptually clean and modular fashion.
Consistent with layered architecture tradition,
each layer in our distributed coordination architecture
will depend on and build on only the layers below it
to add a minimal increment of functionality or abstraction
needed by or useful to the layers above it.

A key goal of this architecture
is to tolerate not only an asynchronous network
but also Byzantine node behavior.
The Byzantine consensus problem is traditionally addressed
using protocols specifically designed for this purpose~\cite{castro99practical},
which are fairly different throughout
from their non-Byzantine counterparts
such as Paxos~\cite{lamport98parttime,lamport01paxos}.
The architecture introduced here, in contrast,
shows how the application of relatively standard protection tools
in appropriate architectural layers,
such as cryptographic algorithms,
Shamir secret sharing~\cite{shamir79share,stadler96publicly,schoenmakers99simple},
and PeerReview~\cite{haeberlen07peerreview,haeberlen10accountable},
can make the QSC protocol described above Byzantine fault tolerant
without any fundamental changes to the consensus protocol's
basic logic or operation.

\begin{figure}[t]
\begin{center}
\begin{footnotesize}
\begin{tabular}{|c|l|}
\multicolumn{1}{c}{\bf Layer}	& \multicolumn{1}{l}{\bf Description} \\
\hline
consensus	& single globally-consistent historical timeline \\
\hline
randomness	& unpredictable, unbiasable public randomness \\
\hline
time release	& withholds information until designated time \\
\hline
threshold time	& communication-driven global notion of time \\
\hline
witnessing	& threshold certification that nodes saw messages \\
\hline
causality	& ensures nodes have complete causal history views \\
\hline
real time	& labeling events with approximate wall-clock time \\
\hline
messaging	& direct messaging between pairs of nodes \\
\hline
\end{tabular}
\end{footnotesize}
\end{center}
\caption{Layered architecture for threshold logical time and consensus
	atop asynchronous networks}
\label{fig:arch}
\end{figure}

Figure~\ref{fig:arch} briefly summarizes the layers 
of the distributed coordination architecture we develop here.
While important interdependencies
between the functional abstractions implemented by the layers
motivate this layering,
there is nothing fundamental or necessary
about a layered approach or this particular layering:
many other decompositions are certainly possible.

As usual, layering achieves modularity and simplicity of each layer
at a potential risk of introducing implementation inefficiencies
due to cross-layer coordination,
or potentially increasing the complexity of the complete system
over a tightly-focused and carefully-optimized  ``monolithic'' design.
Many ``cross-layer'' optimizations and simplifications
are likely to be possible and desirable in practical implementations.
This layering scheme is intended to be a conceptual model
to simplify reasoning about complex distributed coordination processes,
not a prescription for an optimal implementation.

While this architecture is driven by the goal of developing a
clean, modular, efficient, and practical approach
to asynchronous Byzantine consensus,
many of the lower layers that the architecture incorporates
can also serve other, general purposes
even in distributed systems that may not necessarily require consensus.
Shorter stacks comprised of only a subset of the layers described here
may be useful in such situations.

\xxx{ rename vector time layer to causality layer? }

\xxx{ motivations for this deep layering:
(a) keeping each layer conceptually simple, independent,
easy-to-understand separately, and nearly single-purpose;
(b) ``design for verification'' (either pencil-and-paper or mechanically checkable). }

\subsection{Basic elements of consensus}

Before developing the architecture layer-by-layer,
we first break down the architecture's layers
into three functional categories
representing three basic elements of consnesus:
choice, timing, and rapport between participants.

\paragraph{Choice:}
Consensus requires choosing making a choice among alternatives:
typically by choosing among either {\em leaders} or among {\em proposals}.
Leader-driven protocols such as Paxos and PBFT first choose a leader
and that leader drives the choices until the leader fails
(the detection of which typically requires timeouts, violating asynchrony),
resulting in a view change.
In randomness-driven consensus protocols such as Bitcoin and this,
participants first form {\em potential} choices for each round,
then we use a source of randomness to choose among them.
Supporting this choice among proposals is the purpose of
the randomness layer in our archiecture,
which in turn builds on the time release layer immediately below it.

\paragraph{Timing:}

In any consensus algorithm, nodes need to know {\em when} to make a choice
(or when a choice has been made),
either among proposals for a decision or among potential (new) leaders.
Much of the complexity of leader-based protocols is due to the difficulty
of coordinating the numerous timing-related decisions nodes must make:
when a proposal is accepted, when a proposal is committed,
when a node decides that a leader has failed,
when {\em enough} nodes have decided this to trigger a view change,
when a new leader knows enough to take over from the last one, etc.
Asynchronous consensus protocols fundamentally need to be threshold-based
rather than timeout-based in their timing decisions,
but while simple in concept (simply wait for ``enough'' messages to arrive),
the question of how many messages of what kinds are ``enough''
often tends to be complex.
Our architecture uses threshold logical time to decompose
all the main timing and progress decisions into a separate layer --
the threshold time layer --
that uses simple threshold logic to form a global logical clock
to drive all key timing decisions in the layers above it.

\paragraph{Rapport:}

Consensus participants need not only ``raw communication'' with each other,
but also in practice need a way to know
{\em what other participants know} at a given point.
This mutual understanding is often required for a node to know
when a sufficient number of {\em other} nodes know ``enough''
so that an important fact (such as a proposal's acceptance or commitment)
will not be forgotten by the group even if individual nodes fail.
While monolithic consensus protocols use integrated, ad hoc mechanisms
to achieve the inter-node rapport needed for consensus,
our architecture instead decomposes rapport-establishment functions
into separate lower layers that can be easily understood
and cleanly implemented.

In particular, three layers of our architecture are devoted
to three complementary forms of rapport-building.
Our {\em witnessing} layer enables nodes to learn
when a threshold of participants have seen and validated
a particular message or historical event.
Our {\em causality} layer enables nodes to reason causally about history
and determine precisely what events other nodes had seen
{\em before} a given message or event.
Finally,
our {\em gossip} layer ensures that participants
can exchange information and build rapport indirectly as well as directly,
so that correct nodes with slow or unreliable connectivity
may still be included as reliably and securely as possible in consensus.

\subsection{Four contrasting notions of time}

While this paper's central novel contribution
is the notion of asynchronous threshold time
and a distributed coordination and consensus architecture built on it,
this architecture also internally leverages and builds upon
other classic, complementary notions of time.
We utilize four different conceptual notions of time, in fact,
in different elements and for different purposes in the architecture:

\begin{itemize}
\item	{\bf Real time:}
	Although consensus in our architecture
	is driven asynchronously purely by communication
	and requires no timeouts,
	we nevertheless use real or ``wall-clock'' time,
	as measured by each node's system clock,
	to label blocks with the date and time they were committed,
	and to allow the {\em timed release} of contributions
	after designated moments in real time as described below.
	We assume that correct nodes' system clocks are roughly synchronized
	purely for these block-labeling and timed-release purposes,
	but Byzantine nodes' clocks may behave arbitrarily.

\item	{\bf Node-local log time:}
	For coordination and accountability purposes
	each node maintains its own tamper-evident append-only log
	of all the nondeterministic events it observes,
	such as message arrivals, as described below.
	Each such event increments
	the node's local {\em log time} by one,
	independent of real wall-clock time or other nodes' log timelines.

\item	{\bf Vector time:}
	As nodes communicate and learn about new events in other nodes' logs,
	each node $i$ maintains an $N$-element vector
	of the most recent local log times it knows about across all nodes.
	This {\em vector time}~\cite{fischer82sacrificing,liskov86highly,mattern89virtual,fidge91logical}
	represents the exact set of historical events across all nodes
	that are {\em causally prior} to the present moment at node $i$.
	Our architecture uses vector time
	to enable nodes to reason about what other nodes saw or knew
	at specific historical moments,
	and for systematic accountability
	via accountable state
	machines~\cite{haeberlen07peerreview,haeberlen10accountable}.

\item	{\bf Threshold logical time:}
	Finally, our consensensus architecture both provides
	and internally uses threshold logical time
	as a global metric of asynchronous communication progress
	across the (fastest threshold subset of) all $N$ participants.
\end{itemize}

\subsection{The consensus architecture by layer}

We now briefly describe the functional purpose
of each layer in the consensus architecture.
For clarity,
we also point out at least one simplistic potential
``baseline'' implementation approach for each layer.
These baseline implementations approaches are meant
only to be illustrative,
and would meet neither our efficiency nor security objectives in practice.
We defer the description of more practical and secure,
but also inevitably more complex,
implementations of these layers to Section~\ref{sec:byz}.

\xxx{ probably better to go top-down rather than bottom-up }

\paragraph{Messaging layer:}

The messaging layer represents
the baseline functionality this architecture builds on,
namely a primitive point-to-point communication capability
allowing message transmission directly between pairs of nodes.
In Internet-based deployments,
the messaging layer in our architecture typically maps to
connections via TCP, TLS, or another point-to-point overlay protocol.

This layer effectively represents the underlying network infrastructure
that this paper's architecture build on top of,
and thus is not implemented in this architecture
but instead represents the underlying network API (e.g., TCP/IP)
that the architecture builds on.

We do not assume the underlying messaging layer supports broadcast or multicast,
but can make use of such a facility if available.
If no broadcast/multicast is available in the underlying messaging layer,
then a broadcast/multicast simply means $N$ point-to-point transmissions
to each of the $N$ participants.

\paragraph{Real time layer:}
\label{sec:arch:realtime}

The optional real time layer enables the asynchronous,
self-timed group of nodes to interact correctly with wall-clock time
to enable time-based applications such as those
described in Section~\ref{sec:app:time}.
Upon transmitting a new message,
each node includes a wall-clock timestamp in the message
representing its local clock at the time of transmission.
Upon receiving a timestamp-labeled message,
correct nodes delay internal delivery of the message if necessary
until the receiver ``agrees'' with the sender
that the message's indicated timestamp has passed,
as detailed in Section~\ref{sec:time}.

\xxx{ insert optional wall-clock time layer here? 
Messages labeled with a particular wall-clock time on transmission
get delayed at the receiving end if necessary
until the designated wall-clock time passes by the receiver's clock.
This way, if the sender's or receiver's wall-clock time are way off,
this manifests either as a long perceived (virtual) wall-clock delay,
or a long actually-imposed delay if the receiver's clock is ahead.
In either case, it causes upper-level threshold protocols built on it
to ensure that a consensus proposal labeled with a given wall-clock time
cannot be committed until a threshold of correct nodes all
consider their wall-clock time to be no less than
the proposed block's wall-clock time label.
The time-release layer below can build on this to allow
time release based on wall-clock as well as threshold logical time...
}

\paragraph{Causality layer:}
\label{sec:arch:causality}

This layer provides ``rapport'' among nodes as discussed above,
by ensuring that whenever one node receives a message from another,
the receiver knows or effectively learns not just the message's content
but {\em everything the sender had observed} upon sending the message.

A simple implementation of this layer might simply tag all transmitted messages
with vector timestamps~\cite{fischer82sacrificing,liskov86highly,mattern89virtual,fidge91logical},
to define a precise ``happens-before'' causality relationship between events,
then use these vector timestamps
to delay the internal delivery of received messages (much as TCP does)
until causally prior messages have been received and delivered.
This simplistic approach works if nodes never fail
and messages are always eventually delivered,
but must be refined and augmented in practice to handle failures.

A more robust solution to causal delivery is for this layer
to build on a reliable broadcast protocol~\cite{lynch96distributed,cachin11introduction},
which ensure a message's eventual delivery to all nodes
provided not too many nodes fail or misbehave,
but typically require each message to be rebroadcast by $O(N)$ nodes.
A more practically efficient approach to achieving this robustness
is to build on gossip protocols~\cite{demers87epidemic,li06bar},
which handle only sparsely-connected networks
and are easily adapted to enforce causal ordering
at the level of pairwise interactions between nodes.
Section~\ref{sec:causal} discusses these approaches in more detail.


\paragraph{Witnessing:}

The witnessing layer allows a node $i$ that sent some message $m$
to learn -- and further indicate to other nodes -- 
when some threshold $T_w$ of participants
have received and confirmed seeing $m$.
The witnessing layer thus performs a function
analogous to acknowledgment in reliable point-to-point protocols like TCP,
but generalized to group communication contexts
in which many of the $N$ participants are expected to receive and witness
(acknowledge) each message that any group member receives.

Once a node has collected $T_w$ witness acknowledgments on a message $m$,
we say that the message has been {\em witnessed}.
This layer delivers received messages locally to upper layers
only once they are witnessed.
The important property this behavior enforces
is that once a higher layer on any node $i$
has received a message $m$ from another node via the witnessing layer,
$i$ knows that at least $T_w$ nodes total have seen and validated $m$,
not just $i$ itself.
Upper layers can control when and under what conditions
the witnessing layer starts or stops witnessing messages of specific classes
(\eg, messages labeled with particular identifiers in their headers),
to enforce semantic and temporal properties defined by the upper layer.

A trivial implementation of this layer,
which assumes that all nodes are well-behaved,
is simply for each node to reply with an acknowledgment
to each new upper-layer protocol message the node receives,
just as TCP/IP and numerous other classic protocols do.
Byzantine-protected instantiations outlined in Section~\ref{sec:tlc:byz}
use digital signatures
to produce transferable but cryptographically unforgeable ``evidence''
of message receipt,
and use threshold signing~\cite{shoup00practical,boldyreva03threshold}
or witness cosigning~\cite{syta16keeping,ford16apple,nikitin17chainiac}
to compress the space and verification costs of reducing $O(N)$
witness cosignatures on the same message.
In addition, Byzantine-protected implementations of this layer
can offer {\em proactive} protection against equivocation
and other detectable forms of misbehavior,
because honest nodes will not witness-cosign invalid or equivocating messages.

\paragraph{Threshold time:}

The threshold time layer implements a global notion of time
operating in lock-step across all the $N$ nodes,
in contrast with the node-local sequence numbers and clocks
implemented by the vector time layer.
In essence, at any given threshold time $t$,
the threshold time layer at each node $i$
waits until it has received time $t$ messages
from a threshold of $T_m$ unique nodes
before advancing to time $t+1$.

Since the threshold time layer builds upon the witnessing layer,
the collection of $T_m$ messages a node $i$ needs to advance to time $t+1$
is in fact a collection of $T_m$ {\em witnessed} messages,
each of which has been witnessed (acknowledged or signed)
by at least $T_w$ nodes.
In addition, the threshold time layer at any node $i$
uses its control over the witnessing layer to start witnessing messages
labeled with time $t$ only upon reaching time $t$ and not before,
and to stop witnessing time $t$ messages upon reaching time $t+1$,
ensuring in effect that messages witnessed by correct nodes at time $t$
were witnessed {\em during} logical time $t$ and not before or after.
This interaction between the witnessing and threshold time layer
ensures the important property that upon advancing to time $t+1$,
each node $i$ knows that at least $T_m$ messages from time $t$
were each seen (and witnessed) by at least $T_w$ nodes
during threshold time $t$.

\xxx{ this probably needs to be moved to the Byzantine protection section later, leaving only a simplified non-Byzantine version here...}

Since each node may reach its condition for advancing from $t$ to $t+1$
at different wall-clock times,
logical time advancement generally occurs
at varying real times on different nodes.
In a Byzantine consensus context where $N \ge 3f+1$
and $T_m = T_w = 2f+1$,
however,
we can divide wall-clock time into {\em periods} demarked by the moment
at which exactly $f+1$ correct nodes have reached a given threshold time $t$.
That is, wall-clock time period $t$ starts the moment
any set of exactly $f+1$ correct nodes have reached threshold time $t$,
and ends once any set of exactly $f+1$ correct nodes reach threshold time $t+1$.
Because a majority of ($f+1$) correct nodes must witness a time $t$ message
in order for it to be successfully witnessed and delivered
to the threshold time and higher layers,
and no correct node will witness a time $t$ message after advancing to $t+1$,
this means that a message formulated after the end of global period $t$
can never be successfully witnessed by the required threshold of $T_w$ nodes,
and therefore will never be delivered to upper layers on any correct node.
\xxx{ unpack and illustrate this reasoning. }

\paragraph{Time release:}
\label{sec:arch:release}

This layer schedules information to be revealed at a later time,
which might be defined either based on a threshold time,
as needed by QSC to protect lottery tickets,
or based on a target wall-clock time,
as needed by applications such as smart contracts
and time vaults (see Section~\ref{sec:app:time:vault}).

\xxx{PJ: It seems~\cite{cachin01secure} calls this property ``secure causality''.}

In a trivial implementation for a non-Byzantine environment,
each node simply labels information with the threshold time
it is supposed to be released,
and the (trusted) time release layer implementation at each node $i$
is responsible for delaying the release of that information to upper layers
until node $i$ has reached the designated release time $t$.
This simple implementation approach might be suitable
in a cluster, cloud, or data center context
in which all the nodes' implementations of this layer are under control
of the same administrative authority anyway,
or in a hardware-attested context such as within
Intel SGX enclaves~\cite{intel14software}.

Byzantine-protected implementations of this layer
instead typically use
verifiable secret sharing (VSS)~\cite{shamir79share,stadler96publicly,schoenmakers99simple},
together with threshold identiy-based encryption~\cite{boneh01identity,baek04identity,waters05efficient,kate10distributed}
to encrypt the time-release information such that a threshold of nodes
must release shares of the decryption key
upon advancing to the appropriate time,
enabling any node to be able to learn the encrypted information.

\paragraph{Public Randomness:}

This layer builds on cryptographic commitment and time release layer
to provide unpredictable, bias-resistant public randomness
at each threshold logical time-step $s$.
It is needed both by the Byzantine-hardened QSC consensus protocol,
and useful in practice for many other purposes 
outlined in Section~\ref{sec:app:random}.

A trivial implementation is just to pick a random number
and transmit it via the time release layer.
Practical Byzantine-protected implementations typically
generate public randomness via
secret sharing~\cite{cachin05random,syta17scalable},
such as the PVSS-based approach
outlined in Section~\ref{sec:byz:pvss},
or using a more efficient asynchronous randomn beacon set up using DKG
as discussed in Section~\ref{sec:dkg}.

\paragraph{Consensus:}

The consensus layer, finally,
builds on the abstractions provided by the lower layers
to implement robust, efficient consensus
enabling the group to agree on a serialized history.
QSC3 (Section~\ref{sec:cons})
achieves this in a fail-stop threat model,
while QSC4 (Section~\ref{sec:byz:pvss})
provides protection against Byzantine nodes.
There are certainly many other ways to implement the consensus layer, however,
which will embody different sets of tradeoffs
and dependencies on lower layers.
Again, this layering scheme is suggested as a conceptual reference,
not a prescription for a specific implementation of any or all the layers.

%% file: stamp.tex
\section{Threshold Timestamping}
\label{sec:stamp}

Goal here is to answer
{\em existence} and 
{\em temporal precedence} questions,
as in timeline entanglement~\cite{maniatis02secure}.
But purely opportunistic timeline entanglement
does not provide any guarantee 
that the system {\em will} be able to provide an answer
with respect to any two events,
even when those two events are well-separated in time.
Example: two subsets of nodes might be well-entangled with each other,
but rarely or never entangle across the subsets.
Incorporating thresholding logic
gives us a way to {\em guarantee} that for any two events,
if separated across (how many?) epochs,
we will {\em definitely} be able to create a Merkle proof
that one preceeds the other.

\subsection{Fairness and Starvation Avoidance}

(maybe should be in witnessed TLC section...)

...starvation problem:
slow nodes might never get to contribute.
We'd like to ensure that even extremely slow nodes
get to contribute and entangle their timelines regularly
at least once every $k$ epochs for some finite $k$.

Solution: ensure that when a node is trying to send a message labeled time $t$
and receives evidence that ``time has moved on'' to time $t' > t$,
the node does not need to start over.
Instead, the node should be able
to keep the witness acknowledgments or signatures it has for $t$
and just keep using the new signatures it is receiving,
keeping track of which (potentially-newer) time each ack/sig is for,
and when it has a threshold,
just publish that threshold as its contribution
to a {\em range} of epochs.

Implementation 1: simply use individual signatures.

Implementation 2: use dynamically-aggregable signatures,
such as BLS signatures.
Then, just need to keep one aggregate signature per timestep,
each with a bitmask.
In this case, need to gossip signatures individually during timesteps,
but can 

Implementation 3: use dynamically-aggregable signatures,
but gossip with integer-vectors.
In this case, even during gossip,
only ever need to send one aggregate sig at a time,
together with an integer vector of the number of duplicates of each.

How big do the integers need to be?
If we impose the rule that we only aggregate
when it's useful -- i.e., when at least one previously-zero vector element
becomes nonzero in a node's accumulated aggregate signature.
This means there can be at most $n$ useful aggregations,
and any vector element can hence have been doubled at most $n$ times.
We can impose the invariant that in any aggregate
with $k$ nonzero vector elements,
each vector element must have a value of at most $2^{k-1}$.
Thus, each vector element needs at most $n$ bits for $n$ nodes total.

%% file: opt.tex
\section{Protocol Optimizations and Variations}
\label{sec:opt}

\xxx{ merge into appropriate consensus architecture layer sections}

\subsection{Optimized Gossip}

Optimizing for minimum number of messages without pairwise chatter: use matrix
clocks to bundle up everything that might be new to a target node when sending
to them.

Optimizing for bandwidth-efficiency in contexts where pairwise relationships
are bidirectional channels (e.g., TCP connections): lightweight
offer/request/transfer protocol, UseNet-like, to avoid transferring big data
chunks unnecessarily.  

Speculative transmission start: in latter case, perhaps still use matrix clocks
and randomized speculative transmission to get transfers started as quickly as
possible while maximizing chance that the speculatively transmitted data
chunk(s) will be novel to the target.

(opportunistic gossip? driven by collective signing?)

Ensures that in the case of a triangle inequality violation (TIV),
where the a link $AC$ proves slower than an indirect path $ABC$
(even just briefly due to a transient failure or slowdown of the $AC$ link),
the nodes can opportunistically take advantage of the indirect path.

\subsection{Blockchain Consensus}

Iteration.  Previous block selection.  Proposing provisional blocks so that a
block gets added even when consensus doesn’t succeed (or it may not be
universally known that consensus succeeded).

\subsection{Pipelinined Consensus}

Can initiate a new consensus round in each time-step.  Each is resolved only at
the Reveal phase at the end.

At the proposal phase, don’t yet know the predecessor of the proposed block.
But the proposed block fixes everything necessary to choose the predecessor
(i.e., the exact set of eligible predecessors), other than the randomness that
gets used to choose the precise predecessor.  The precise predecessor is chosen
later when the randomness for the potential predecessors are revealed; then we
just take the predecessor with the numerically smallest ticket.  (Which might
or might not be committed by itself.)

Application specifically to Fat and Thin Consensus.

Must allow nodes to propose blocks that may innocently conflict with
transactions posted within a window of recent time-steps.  The
innocently-conflicting transactions are determined only once a committed
ordering is established and deterministically evaluated as aborted
transactions.  (Cite deterministic databases etc.)

\subsection{Fairness, Censorship, and Vector Consensus}

An asynchronchronous network adversary can by definition
censor all the blocks proposed by up to $f$ honest nodes.
But how can we minimize the advantage of censoring adversaries in practice?

First, by ensuring that when honest pairs of nodes
are well-connected and not vulnerable to delays or DoS attacks,
those honest nodes at least cannot be censored.

Second, by ensuring that the adversary's advantage is limited:
\eg, even if the adversary can get an outsize proportion of the outputs,
it can't get a {\em majority}...
Vector consensus: see "Byzantine consensus in asynchronous message-passing systems" and within that: (Doudou et al., 1999).

\subsection{Threshold Key Generation and Regeneration}

Initially, assume manual “genesis block” creation event that involves all
parties or a manually-chosen threshold set of initial parties.

Then we periodically need to regenerate groups as nodes come and go, or as
individual nodes might be found to be compromised and need to be expelled or
replaced.  The membership might even be automatically-rotating, for example if
it is produced from proof-of-work from a ByzCoin-like hybrid consensus
mechanism [ByzCoin, Pass/Shi] or a proof-of-stake mechanism [], explored in
more detail later.

At each regeneration event we have two options: either regenerate only the
individual shares while keeping the group’s composite public key constant, or
generate a fresh public joint public key as well as all shares.  [summarize
upsides and downsides of each, and why we probably want both, the latter less
frequent than the former.]

In either case, we have a slight chicken-and-egg problem that we’d in principle
like all parties to participate in the DKG or DKR (re-sharing) process, but if
we expect all parties to be present, we risk availability and create
vulnerabilities to DoS attack.  Thus, we’d like to require only a threshold,
e.g., 2f+1, of parties to provide contributory polynomials and shares - but
then we have the problem of deciding which particular threshold set of 2f+1
polynomials to use, since the choice clearly affects the final distributed key
or sharing and we need one but only one (and security might be compromised
through related-key attacks if more than one distinct set of related
polynomials get used?).  Hence we have a problem needing consensus, in order to
set up for the consensus algorithm we want to run.

One solution is to increase the sophistication of the DKG protocol [cite and
summarize Goldberg et al approach].  But if this DKG or DKR is a periodic
process that we normally expect to run merely to set up a next-generation
shared key or consensus group while a previous-generation key and consensus
group is still in operation, then we can use the previous consensus group to
agree on one and only one set of 2f+1 standard key-sharing polynomials for the
next-generation group.

Also, for efficient public verifiability, can use XXX (the verification
optimization based on erasure codes)…

\subsection{Permissionless Membership Protocols}

Membership via Proof-of-Work 

Membership via Proof-of-Stake

%% file: proto.tex
\section{Protocol Details}
\label{sec:proto}

(for long tech report version, not publication version)

Assume links are TLS or similar encrypted, authenticated, stream connections.

protocol messages on a given link:
(a) time-increment message proposal, containing
	new upper-level message and bitmap of
	threshold of messages received from last timestep,
	hash of timestep signature share,
	hash of random beacon share;
(b) i-want message, indicating updated vector of sequence number ranges
	of desired and not yet received messages from other nodes;
(c) here-is message, containing a message or sig forwarded from some node

When I receive a new message via gossip,
via any route direct or indirect,
I remember (for the rest of this time-step?)
which node I received it from,
in order to return (signature) responses along the same path?

But I also always (try to) return a signature
on the direct link from the message sender regardless...

bandwidth-conservative policy: 
request sequence numbers only when I know they exist and want them.

latency-conservative policy:
requst all future sequence numbers (upper bounds are infinite)
so that peers start sending new messages whenever available.

%% file: analys.tex
\section{Formal Development of TLC}
\label{sec:analys}

{\em In preparation. }

\xxx{

We define TLC not as a single protocol but a family of protocols.
A particular instance \tlc{\tm}{\tw}{n}
is parameterized by 
a message threshold $\tm$,
witness threshold $\tw$,
and total number of nodes $n$,
where $0 < \tm \le n$ and $0 \le \ta \le n$.
Intuitively, $\tm$ is the number of qualified messages each node must collect
from time step $s$ in order to progress to step $s+1$,
and $\tw$ is the number of nodes that must first witness --
\ie, acknowledge having seen -- each such message
in order for it to count towards the $\tm$ threshold.
We say that an instance of TLC is {\em witnessed} if $\tw>0$
and is {\em unwitnessed} otherwise.

For simplicity, we will assume for now that all nodes are correct
and leave handling of Byzantine node failures until later
in Section~\ref{sec:byz}.
To be clear, we allow the {\em network} to behave adversarially
in its message delivery scheduling,
but merely assume for now that none of the $N$ {\em nodes} are adversarial.

We introduce the following three properties to analyze progress in terms
of the logical time:
\begin{compactenum}
    \item {\em Monotonicity}: Logical time only advances.
    \item {\em Pacing}: Logical time never advances ``too quickly''.
    \item {\em Liveness}: Logical time advances regularly.
\end{compactenum}
\xxx{TODO: We probably need proper technical definitions for those properties.}
\xxx{Investigate technical definitions by Cachin~\cite{ cachin01secure,cachin05random}, Loss~\cite{loss18combining},
and Abraham~\cite[page 4]{abraham18vaba}}

\xxx{Add lock-step time-bracketing property for witnessed TLC
	with message and witness thresholds being a majority of correct nodes? }

The rest of this section is organized as follows: We first explain
unacknowledged and then acknowledged TLC. For clarity, we will first describe
TLC in a simplified threat model in which the adversary cannot corrupt nodes
($f_c=0$), but can indefinitely delay messages ($f_d \ge 0$). We address
Byzantine node failures later in Section~\ref{sec:tlc:byz}.

\com{
\xxx{ gossip? }
We will also simplistically assume for now that
whenever any node $i$ broadcasts any message,
that message includes $i$'s complete causal history:
an ordered list of all the messages $i$ ever received up to this moment.
While  ...
}

\input{a-threat}

\input{a-unack}

\input{a-acked}

\input{a-cons}

}

%% file: a-threat.tex
\subsection{Network and Threat Model}
\label{sec:model}

Consistent with most consensus literature,
we assume a well-known, fixed set of $n$ participating nodes.
In practice this consensus group must be chosen somehow,
and periodically refreshed or reconfigured
for long-term availability and security.
In an open, ``permissionless'' blockchain setting, for example,
consensus groups might be chosen via a rotating window
of proof-of-work miners~\cite{kokoris16enhancing,pass17hybrid},
or by sampling subsets of larger collectives
for ``scale-out''~\cite{kokoris17omniledger,luu16secure}.
Consensus group selection and maintenance
is outside this paper's scope, however.

\subsubsection{Cryptographic Tools}

We assume each node $i$
holds a signing key-pair $(K_i,k_i)$,
where node $i$ uses its secret key $k_i$ to sign messages it sends,
while all nodes know and use $i$'s public key $K_i$ to verify signatures.
We leave key distribution out of scope.

We assume a publicly-verifiable threshold secret sharing
scheme~\cite{shamir79share,stadler96publicly,schoenmakers99simple},
which we use to provide a public source of
randomness~\cite{cachin05random,syta17scalable}
that a threshold $T_r$ of nodes can reveal cooperatively,
while fewer than $T_r$ can neither learn any information about nor bias.
For simplicity, we assume for now that the secret-sharing scheme has been setup
by a trusted dealer
or some existing distributed key generation (DKG)
protocol~\cite{cachin02asynchronous,zhou05apss,kate09distributed},
but we revisit DKG in the TLC framework
later in Section~\ref{sec:dkg}.

\subsubsection{Threat Model}
\label{sec:threat}

Our threat model decouples the adversary's power
to attack a protocol's {\em progress} (liveness and performance)
from its power to attack {\em integrity} (node and message state),
maximizing each attack power independently.
This decoupling is analogous to that of UpRight~\cite{clement09upright},
which distinguishes between {\em failures of omission},
or failure to produce or deliver an expected message,
from {\em failures of comission},
or state corruption.
Despite this decoupling
we still assume one coordinated adversary,
not two:
\eg,
the adversary can use information learned from attacks on integrity
in subsequent attacks on progress.

Our adversary attacks progress
via power to schedule and delay message delivery.
Asynchronous network models traditionally require
that the network {\em eventually} deliver all messages,
but permanent node or link failures,
denial-of-service, and routing attacks~\cite{apostolaki16hijacking}
realistically can delay communication indefinitely
between at least some honest nodes.
In addition, asynchronous network models often assume
message delivery is arbitrary but non-adversarial:
\eg, that the ``scheduler is oblivious to the
Byzantine process behavior''~\cite{friedman05simple}.
Advanced network adversaries today
such as compromised ISPs or local-area WiFi routers, however,
realistically can delay, reorder, and replay messages adversarially
based on all available information
including what they learn from corrupted nodes.

Reflecting this reality,
our adversary can choose any set $S_d$ of up to $f_d$ (typically $f$) nodes
whose messages it may {\em delay indefinitely};
other nodes' messages it must deliver eventually
on some adversary-chosen schedule.
When any node $i$ sends a message,
if $i \not\in S_d$, the adversary must 
{\em schedule} the message
by committing to some arbitrary future delivery deadline
the adversary chooses adaptively.
If $i \in S_d$, however,
the adversary need not (yet) schedule the message for delivery at all,
but can instead hold it indefinitely.
Our network adversary is {\em constantly-adaptive},
in that it can change its indefinite-delay set at any time.
Whenever the adversary changes its set from $S_d$ to $S'_d$,
the adversary must immediately schedule
and commit to some delivery deadline for
in-flight messages of nodes leaving the indefinite-delay set ($S_d - S'_d$);
otherwise the adversary could trivially
delay {\em all} messages indefinitely.
Once a message is scheduled for delivery,
the adversary can advance the message's delivery deadline at any time
(\eg, to reorder messages adaptively),
but cannot further delay delivery beyond the message's committed deadline.

The adversary attacks integrity via the power
to {\em corrupt} nodes, and messages they send, adaptively.
All $N$ participating nodes start out {\em honest},
following the prescribed protocol.
The adversary can corrupt honest nodes adaptively at any time,
choosing when and which honest node(s) to corrupt next
based on any information it has learned
up to the point of corruption.
Once corrupted, the adversary learns all the node's secrets
including private keys,
and can produce arbitrary messages signed by the node.
Because leaked secrets cannot be un-leaked,
the node remain corrupted for the consensus group's remaining lifetime.
While periodic group reconfiguration and key refresh
can enable recovery from corruption in practice,
we leave such mechanisms out of scope.
The adversary can corrupt at most $f_c$ nodes,
a parameter usually set to a global fault-tolerance threshold $f$.
The adversary can replace in-flight messages
sent by the corrupted node,
up to the moment of delivery to honest nodes.
This implies that the adversary can {\em equivocate}
by delivering different correctly-signed messages
to different honest receivers.

Corruption power alone affects only node and message state, not timing.
Even corrupted nodes must produce {\em some} message
whenever the protocol requires one,
although the adversary's progress-attack power above
can delay its delivery indefinitely.
We define a message's {\em label} as its metadata uniquely identifying
the message's role and logical position in the protocol,
\ie, a type of message other nodes may expect and wait for.
Changing a message's label from $l$ to $l'$ constitutes
both an attack on integrity (a failure of comission)
and an attack on progress (a failure of omission),
as other nodes expecting a message with label $l$ may never see one.
To model this reality, our attacker can corrupt message labels
only if it also keeps the sending node in its indefinite-delay set
until and unless the corrupt node subsequently delivers to all recipients
{\em some} message with the orginally-expected label $l$.

Because the indefinite-delay set $S_d$
is independent of the corrupted node set $S_c$,
the adversary has ``attack powers'' over up to $2f$ nodes:
state corruption power over up to $f_c=f$ nodes and
indefinite-delay power over up to $f_d=f$ additional nodes,
with constantly-adaptive control over the latter set.
Constraining our adversary to set $S_d$ equal
to the set of corrupted nodes $S_c$ at all times
reduces its power to that of conventional adaptive adversaries
who can indefinitely delay only messages from corrupt nodes.
Allowing the adversary indefinite delay power ($f_d>0$)
without corruption power ($f_c=0$)
emulates a classic asynchronous network with $f_d$ fail-stop nodes.

While the above describes our main ``target'' threat model,
we will use weakened versions of this threat model
in developing straw-man or simplified protocols below for expository purposes,
stating in each case any special constraints we place on the adversary.

%% file: a-unack.tex
\subsection{Unacknowledged TLC (fail-stop case)}

\com{
\begin{algorithm}[ht]
\begin{algorithmic}
    \State $\lts \gets 0$
    \While {true}
        \State $M_\lts \gets \{ \msg_{\lts i} \}$
        \State broadcast($\msg_{\lts i}$)
        \While {true}
            \State $\msg \gets$ receive()
            \If {$\msg.\lts = \lts \wedge \msg \notin M_\lts$}
                \State $M_\lts \gets M_\lts \cup \{ \msg \} $
                \If {$|M_\lts| = \tm$}
                    \State $\lts \gets \lts+1$
                    \State \textbf{break}
                \EndIf
            \EndIf
            \If {$\msg.\lts > \lts$}
                \State $\lts \gets \msg.\lts$
                \State \textbf{break}
            \EndIf
        \EndWhile
    \EndWhile
\end{algorithmic}
\label{alg:utlc}
\caption{Unacknowledged TLC for node $i$}
\end{algorithm}
}

In unacknowledged TLC, where $\ta=0$, a node $i$ broadcasts an
application-defined message $m_{\lts i}$ every time it reaches a new time slot
$\lts \geq 0$. At the beginning of time, all nodes have both the right and
obligation to broadcast an initial message labeled $\lts=0$.

Each node $i$, after having sent a message $m_{\lts i}$ for a given time slot
$\lts$, has the following two options to have both the right and obligation to
advance its logical time:
\begin{compactenum}

    \item Receive messages $m_{\lts j}$ from at least $\tm$ distinct nodes
        $j \neq i$ to advance from $\lts$ to $\lts+1$.

    \item Receive a message $m_{\lts' j}$ from at least one other node~$j$ to
        advance to $\lts' > \lts$ while skipping any intervening time slots
        strictly between $\lts$ and $\lts'$ that $i$ might have missed.

\end{compactenum}
While the first rule realizes regular collaborative time progression by at least
$t_m$ nodes, the second one introduces a viral time synchronization mechanism,
as with Lamport clocks, enabling nodes to catch-up quickly with their peers if
necessary.

\xxx{ rename Raw TLC or Basic TLC or something simpler like that? }
\com{

In unacknowledged TLC, where $\ta=0$,
nodes send messages only once
at the start of each consecutive logical time slot
($\lts=0$, $\lts=1$, etc.).
Each node $i$, upon reaching slot $\lts$,
prepares an application-defined message payload $M_{\lts i}$,
labels it explicitly with integer time $\lts$,
and broadcasts the message to all $n$ nodes.

At the beginning of time,
all nodes have both the right and obligation
to broadcast an initial message labeled $\lts=0$.
Upon reaching any time-step $\lts$ and broadcasting a message labeled $\lts$,
each node $i$ must wait to receive messages
labeled with time $\lts$ from at least $\tm$ distinct nodes,
before $i$ can advance its logical time to $\lts+1$.
Once it collects this threshold of $\tm$ messages at time $\lts$,
node $i$ has both the right and obligation
to advance its logical time and broadcast a message labeled $\lts+1$.

As with Lamport clocks, time advancement in TLC is viral.
If node $i$ reaches time $\lts'>\lts$ and broadcasts a message $m_{\lts'i}$,
and another node $j$ receives $m_{\lts'i}$ while still at earlier time $\lts$,
then $j$ immediately ``catches up'' to time $\lts'$.
On this event, node $j$ broadcasts its own message $m_{\lts'j}$ at time $\lts'$,
skipping any intervening time-steps strictly between $\lts$ and $\lts'$
that $j$ might have missed.

In summary, there are two and only two ways a node in unacknowledged TLC
can advance from time $\lts$ to time $\lts'>\lts$:
(a) by receiving at least $\tm$ messages sent by distinct nodes at $\lts'-1$, or
(b) by receiving {\em any} message from another node
that had already reached time $\lts'$.
}

Note that network communication alone drives each node's advancement of logical
time. If the application protocol layered atop TLC does not have anything useful
to send when node $i$ satisfies the conditions to advance from time $\lts$ to
$\lts+1$, then TLC advances time anyways and broadcasts an empty application
message $m_{\lts+1, i}$.

Further note that the adversary can prevent individual nodes from advancing
their notion of time by delaying the delivery of up to $f_d$ nodes' messages
indefinitely, as discussed above in Section~\ref{sec:threat}.

\subsubsection{Time Progress Properties}
\label{sec:unack:properties}

\xxx{ diagram: epochs separated by events of majority of nodes
	reaching next logical time-step }

\com{
TLC ensures that the logical clocks it produce satisfies
three useful {\em progress properties}:
informally,
that logical time only advances ({\em monotonicity}),
that logical time never advances ``too quickly'' ({\em pacing}),
and that logical time regularly advances ({\em liveness}).
}
We now analyze each of the initially introduced time progression properties.

\paragraph{Monotonicity:}
We wish to ensure that for all TLC nodes,
time only ever advances when it changes at all:
\ie, if a node $i$ has logical time $\lts$ at some real-world moment $m$
and has logical time $\lts'$ at some later moment $m'>m$,
then $\lts' \ge \lts$.
The rules of time advancement trivially ensure monotonicity.

\xxx{Notation for real world time $m$ collides with the one for messages}

\paragraph{Pacing:}
We wish to ensure that no node or set of nodes smaller than $\tm$
can ``race ahead'' of the rest.
Specifically, at the moment any node $i$ advances to time $\lts+1$,
we require that there exist at least $\tm$ nodes at time at least $\lts$
(\ie, no more than one time-step behind $i$).
This property is ensured by a trivial inductive argument over time-steps
and the two conditions above for time to advance.
Ensuring this property will become slightly less trivial
later when we consider Byzantine node corruptions, of course.

\com{XXX this property probably needs more work, and not sure if it's needed.
\paragraph{Unison:}
We wish to ensure that no node or set of nodes smaller than $\tm$
can ``lag behind'' the rest.
This property is complementary to pacing but slightly more tricky to define,
because the adversary can clearly schedule a node or minority group
to learn about a logical time advance arbitrarily long after
the rest have moved on;
allowing for this case is the whole point of fault tolerance.
The disfunction we wish to avoid is not for a node to {\em be} behind,
but to {\em progress independently} while {\em remaining} behind,
without ``catching up'' to others at each step.
We therefore define pacing with respect to two consecutive time steps.
Specifically,
if any node $i$ advances to logical time $\lts$ at moment $m$
and then to $\lts'>\lts$ at later moment $m'>m$,
then we require that at moment $m$ there were at least $\tm$ nodes
that had logical time no greater than $\lts'$.
This property again follows trivially from the rules of time advancement,
in the absence of Byzantine node corruption.

\xxx{ is ``lockstep'' a better term than ``unison''? }
}

\paragraph{Liveness:}
We wish to ensure that TLC can never ``get stuck,''
such that all nodes stop advancing forever,
despite the adversary's ability
to schedule messages arbitrarily and indefinitely delay
the broadcasts of up to $f_d$ nodes.
Unacknowledged TLC satisfies this property provided $n - f_d \ge \tm$,
which we can prove by contradiction.

For TLC to get stuck,
there must be some {\em last} moment $m$ in real-world time
when any node makes progress by advancing its logical time.
Let $L$ be the set of {\em live} nodes that are not
in the adversary's indefinite-delay set {\em continuously forever}
from moment $m$ on.
Since at most $f_d$ nodes can be in the indefinite-delay set
continuously forever from $m$, clearly $|L| \ge \tm$.
By moment $m$,
each node $i \in L$ has broadcast some message $M_i$
the last time $i$ advanced its logical time.
Since the nodes in $L$ are not indefinitely delayed forever,
each final broadcast $M_i$ is eventually delivered to all nodes in $L$.
Let $m'$ be the moment the last of these final broadcasts
is delivered to the last recipient in $L$.
If at $m$ all nodes in $L$ were at the same logical time $\lts$,
then by $m'$ each node in $L$
has received a threshold $\tm$ of time $\lts$ messages
and will advance to time $\lts+1$.
If at $m$ the nodes in $L$ were at varying logical times,
then by $m'$ the final broadcast of some node $n \in L$
that was at the highest logical time $\lts$ at $m$
will have reached some other node $n' \in L$
that was at a lower logical time $\lts'<\lts$ at $m$,
causing $n'$ to advance its time to $\lts$ virally.
Either way, by $m'$ at least one node in $L$
has advanced its logical time beyond its state at moment $m$,
contradicting the original assumption that no node's logical time
increases after moment $m$,
and thus proving TLC's liveness.

Because ``final'' messages from all nodes in $L$
must be in flight (if not already delivered) by moment $m$,
and the adversary must schedule those messages for delivery
if a node leaves the indefinite-delay set even momentarily
(Section~\ref{sec:threat}),
the adversary must eventually deliver these messages
despite any changes to its indefinite-delay set.
Further, the set $L$ can be different from one progress event to the next,
so changing the indefinite-delay set continuously
does not help the adversary across progress events.

%% file: a-acked.tex
\subsection{Acknowledged TLC (fail-stop case)}
\label{sec:acked}

\xxx{ renamed Witnessed TLC (WTLC)? }

\xxx{ architecture diagram:
\begin{center}
\begin{tabular}{|c|}
\hline
Upper Level Protocol (ULP) \\
\hline
Threshold Logical Clock Protocol (TLCP?) \\
\hline
Witness Cosigning Protocol (WCP?) \\
\hline
Overlay/Network Protocol (ONP) \\
\hline
\end{tabular}
\end{center}
}

\xxx{ in explanation, the ``time frontier'' for time $t$:
	the varying points in real time at which the various nodes
	reach time $t$.
	If we sort these chronologically,
	then the moment the $f+1$st node crosses into time $t$
	represents a global {\em epoch boundary}
	defining the start of epoch $t$.
}

\xxx{ one way to think of the protocol: ``push-pull''.
	first push my contribution to a threshold of nodes,
	then my contribution is available to be pulled by nodoes
	as evidence to make progress to the next time step.
}

The unacknowledged TLC protocol above
makes it easy to reason about the amount of information that propagates
{\em to} a particular node across logical time-steps.
That is, whenever any node $i$ reaches some logical time $\lts+1$,
we know that $i$ has ``seen'' and learned information from
the viewpoints of at least $\tm$ nodes at prior logical time $\lts$.
However, in unacknowledged TLC it is not so trivial to reason
about information propagation {\em from} a particular node into the future
(though we will develop a way to do this later in Section~\ref{sec:thin}).
That is, once any node $i$ has reached and broadcast a message at time $\lts$,
how many nodes will have seen that message by time $\lts+1$?

\xxx{TODO: why is that important}

To make it easy to reason in {\em both} directions across logical time-steps,
acknowledged TLC requires nodes to wait longer before advancing time
to gather information on who has seen which messages.
In an acknowledged TLC instance \tlc{\tm}{\ta}{n} where $\ta>0$,
each node at time $\lts$ waits
not only to receive $\tm$ messages from distinct nodes at time $\lts$,
but also to receive at least $\ta$ {\em acknowledgments}
of each of these messages from distinct nodes.
Thus, for any node $i$ to progress from $\lts$ to $\lts+1$,
not only must $i$ have seen $\tm$ messages from time $\lts$,
but $i$ must know that not only $i$ itself but a threshold of $\ta$ nodes
have seen each of those $\tm$ messages.
Since each of these $\tm$ messages requires $\ta$ acknowledgments,
$i$ must effectively build a $\tm \times \ta$ matrix of acknowledgments
before it is allowed to advance to time slot $\lts+1$.

When any node $i$ reaches logical time slot $\lts$, 
it broadcasts a time-progress message $\msg_{\lts i}$ with an upper-level payload
as in unacknowledged TLC,
then starts waiting for the messages and acknowledgments of other nodes.
Until $i$ advances to $\lts+1$ it broadcasts an acknowledgment $\ack_{\lts ij}$
for any message $\msg_{\lts j}$ it receives from any other node $j \neq i$ that
$i$ has not already seen.
Node $i$ continues acknowledging all new messages $\msg_{\lts j}$ it sees
even after having received $\tm$ messages at time $\lts$,
because $i$ does not know who will see which of its acknowledgments first.
Further, $i$ collects {\em all} acknowledgments $\ack_{\lts ij}$
for {\em all} messages $\msg_{\lts j}$,
because $i$ does not know for precisely which set of $\tm$ messages
it will first be able to collect $\ta$ acknowledgments each.
Even if some message $\msg_{\lts j}$ arrives early at $i$,
its $\ta$ supporting acknowledgments might arrive late.
The acknowledgments $i$ collects supporting each of the $\tm$ messages
can be from different sets of nodes:
each message $\msg$ need only
have supporting acknowledgments from {\em any} set of at least $\ta$ nodes.

Once a node $i$ collects $\tm$ messages
with a ``complete matrix'' of $\tm \times \ta$ supporting acknowledgments,
$i$ progresses to time slot $\lts+1$
and stops either collecting or sending further acknowledgments at slot $\lts$.
Other nodes will thus no longer be able to get new acknowledgments from $i$,
and may be unable to build a complete acknowledgment matrix at slot $\lts$
once too many other nodes have reached slot $\lts+1$.
This is not a problem, however,
because any lagging node catches up to slot $\lts+1$ virally
upon seeing any time slot $\lts+1$ broadcast,
as in unacknowledged TLC.

\subsubsection{Time Progress Properties}
\label{sec:acked:properties}

\xxx{	Ensures that if any node receives a message $\msg$ labeled time $\lts$,
	then $\msg$ was formulated and sent at a time when:
	(a) at least $\msg$ nodes had reached time $\lts-1$, and
	(b) strictly fewer than $\tm$ nodes had reached time slot $\lts+1$.
	That is, $\msg$ was formulated and sent during slot $\lts$.
}

Acknowledged TLC satisfies the same key properties
of monotonicity, pacing, and liveness
as discussed in Section~\ref{sec:unack:properties}.
The arguments for the first three properties
are identical for unacknowledged and acknowledged TLC,
because the latter only adds new constraints
on the advancement of logical time,
\ie, ``slowing time further.''
We must adjust the liveness argument slightly, however.

\paragraph{Liveness:}
TLC satisfies liveness provided $\tm \le n-f_d$ and $\ta \le n-f_d$.
Acknowledged TLC has two types of {\em progress events}.
A node makes {\em time progress}
when it advances from time $\lts$ to $\lts+1$
and broadcasts an upper-layer message at $\lts+1$.
A node at time $\lts$ makes {\em message progress}
when it receives a time $\lts$ message it hasn't yet seen
and broadcasts an acknowledgment to it.
Since a node can make message progress at most $n$ times consecutively
before it must instead make time progress,
showing that a node must regularly encounter {\em any} progress event
suffices to ensure that it regularly makes time progress.

For TLC to get stuck,
there must be some moment $\msg$ of the last progress event any node reaches.
Let $L$ be the set of live nodes
not in the adversary's indefinite-delay set continuously from $\msg$ on forever.
By the above requirements, $|L| \ge \tm$ and $|L| \ge \ta$.
Let $\msg'$ be the moment the last message broadcast by any node in $L$
is delivered to its final recipient in $L$.
If at $\msg$ there were nodes at different logical times $\lts' < \lts$,
then by $\msg'$ the last broadcast of some node $i$ at $\lts$
will reach some node $j$ at $\lts'$
and virally advance $j$ to time $\lts$.
If at $\msg$ all nodes in $L$ are at the same time $\lts$
and {\em each} have received and acknowledged
time $\lts$ messages from {\em all} nodes in $L$,
then by $\msg'$ all nodes in $L$ have collected
a complete $\tm \times \ta$ matrix of acknowledgments
and progressed to time $\lts+1$.
Otherwise, at $\msg$ all nodes in $L$ are at the same time $\lts$
but there is some node $i$ in $L$ whose time $\lts$ message
has not yet been acknowledged by some other node $j$ in $L$.
By moment $\msg'$ node $j$ will receive this message from $i$,
thus making message progress.
In all of these cases, at least one node
eventually makes either time or message progress,
contradicting the assumption
and ensuring liveness in acknowledged TLC.

%% file: a-cons.tex
\subsection{Consensus Protocol Analysis}

\subsubsection{QSCW}

\subsubsection{QSC4}

\subsubsection{QSC3}

\paragraph{Correctness:}

\paragraph{Chain quality:}

\paragraph{Probability of success without equivocation:}

\xxx{	a subtlety/optimization that maybe belongs in the full formal analysis:

Node $i$ has the information to evaluate these conditions by step $s+2$,
even though the round does not actually complete until $s+3$.
A node $i$ thus concludes a consensus round
to have definitely committed by time $s+3$
if it has learned information from {\em any} node -- itself or another --
that these conditions were satisfied by time $s+2$.
There are many such nodes $j$ that might have observed this at $s+2$
and from whom node $i$ might subsequently learn it from --
but the probability that $i$ learns this fact from
{\em at least one} such node $j$ is conservatively
no less than the probability that {\em any particular} node $j$
observes it directly at time $s+2$.
For simplicity, we thus focus on the latter event's probability.

To simplify the analysis further
we will consider a more specific event that is sufficient,
but not strictly necessary,
for a particular node $j$ to observe the necessary conditions at $t+2$:
namely, that the time $t$ proposal
with the {\em globally lowest} lottery ticket value
is double-certified and that fact is observed by node $j$ by $t+2$.
By TLC's protocol construction,
node $j$ waits for a threshold number of time $t+1$ messages
to be threshold certified and collected by $j$ before advancing to $t+2$.
By the same logic as above,
the probability that at least one of these time $t+1$ messages 
observes the certification of the globally-lowest proposal $p$
is no less than the probability that
{\em any particular node} $k$ in the same set observes $p$ to be certified.
Since node $k$ in turn collected a threshold of certified time $t$ proposals
in order to advance to $t+1$,
and the probability of each proposal having the globally lowest lottery ticket
is uniformly random
because all lottery tickets are drawn from the same distribution,
the probability of the original node $i$ observing success
is at least the threshold fraction:
$1/2$ for non-Byzantine and $2/3$ for Byzantine consensus.

We assume here that the lottery ticket entropy or value distribution
is large enough that the probability
of two randomly-chosen lottery tickets colliding is negligible.
This assumption is not necessary:
compensating for its absence would merely involve adjusting
the overall probability of success accordingly,
since a collision between two lowest-valued lottery tickets
simply increases the probability or consensus failure accordingly.
The ``less than or equal'' condition
for a potentially-uncertified time $t$ proposal $p'$
to prevent a certified time $t$ proposal $p$
from being considered definitely committed
ensures that the hopefully-unlikely case of a lottery tie,
neither winning proposal can be considered committed by any node.
}

\paragraph{Probability of success with equivocation:}

At t+1, with all f faulty nodes equivocating,
we could produce (2f+1) different equivocating views of t+0,
each for the benefit of one of the honest nodes,
each view containing f unique equivocating proposals from the faulty nodes,
for a total of up to (f)(2f+1) different faulty proposals via equivocation.
However, an attack this aggressive will result in the equivocation
being detected immediately at time t+2,
because each t+2 view would have to include more than one
(in fact f+1) of these different equivocating views.
Thus, none of the (f)(2f+1) proposals would be deemed eligible
by any honest node at time t+2,
and the attack would accomplish nothing
but eliminate the equivocating nodes.

A slightly less agressive equivocation attack
is for the f faulty nodes to conspire in creating
only (f+1) different equivocating views of t+0,
each again with a different set of f proposals from the faulty nodes,
for a total of up to (f)(f+1) different faulty proposals via equivocation.
At t+1, the network adversary arranges that a set of f honest nodes
see honly honest proposals from time 0,
and hence are ``uncolored'' by any of the equivocating views
and thus compatible with all of them.
The other f+1 honest nodes at t+1 each see a different equivocating view,
each producing a self-consistent viewpoing of f unique equivocating proposals,
one honest proposal "colored" by this view,
and the f "uncolored" honest proposals.
Further, the network adversary can arrange
that each of the (f)(f+1) resulting adversary proposals
appears eligible to (different subsets of) correct nodes at t+2,
each of these based on evidence that each of the f faulty nodes
(wearing their correctly-colored hats)
and the one appropriately-colored honest node
had seen each of the equivocating adversarial proposals by t+1.

Although at most one proposal can only emerge as definitely committed by t+3,
the existence of these (f)(f+1) equivocating but apparently eligible
adversary proposals at t+3
can increase the adversary's chances of one of its proposals winning the lottery
to (f)(f+1)/(f+1)(f+1), or f/(f+1),
leaving the f+1 correct proposals in a time t+3 view
only a 1/(f+1) chance of winning.
Further, this aggressive attack would guarantee that all f faulty nodes
are exposed as equivocating and eliminated by t+3,
so the adversary would expend its entire DoS attack capability in one round.

The adversary could of course equivocate with fewer of its nodes,
thereby reserving the rest for future attacks,
but this would correspondingly decrease the DoS attack's impact.
For example, if the adversary uses only f/2 nodes to equivocate,
then it can use them to support (f/2)(f+1) eligible equivocating proposals
in (f+1) views together with (f/2+f) uncolored correct proposals.
Maximizing the adversary's equivocating proposals in a t+3 view
would include the (f/2)(f+1) equivocating adversary proposals
together with the f+1 correct proposals
of the honest nodes that saw and accepted these views at t+2,
giving the adversary a probability of (f/2)/(f/2+1)
that one of its eligible equivocating proposals wins at t+3.

%% file: eval.tex
\section{Experimental Evaluation}
\label{sec:eval}

{\em In preparation.}

%% file: rel.tex
\section{Related Work}

This section summarizes related work,
focusing first on TLC in relation to classic logical clocks,
then on QSC in relation to the large body of prior work on consensus.

\subsection{Logical Clocks and Virtual Time}

Threshold logical clocks are of course inspired by
classic notions of logical time,
such as Lamport clocks~\cite{lamport78time},
vector clocks~\cite{fischer82sacrificing,liskov86highly,mattern89virtual,fidge91logical}, and
matrix clocks~\cite{wuu84efficient,sarin87discarding,ruget94cheaper,drummond03reducing,raynal92about}.
We even use vector and matrix clocks as building blocks
in implementing TLC.


Prior work has used logical clocks and virtual time for purposes
such as discrete event simulation and rollback~\cite{jefferson85virtual},
verifying cache coherence protocols~\cite{plakal98lamport},
and temporal proofs for digital ledgers~\cite{hughes17radix}.
We are not aware of prior work defining a threshold logical clock abstraction
or using it to build asynchronous consensus
or distributed key generation protocols,
however.

\com{ Virtual Time:
\url{http://homepage.divms.uiowa.edu/~ghosh/VirtualTime.pdf}
- uses Lamport clock time for consistent rollbacm and failure recovery,
e.g., for discrete event simulation, distributed database concurrency control,
virtual circuit implementation.
}

\com{ Lamport Clocks: Verifying a Directory Cache-Coherence Protocol 
\url{http://research.cs.wisc.edu/multifacet/papers/spaa98_lamport_pdf.pdf}
- extends lamport clocks with processor ID to ensure a total order
}

\com{ Tempo logical clocks
\url{https://papers.radixdlt.com/tempo/latest/#logical-clocks}:
a DLT design that uses vector clocks to search for temporal proofs.
But unclear whether or how it works or what exactly it accomplishes,
what happens when it can't find a temporal proof between incomparable events,
etc.
}

Conceptually analogous to TLC,
Awerbuch's {\em synchronizers}~\cite{awerbuch85complexity}
are intended to simplify the design of distributed algorithms
by presenting a synchronous abstraction atop an asynchronous network.
Awerbuch's synchronizers assume a fully-reliable system, however,
tolerating neither availability nor correctness failures in participants.
TLC's purpose might therefore be reasonably described as building
{\em fault-tolerant synchronizers}.

The basic threshold communication patterns TLC employs
have appeared in numerous protocols in various forms,
such as classic reliable broadcast
algorithms~\cite{bracha84asynchronous,bracha85asynchronous,reiter94secure}.
Witnessed TLC is inspired by
threshold signature schemes~\cite{shoup00practical,boldyreva03threshold},
signed {\em echo broadcast}~\cite{reiter94secure,cachin01secure,abraham18vaba},
and
witness cosigning protocols~\cite{syta16keeping,ford16apple,nikitin17chainiac}.
We are not aware of prior work to develop or use a form of logical clock
based on these threshold primitives, however.


\subsection{Asynchronous Consensus Protocols}

The FLP theorem~\cite{fischer85impossibility}
implies that consensus protocols
must sacrifice one of safety, liveness, asynchrony, or determinism.
Paxos~\cite{lamport98parttime,lamport01paxos}
and its leader-based derivatives for
fail-stop~\cite{boichat03deconstructing,ongaro14search,renesse15paxos,howard15raft}
and Byzantine consensus~\cite{castro99practical,kotla09zyzzyva,clement09making,clement09upright,bessani14state,aublin15bft,yin19hotstuff}
sacrifice asynchrony by relying on timeouts to ensure progress,
leaving them vulnerable to performance and DoS
attacks~\cite{clement09making,amir11byzantine}.
QSC instead sacrifices determinism
and uses randomness.

Consensus protocols have employed randomness in many ways.
Some use private coins that nodes flip independently,
but require time exponential in group
size~\cite{ben-or83another,bracha84asynchronous}.
Others assume that the network embodies randomness
in the form of a {\em fair scheduler}~\cite{bracha85asynchronous}.
More practical randomized consensus protocols
handling arbitrary asynchrony
typically rely on shared coins~\cite{rabin83randomized,ben-or85fast,canetti93fast,cachin01secure,cachin05random,friedman05simple,correia06consensus,correia11byzantine,mostefaoui14signature,miller16honey,duan18beat,abraham19vaba}.
Current practical methods of setting up shared coins, however,
assume a trusted dealer~\cite{chor85verifiable,herzberg95proactive,cachin02asynchronous},
a partially-synchronous network~\cite{gennaro07secure,kate12distributed},
a weakened fault tolerance
threshold~\cite{feldman88optimal,canetti93fast,canetti98fast},
or weakened termination
guarantees~\cite{canetti93fast,canetti98fast,bangalore18almost},
due to the ``chicken-and-egg'' problem discussed
in Section~\ref{sec:dkg:challenges}.

\com{
QSC therefore appears be the first practical asynchronous consensus protocol
depending on neither weakened asynchrony
nor shared coins whose setup circularly depends on consensus.
}
With fail-stop nodes, in contrast,
QSC requires only private randomness and private communication channels
(Section~\ref{sec:cons:odds}).
With Byzantine nodes,
QSC relies on leader-driven publicly-verifiable randomness,
which a public randomness protocol like RandHound~\cite{syta17scalable}
can implement without requiring consensus (Section~\ref{sec:byz:pvss}).


QSC's ``genetic consensus'' approach (Section~\ref{sec:cons:genetic}),
where each node maintains its own history
but randomly adopts those of others so as to converge statistically,
is partly inspired by randomized blockchain consensus protocols
such as Bitcoin~\cite{nakamoto08bitcoin},
Algorand~\cite{gilad17algorand},
and DFINITY~\cite{hanke18dfinity,abraham18dfinity}.
These prior blockchain protocols rely on synchrony assumptions, however,
such as the essential block interval parameter that paces
Bitcoin's proof-of-work~\cite{gervais16security}.
QSC in a sense provides Bitcoin-like genetic consensus
using TLC for fully-asynchronous pacing.

QSC builds on the classic techniques of
tamper-evident logging~\cite{schneier99secure,crosby09efficient},
timeline entanglement~\cite{maniatis02secure},
and accountable state
machines~\cite{haeberlen07peerreview,haeberlen10accountable}
for general protection against Byzantine node behavior.
Several recent DAG-based blockchain consensus
protocols~\cite{lewenberg15inclusive,baird16hashgraph,popov18tangle,danezis18blockmania}
reinvent specialized variants of these techniques.

\xxx{ \cite{feldman88optimal,feldman88thesis} presents a synchronous scheme
for bootstrapping common coins for Byzantine agreement,
and refers to an asynchronous analog for $n > 4f$,
which was never published.
Rabin and Canetti~\cite{canetti93fast,canetti98fast}
build this approach into a protocol for $n > 3f$,
but which terminates only probabilistically.
}

\xxx{
Combining Asynchronous and Synchronous Byzantine Agreement: The Best of Both
Worlds~\cite{loss18combining}:
provides a framework for combining synchronous and asynchronous
Byzantine consensus protocols
into a hybrid protocol providing
the responsiveness of the asynchronous protocols
if fewer than $1/4$ of the participants are malicious,
but can tolerate just under $1/2$ malicious participants
with synchrony assumptions and timeouts.
}

\xxx{
potentially relevant para from our RandHound paper:
An important observation by Gennaro et al. \cite{gennaro99secure} is that in many distributed key
generation protocols [47] an attacker can observe public values of honest
participants. To mitigate this attack, the authors propose to delay the
disclosure of the protocol’s public values after a “point-of-no-return” at
which point the attacker cannot influence the output anymore. We also use the
concept of a “point-of-no-return” to prevent an adversary from biasing the
output. However, their assumption of a fully synchronous network is unrealistic
for real-world scenarios.
}

\xxx{
DKG, beacons, etc.?
}


%% file: concl.tex
\section{Conclusion}

This paper has introduced a new type of logical clock abstraction,
which appears to be quite useful for simplifying the design
and implementation of asynchronous distributed coordination systems
such as consensus protocols, beacons, and other high-reliability services.
The concept is currently preliminary and still requires
robust implementations as well as detailed
formal and experimental analysis.
Nevertheless,
the approach seems interesting for its conceptual modularity,
for the simplicity with which it implements asynchronous consensus
given the appropriate set of abstractions to build on,
and for enabling asynchronous verifiable secret sharing
and distributed key generation
without assuming trusted dealers or common coins.
The non-Byzantine QSC3, in particular,
may represent a viable asynchronous competitor
to the venerable Paxos and its many variants,
in terms of both simplicity and practicality.

\subsection*{Acknowledgments}

This preliminary idea paper benefitted in many ways
from discussion with numerous colleagues in recent months:
in particular Philipp Jovanovic,
Ewa Syta,
Eleftherios Kokoris-Kogias,
Enis Ceyhun Alp,
Manuel Jos\'{e} Ribeiro Vidigueira,
Nicolas Gailly,
Cristina Basescu,
Timo Hanke,
Mahnush Movahedi,
and Dominic Williams.
Manuel Vidigueira, in particular,
helped with an excellent semester project prototyping TLC
and obtaining early experimental results.

This ongoing research was facilitated in part by financial support from
DFINITY, AXA, Handshake, and EPFL.
DFINITY's support in paticular,
which funded a joint project to analyze, improve, and formalize
its consensus protocol,
provided a key early impetus to explore randomized consensus protocols further.


%% file: go.tex
\section{TLC and QSC Model in Go}
\label{sec:go}

This appendix lists complete source code
for a working model of Threshold Logical Clocks
and Que Sera Consensus
in the Go language~\cite{golang}.
The model implements consensus only in the fail-stop (not Byzantine) model,
and it implements nodes as goroutines communicating via shared memory
instead of real network connections.
It is less than 250 code lines as counted by \texttt{cloc}~\cite{cloc}.
Despite its simplicity and limitations,
this implementation demonstrates all the fundamental elements of TLC and QSC.
The latest version of this model may be found at
\url{https://github.com/dedis/tlc/tree/master/go/model}.

\long\def\gomodel{

\begin{tiny}

\subsection{\texttt{qsc.go}: Que Sera Consensus}
\lstinputlisting[columns=flexible,tabsize=2]{src/go/qsc.go}

\subsection{\texttt{tlc.go}: Threshold Logical Clocks}
\lstinputlisting[columns=flexible,tabsize=2]{src/go/tlc.go}

\subsection{\texttt{node.go}: Per-Node State Definitions}
\lstinputlisting[columns=flexible,tabsize=2]{src/go/node.go}

\subsection{\texttt{set.go}: Message Sets}
\lstinputlisting[columns=flexible,tabsize=2]{src/go/set.go}

\subsection{\texttt{model\_test.go}: Testing the Model}
\lstinputlisting[columns=flexible,tabsize=2]{src/go/model_test.go}

\end{tiny}

}

\arxiv{
	\lstset{language=Go}
	\gomodel
}{
	\gomodel
}

%% file: spin.tex
\section{Promela Model for Spin Checker}
\label{sec:spin}

This section contains a Promela model of the basic logic of TLC and QSC,
which supports exhaustive verification of the state space using
the Spin model checker.
This implementation currently models only non-Byzantine node behavior and
verifies only safety and not liveness or statistical progress guarantees.
To avoid state space explosion, it models messages merely as
shared-memory interactions.
The model may be exhaustively checked (in a couple minutes)
using the \texttt{run.sh} script below.

\subsection{\texttt{qsc.pml}: Promela model of QSC}

\begin{tiny}
\begin{lstlisting}[language=Promela,columns=flexible,tabsize=2]

#define N	4		// total number of nodes
#define Fa	1		// max number of availability failures
#define Fcu	1		// max number of unknown correctness failures
#define T	(Fa+Fcu+1)	// consensus threshold required

#define STEPS	3		// TLC time-steps per consensus round
#define ROUNDS	2		// number of consensus rounds to run
#define TICKETS	3		// proposal lottery ticket space

typedef Round {
	bit sent[STEPS];	// whether we've sent yet each time step
	byte ticket;		// lottery ticket assigned to proposal at t+0
	byte seen[STEPS];	// bitmask of msgs we've seen from each step
	byte prsn[STEPS];	// bitmaks of proposals we've seen after each
	byte best[STEPS];
	byte btkt[STEPS];
	byte picked;		// which proposal this node picked this round
	bit done;		// set to true when round complete
}

typedef Node {
	Round round[ROUNDS];	// each node's per-consensus-round information
}

Node node[N];			// all state of each node

// Calculate n number of bits set in byte v
inline nset(v, n) {
	atomic {
		int i;

		n = 0;
		for (i : 0 .. 7) {
			if
			:: ((v & (1 << j)) != 0) -> n++;
			:: else -> skip;
			fi
		}
	}
}

proctype NodeProc(byte n) {
	byte rnd, tkt, step, seen, scnt, prsn, best, btkt, nn;
	byte belig, betkt, beseen, k;
	//bool correct = (n < T);

	//printf("Node %d correct %d\n", n, correct);

	for (rnd : 0 .. ROUNDS-1) {

		atomic {

		// select a "random" (here just arbitrary) ticket
		select (tkt : 1 .. TICKETS);
		node[n].round[rnd].ticket = tkt;

		// we've already seen our own proposal
		prsn  = 1 << n;

		// finding the "best proposal" starts with our own...
		best = n;
		btkt = tkt;

		} // atomic

		// Run the round to completion
		for (step : 0 .. STEPS-1) {

			// "send" the broadcast for this time-step
			node[n].round[rnd].sent[step] = 1;

			// collect a threshold of other nodes' broadcasts
			seen = 1 << n;		// we've already seen our own
			scnt = 1;
			do
			::	// Pick another node to try to 'receive' from
				select (nn : 1 .. N); nn--;
				if
				:: ((seen & (1 << nn)) == 0) && 
				    (node[nn].round[rnd].sent[step] != 0) ->

					atomic {

					//printf("%d received from %d\n", n, nn);
					seen = seen | (1 << nn);
					scnt++;

					// Track the best proposal we've seen
					if
					:: step == 0 ->
						prsn = prsn | (1 << nn);
						if
						:: node[nn].round[rnd].ticket < btkt ->
							best = nn;
							btkt = node[nn].round[rnd].ticket;
						:: node[nn].round[rnd].ticket == btkt ->
							best = 255;	// means tied
						:: else -> skip
						fi

					// Track proposals we've seen indirectly
					:: step > 0 ->
						prsn = prsn | node[nn].round[rnd].prsn[step-1];
						if
						:: node[nn].round[rnd].btkt[step-1] < btkt ->
							best = node[nn].round[rnd].best[step-1];
							btkt  = node[nn].round[rnd].btkt[step-1];
						:: (node[nn].round[rnd].btkt[step-1] == btkt) && (node[nn].round[rnd].best[step-1] != best) ->
							best = 255;	// tied
						:: else -> skip
						fi
					fi

					} // atomic

				:: else -> skip
				fi

				// Threshold test: have we seen enough?
				if
				:: scnt >= T -> break;
				:: else -> skip;
				fi
			od

			atomic {

			// Record what we've seen for the benefit of others
			node[n].round[rnd].seen[step] = seen;
			node[n].round[rnd].prsn[step] = prsn;
			node[n].round[rnd].best[step] = best;
			node[n].round[rnd].btkt[step] = btkt;

			printf("%d step %d complete: seen %x best %d ticket %d\n", n, step, seen, best, btkt);

			} // atomic
		}

		atomic {

		// Find the best propposal we can determine to be eligible.
		// We deem a proposal to be eligible if we can see that
		// it was seen by at least f+1 nodes by time t+1.
		// This ensures that ALL nodes at least know of its existence
		// (though not necessarily its eligibility) by t+2.
		belig = 255;	// start with a fake 'tie' state
		betkt = 255;	// worst possible ticket value
		beseen = 0;
		for (nn : 0 .. N-1) {

			// determine number of nodes that knew of nn's proposal
			// by time t+2.
			int nnseen = 0;
			for (k : 0 .. N-1) {
				if
				:: ((node[n].round[rnd].seen[2] & (1 << k)) != 0) && ((node[k].round[rnd].prsn[1] & (1 << nn)) != 0) -> nnseen++;
				:: else -> skip
				fi
			}
			//printf("%d from %d nnseen %d\n", n, nn, nnseen);

			if
			:: (nnseen >= Fa+1) &&	// nn's proposal is eligible
			   (node[nn].round[rnd].ticket < betkt) -> // is better
				belig = nn;
				betkt = node[nn].round[rnd].ticket;
				beseen = nnseen;
				//printf("%d new belig %d ticket %d seen %d\n", n, belig, betkt, beseen);
			:: (nnseen >= Fa+1) &&	// nn's proposal is eligible
			   (node[nn].round[rnd].ticket == betkt) -> // is tied
				belig = 255;
				beseen = 0;
			:: else -> skip
			fi
		}
		printf("%d best eligible proposal %d ticket %d seen by %d\n", n, belig, betkt, beseen);

		// we should have found at least one eligible proposal!
		assert(betkt < 255);

		// The round is now complete in terms of picking a proposal.
		node[n].round[rnd].picked = belig;
		node[n].round[rnd].done = 1;

		// Can we determine a proposal to be definitely committed?
		// To do so, we must be able to see that:
		//
		// 1. it was seen by t+2 by ALL nodes we have info from.
		// 2. we know of no other proposal competitive with it.
		// 
		// #1 ensures ALL nodes will judge this proposal as eligible;
		// #2 ensures no node could judge another proposal as eligible.
		if
		:: (belig < 255) && (beseen >= T) && (belig == best) ->
			printf("%d round %d definitely committed\n", n, rnd);

			// Verify that what we decided doesn't conflict with
			// the proposal any other node chooses.
			select (nn : 1 .. N); nn--;
			assert(!node[nn].round[rnd].done ||
				(node[nn].round[rnd].picked == belig));

		:: (belig < 255) && (beseen < T) ->
			printf("%d round %d failed due to threshold\n", n, rnd);

		:: (belig < 255) && (belig != best) ->
			printf("%d round %d failed due to spoiler\n", n, rnd);

		:: (belig == 255) ->
			printf("%d round %d failed due to tie\n", n, rnd);
		fi

		} // atomic
	}
}

init {
	atomic {
		int i;
		for (i : 0 .. N-1) {
			run NodeProc(i)
		}
	}
}

\end{lstlisting}
\end{tiny}

\subsection{\texttt{run.sh}: Model checking script}

\begin{tiny}
\begin{lstlisting}[language=Bash,columns=flexible,tabsize=2]
#!/bin/sh
# Exhaustively analyze the QSC model using the Spin model checker.
spin -a qsc.pml
gcc -O2 -DSAFETY -DBITSTATE -o pan pan.c 
./pan -m20000
\end{lstlisting}
\end{tiny}

%% file: tlc_properties.tex
\section{TLC properties}
\label{sec:prop}

\xxx{ Some potential sets of properties for TLC that might be useful for the earlier (TLC) sections. }\\

\xxx{ (The following is taken from my semester project report. Copied here in case it's useful.)}\\

Define a (T$_{\text{m}}$, T$_{\text{a}}$) \emph{threshold set} to be any set that includes T$_{\text{m}}$ messages broadcast in the same round, where each has at least T$_{\text{a}}$ different acknowledgements also from the same round. Under a TLC instance with parameters (T$_{\text{m}}$, T$_{\text{a}}$), we refer to it simply as a \emph{threshold set}.\\

A potential set of properties for TLC is:

\begin{itemize}
	\item \textbf{TLC1: Validity:} If every correct process $p$ broadcasts a message $m$ at round t, then every correct process $q$ will eventually deliver a threshold set for round t.
	\item \textbf{TLC2: No round duplication:} No correct process delivers more than one message for the same process and round.
	\item \textbf{TLC3: Limited round broadcast:} No correct process broadcasts more than one message per round.
	\item \textbf{TLC4: Authenticity:} If a correct process $p$ delivers a message $m$ with source $q$ and broadcast round $t$, then $m$ was previously broadcast by $q$ with round $t$.
	\item \textbf{TLC5: Forced round increase:} If a correct process $p$ delivers (or broadcasts) a message $m$ with round t, and process $p$ has previously delivered a threshold set with round t', then $t > t'$.
	\item \textbf{TLC6: Pacing:} If a correct process $p$ delivers (or broadcasts) a message $m$ with round t, then $p$ has delivered a threshold set in every previous round.
\end{itemize}

In order words, validity captures the notion of \emph{liveness}, even in the presence of failures. If all correct nodes act in a symmetric way (as TLC is supposed to be used) and the adversarial assumptions hold, then TLC ensures that progress will be made (the delivery of threshold sets).

The second and third properties formalize the idea that \emph{correct} nodes are only supposed to send one message per round. The fourth property, authenticity, is a naturally desirable property for broadcast primitives.

The last two properties are what make TLC a group synchronization primitive. TLC5 intuitively forces nodes to advance their clock time in order to communicate more information. This can be used directly to imply a causal ordering between messages: for any two messages delivered or broadcast by the same correct node, the one with the lowest round time is the oldest one. TLC6, on the other hand, keeps nodes from advancing time \emph{too} rapidly. As honest nodes are have to participate and receive messages from every previous round, some types of byzantine behaviour become inexcusable, like claiming to not have seen certain messages. They can't just ``bury their heads in the sand" and pretend to not have heard the news.\\

\xxx{It might be better to formulate TLC without specifically mentioning acknowledgements (that's \emph{one} of the possible ways to implement TLC, but the theoretical properties should be decoupled from implementation). We can formalize it maybe with something along the lines of: }

\theoremstyle{definition}
\newtheorem{definition}{Definition}[section]

\theoremstyle{definition}
\begin{definition}{\emph{Witnessing}}
If a message $m$ is witnessed by a process $p$ for round $t$, and $p$ is correct, then $m$ is delivered by $p$ in round $t$.
\end{definition}

Intuitively, witnessing corresponds to a \emph{promise} of delivery. Correct processes always keep their promises, and so anyone that receives authentic, verifiable evidence of this promise \emph{knows} that it will be kept, or can use it to prove that a process is not correct if it fails to deliver a message.\\

\xxx{We could then redefine a threshold set this way:}
\theoremstyle{definition}
\begin{definition}{\emph{(T,A) Threshold set}}
is any set of messages broadcast and delivered in the same round $t$, where T of those messages are each witnessed by A different processes in round $t$.
\end{definition}


\xxx{After the meeting on 27/06:}\\

Each node running (T,A)-TLC "clocks" events that happen in the system by appending a (logical) timestamp to them. Events can be external (e.g. another node's message), which the local node has been made aware of through some form of communication.

\theoremstyle{definition}
\begin{definition}{}
We say that an event is confirmed with round r if (at least) A different nodes clocked it with round r.
\end{definition}

\begin{itemize}
	\item \textbf{TLC1: Forced round increase:} For any given node, if there are T events both confirmed and clocked with round $r$, any event clocked later has round $r' > r$.
	\item \textbf{TLC2: Pacing:} For any given node, if an event is clocked with round $r$, then it has T events both confirmed and clocked with the previous round.
\end{itemize}

%% file: main.bbl
\begin{thebibliography}{100}

\bibitem{abraham18dfinity}
Ittai Abraham, Dahlia Malkhi, Kartik Nayak, and Ling Ren.
\newblock \href{https://eprint.iacr.org/2018/1153}{Dfinity Consensus,
  Explored}.
\newblock Cryptology ePrint Archive, Report 2018/1153, November 2018.

\bibitem{abraham18vaba}
Ittai Abraham, Dahlia Malkhi, and Alexander Spiegelman.
\newblock \href{https://arxiv.org/abs/1811.01332}{Validated Asynchronous
  Byzantine Agreement with Optimal Resilience and Asymptotically Optimal Time
  and Word Communication}.
\newblock {\em CoRR}, abs/1811.01332, 2018.

\bibitem{abraham19vaba}
Ittai Abraham, Dahlia Malkhi, and Alexander Spiegelman.
\newblock
  \href{https://dahliamalkhi.files.wordpress.com/2019/06/vaba-podc2019-2.pdf}{Asymptotically
  Optimal Validated Asynchronous Byzantine Agreement}.
\newblock In {\em \bibconf{PODC}{ACM Symposium on Principles of Distributed
  Computing}}, July 2019.

\bibitem{rfc3161}
C.~Adams, P.~Cain, D.~Pinkas, and R.~Zuccherato.
\newblock Internet x.509 public key infrastructure time-stamp protocol (tsp),
  August 2001.
\newblock RFC 3161.

\bibitem{ansi16trusted}
{American National Standards Institute}.
\newblock \href{https://webstore.ansi.org/standards/ascx9/ansix9952016}{ANSI
  X9.95-2016: Trusted Time Stamp Management And Security}, December 2016.

\bibitem{amir11byzantine}
Yair Amir, Brian Coan, Jonathan Kirsch, and John Lane.
\newblock {Prime}: {Byzantine} replication under attack.
\newblock {\em IEEE Transactions on Dependable and Secure Computing},
  8(4):564--577, July 2011.

\bibitem{apostolaki16hijacking}
Maria Apostolaki, Aviv Zohar, and Laurent Vanbever.
\newblock \href{http://arxiv.org/abs/1605.07524}{Hijacking Bitcoin: Large-scale
  Network Attacks on Cryptocurrencies}.
\newblock {\em 38th IEEE Symposium on Security and Privacy}, May 2017.

\bibitem{aspnes03randomized}
James Aspnes.
\newblock
  \href{https://link.springer.com/article/10.1007%2Fs00446-002-0081-5}{Randomized
  protocols for asynchronous consensus}.
\newblock {\em Distributed Computing}, 16(2--3):165--175, September 2003.

\bibitem{aspnes15faster}
James Aspnes.
\newblock
  \href{https://link.springer.com/article/10.1007%2Fs00446-013-0195-y}{Faster
  randomized consensus with an oblivious adversary}.
\newblock {\em Distributed Computing}, 28(1):21--29, February 2015.

\bibitem{aublin15bft}
Pierre-Louis Aublin, Rachid Guerraoui, Nikola Kne\v{z}evi\'{c}, Vivien
  Qu{\'e}ma, and Marko Vukoli\'{c}.
\newblock The next 700 {BFT} protocols.
\newblock {\em ACM Trans. Comput. Syst.}, 32(4):12:1--12:45, January 2015.

\bibitem{aumann96efficient}
Yonatan Aumann and Michael~A. Bender.
\newblock \href{https://doi.org/10.1007/3-540-61440-0_164}{Efficient
  Asynchronous Consensus with the Value-Oblivious Adversary Scheduler}.
\newblock In {\em \bibconf[23rd]{ICALP}{International Colloquium on Automata,
  Languages and Programming}}, July 1996.

\bibitem{aumann05efficient}
Yonatan Aumann and Michael~A. Bender.
\newblock
  \href{https://link.springer.com/article/10.1007%2Fs00446-004-0113-4}{Efficient
  low-contention asynchronous consensus with the value-oblivious adversary
  scheduler}.
\newblock {\em Distributed Computing}, 17(3):191--207, March 2005.

\bibitem{awerbuch85complexity}
Baruch Awerbuch.
\newblock \href{https://dl.acm.org/citation.cfm?id=4227}{Complexity of Network
  Synchronization}.
\newblock {\em Journal of the Association for Computing Machinery},
  32(4):804--823, October 1985.

\bibitem{baek04identity}
Joonsang Baek and Yuliang Zheng.
\newblock
  \href{https://link.springer.com/chapter/10.1007/978-3-540-24632-9_19}{Identity-Based
  Threshold Decryption}.
\newblock In {\em \bibconf[7th]{PKC}{International Workshop on Theory and
  Practice in Public Key Cryptography}}, March 2004.

\bibitem{bagherzandi08multisignatures}
Ali Bagherzandi, Jung~Hee Cheon, and Stanis\l{}aw Jarecki.
\newblock Multisignatures secure under the discrete logarithm assumption and a
  generalized forking lemma.
\newblock In {\em \bibconf[15th]{CCS}{ACM Conference on Computer and
  Communications Security}}, October 2008.

\bibitem{baird16hashgraph}
Leemon Baird.
\newblock
  \href{http://www.swirlds.com/wp-content/uploads/2016/06/2016-05-31-Swirlds-Consensus-Algorithm-TR-2016-01.pdf}{Hashgraph
  Consensus: fair, fast, {Byzantine} fault tolerance}.
\newblock Technical Report TR-2016-01, Swirlds, May 2016.

\bibitem{bangalore18almost}
Laasya Bangalore, Ashish Choudhury, and Arpita Patra.
\newblock \href{https://dl.acm.org/citation.cfm?id=3212735}{Almost-Surely
  Terminating Asynchronous Byzantine Agreement Revisited}.
\newblock In {\em \bibconf{PODC}{Principles of Distributed Computing}}, pages
  295--304, July 2018.

\bibitem{ben-or83another}
Michael Ben-Or.
\newblock Another advantage of free choice: Completely asynchronous agreement
  protocols.
\newblock In {\em Principles of Distributed Computing (PODC)}, August 1983.

\bibitem{ben-or85fast}
Michael Ben-Or.
\newblock \href{https://dl.acm.org/citation.cfm?id=323609}{Fast Asynchronous
  Byzantine Agreement (Extended Abstract)}.
\newblock In {\em \bibconf[4th]{PODC}{Principles of Distributed Computing}},
  pages 149--151, August 1985.

\bibitem{bessani14state}
Alysson Bessani, Joao Sousa, and Eduardo~E.P. Alchieri.
\newblock State machine replication for the masses with {BFT-SMART}.
\newblock In {\em International Conference on Dependable Systems and Networks
  (DSN)}, pages 355--362, June 2014.

\bibitem{bogetoft09secure}
Peter Bogetoft, Dan~Lund Christensen, Ivan Dam\r{g}ard, Martin Geisler, Thomas
  Jakobsen, Mikkel Kr\o{}igaard, Janus~Dam Nielsen, Jesper~Buus Nielsen, Kurt
  Nielsen, Jakob Pagter, Michael Schwartzbach, and Tomas Toft.
\newblock
  \href{https://link.springer.com/chapter/10.1007/978-3-642-03549-4_20}{Secure
  Multiparty Computation Goes Live}.
\newblock In {\em \bibconf[13th]{FC}{International Conference on Financial
  Cryptography and Data Security}}, February 2009.

\bibitem{boichat03deconstructing}
Romain Boichat, Partha Dutta, Svend Frølund, and Rachid Guerraoui.
\newblock Deconstructing {Paxos}.
\newblock {\em ACM SIGACT News}, 34(1), March 2003.

\bibitem{boldyreva03threshold}
Alexandra Boldyreva.
\newblock
  \href{https://link.springer.com/chapter/10.1007/3-540-36288-6_3}{Threshold
  Signatures, Multisignatures and Blind Signatures Based on the
  {Gap-Diffie-Hellman-Group} Signature Scheme}.
\newblock In {\em \bibconf[6th]{PKC}{International Workshop on Practice and
  Theory in Public Key Cryptography}}, January 2003.

\bibitem{boneh18verifiable}
Dan Boneh, Joseph Bonneau, Benedikt B\"unz, and Ben Fisch.
\newblock Verifiable delay functions.
\newblock In {\em \bibconf[38th]{CRYPTO}{Advances in Cryptology}}, August 2018.

\bibitem{boneh18compact}
Dan Boneh, Manu Drijvers, and Gregory Neven.
\newblock \href{https://eprint.iacr.org/2018/483}{Compact Multi-Signatures for
  Smaller Blockchains}.
\newblock In {\em Advances in Cryptology -- ASIACRYPT 2018}, December 2018.

\bibitem{boneh01identity}
Dan Boneh and Matt Franklin.
\newblock Identity-based encryption from the {Weil} pairing.
\newblock In {\em \bibconf[21st]{CRYPTO}{IACR International Cryptology
  Conference}}. 2001.

\bibitem{bonneau15bitcoin}
Joseph Bonneau, Jeremy Clark, and Steven Goldfeder.
\newblock \href{https://eprint.iacr.org/2015/1015.pdf}{On Bitcoin as a public
  randomness source}.
\newblock {IACR} eprint archive, October 2015.

\bibitem{bracha84asynchronous}
Gabriel Bracha.
\newblock \href{https://dl.acm.org/citation.cfm?id=806743}{An asynchronous
  [(n-1)/3]-Resilient Consensus Protocol}.
\newblock In {\em \bibconf[3rd]{PODC}{ACM Symposium on Principles of
  Distributed Computing}}, pages 154--162, August 1984.

\bibitem{bracha85asynchronous}
Gabriel Bracha and Sam Toueg.
\newblock \href{https://dl.acm.org/citation.cfm?id=214134}{Asynchronous
  Consensus and Broadcast Protocols}.
\newblock {\em Journal of the Association for Computing Machinery (JACM)},
  32(4):824--840, October 1985.

\bibitem{c4ads19above}
C4ADS.
\newblock \href{https://www.c4reports.org/aboveusonlystars}{Above Us Only
  Stars: Exposing GPS Spoofing in Russia and Syria}, April 2019.

\bibitem{cachin02asynchronous}
Christian Cachin, Klaus Kursawe, Anna Lysyanskaya, and Reto Strobl.
\newblock \href{https://dl.acm.org/citation.cfm?id=586124}{Asynchronous
  Verifiable Secret Sharing and Proactive Cryptosystems}.
\newblock In {\em \bibconf[9th]{CCS}{ACM Conference on Computer and
  Communications Security}}, November 2002.

\bibitem{cachin01secure}
Christian Cachin, Klaus Kursawe, Frank Petzold, and Victor Shoup.
\newblock \href{http://www.zurich.ibm.com/~cca/papers/abc.pdf}{Secure and
  Efficient Asynchronous Broadcast Protocols}.
\newblock In {\em Advances in Cryptology (CRYPTO)}, August 2001.

\bibitem{cachin05random}
Christian Cachin, Klaus Kursawe, and Victor Shoup.
\newblock Random oracles in constantinople: Practical asynchronous byzantine
  agreement using cryptography.
\newblock {\em Journal of Cryptology}, 18(3):219--246, 2005.

\bibitem{cachin11introduction}
Christian Cachin and Rachid Guerraoui~Lu\'is Rodrigues.
\newblock {\em Introduction to Reliable and Secure Distributed Programming}.
\newblock Springer, February 2011.

\bibitem{calandrino07machine}
Joseph~A. Calandrino, J.~Alex Halderman, and Edward~W. Felten.
\newblock
  \href{https://www.usenix.org/legacy/event/evt07/tech/full_papers/calandrino/calandrino_html/}{Machine-Assisted
  Election Auditing}.
\newblock In {\em \bibconf{ETV}{USENIX/ACCURATE Electronic Voting Technology
  Workshop}}, August 2007.

\bibitem{canetti93fast}
Ran Canetti and Tal Rabin.
\newblock \href{https://dl.acm.org/citation.cfm?id=167105}{Fast Asynchronous
  Byzantine Agreement with Optimal Resilience}.
\newblock In {\em \bibconf[25th]{STOC}{ACM Symposium on Theory of computing}},
  pages 42--51, May 1993.

\bibitem{canetti98fast}
Ran Canetti and Tal Rabin.
\newblock \href{http://people.csail.mit.edu/canetti/materials/cr93.ps}{Fast
  Asynchronous Byzantine Agreement with Optimal Resilience}, September 1998.

\bibitem{cascudo17scrape}
Ignacio Cascudo and Bernardo David.
\newblock
  \href{https://link.springer.com/chapter/10.1007/978-3-319-61204-1_27}{SCRAPE:
  Scalable Randomness Attested by Public Entities}.
\newblock In {\em \bibconf[15th]{ACNS}{International Conference on Applied
  Cryptography and Network Security}}, July 2017.

\bibitem{castro99practical}
Miguel Castro and Barbara Liskov.
\newblock
  \href{http://css.csail.mit.edu/6.824/2014/papers/castro-practicalbft.pdf}{Practical
  {Byzantine} Fault Tolerance}.
\newblock In {\em \bibconf[3rd]{OSDI}{USENIX Symposium on Operating Systems
  Design and Implementation}}, February 1999.

\bibitem{chor85verifiable}
Benny Chor, Shafi Goldwasser, Silvio Micali, and Baruch Awerbuch.
\newblock \href{https://ieeexplore.ieee.org/document/4568164}{Verifiable Secret
  Sharing and Achieving Simultaneity in the Presence of Faults}.
\newblock In {\em \bibconf[26th]{SFCS}{Symposium on Foundations of Computer
  Science}}, October 1985.

\bibitem{clark90architectural}
D.~D. Clark and D.~L. Tennenhouse.
\newblock Architectural considerations for a new generation of protocols.
\newblock In {\em ACM SIGCOMM}, pages 200--208, 1990.

\bibitem{clark10beacon}
Jeremy Clark and Urs Hengartner.
\newblock
  \href{https://www.usenix.org/conference/evtwote-10/use-financial-data-random-beacon}{On
  the Use of Financial Data as a Random Beacon}.
\newblock In {\em \bibconf{EVT/WOTE}{Electronic Voting Technology
  Workshop/Workshop on Trustworthy Elections}}, August 2010.

\bibitem{clement09upright}
Allen Clement, Manos Kapritsos, Sangmin Lee, Yang Wang, Lorenzo Alvisi, Mike
  Dahlin, and Taylor Rich\'{e}.
\newblock {UpRight} cluster services.
\newblock In {\em ACM Symposium on Operating Systems Principles (SOSP)},
  October 2009.

\bibitem{clement09making}
Allen Clement, Edmund~L Wong, Lorenzo Alvisi, Michael Dahlin, and Mirco
  Marchetti.
\newblock
  \href{https://www.usenix.org/legacy/events/nsdi09/tech/full_papers/clement/clement.pdf}{Making
  {Byzantine} Fault Tolerant Systems Tolerate Byzantine Faults}.
\newblock In {\em 6th USENIX Symposium on Networked Systems Design and
  Implementation}, April 2009.

\bibitem{conner-simons19programmers}
Adam Conner-Simons.
\newblock
  \href{https://www.csail.mit.edu/news/programmers-solve-mits-20-year-old-cryptographic-puzzle}{Programmers
  solve MIT’s 20-year-old cryptographic puzzle}, April 2019.

\bibitem{correia06consensus}
Miguel Correia, Nuno~Ferreira Neves, and Paulo Veríssimo.
\newblock From consensus to atomic broadcast: Time-free {Byzantine}-resistant
  protocols without signatures.
\newblock {\em The Computer Journal}, 49(1), January 2006.

\bibitem{correia11byzantine}
Miguel Correia, Giuliana~Santos Veronese, Nuno~Ferreira Neves, and Paulo
  Verissimo.
\newblock {Byzantine} consensus in asynchronous message-passing systems: a
  survey.
\newblock {\em International Journal of Critical Computer-Based Systems},
  2(2):141--161, July 2011.

\bibitem{cramer00general}
Ronald Cramer, Ivan Dam\r{g}ard, and Ueli Maurer.
\newblock
  \href{https://link.springer.com/chapter/10.1007/3-540-45539-6_22}{General
  Secure Multi-party Computation from any Linear Secret-Sharing Scheme}.
\newblock In {\em Eurocrypt}, May 2000.

\bibitem{crosby09efficient}
Scott~A. Crosby and Dan~S. Wallach.
\newblock Efficient data structures for tamper-evident logging.
\newblock In {\em USENIX Security Symposium}, August 2009.

\bibitem{czernik18frontrunning}
Matt Czernik.
\newblock
  \href{https://medium.com/@matt.czernik/on-blockchain-frontrunning-part-i-cut-the-line-or-make-a-new-one-b33850663b55}{On
  Blockchain Frontrunning}, February 2018.

\bibitem{dana97global}
Peter~H. Dana.
\newblock
  \href{https://link.springer.com/article/10.1023%2FA%3A1007906014916}{Global
  Positioning System (GPS) Time Dissemination for Real-Time Applications}.
\newblock {\em Real Time Systems}, 12(1):9--40, January 1997.

\bibitem{danezis18blockmania}
George Danezis and David Hrycyszyn.
\newblock Blockmania: from block dags to consensus.
\newblock {\em arXiv preprint arXiv:1809.01620}, 2018.

\bibitem{cloc}
Al~Danial.
\newblock {Counting Lines of Code}.
\newblock \url{http://cloc.sourceforge.net/}.

\bibitem{davis96kerberos}
Donald~T. Davis, Daniel~E. Geer, and Theodore Ts'o.
\newblock
  \href{https://www.usenix.org/legacy/publications/compsystems/1996/win_davis.pdf}{Kerberos
  With Clocks Adrift: History, Protocols, and Implementation}.
\newblock {\em Computing systems}, 9(1):29--46, 1996.

\bibitem{demers87epidemic}
Alan Demers et~al.
\newblock Epidemic algorithms for replicated database maintenance.
\newblock In {\em \bibconf[6th]{PODC}{ACM Symposium on Principles of
  Distributed Computing}}, pages 1--12, 1987.

\bibitem{desmedt89threshold}
Yvo Desmedt and Yair Frankel.
\newblock Threshold cryptosystems.
\newblock In {\em \bibconf{CRYPTO}{Advances in Cryptology}}, August 1989.

\bibitem{dowlen08political}
Oliver Dowlen.
\newblock {\em The Political Potential of Sortition: A Study of the Random
  Selection of Citizens for Public Office}.
\newblock Imprint Academic, August 2008.

\bibitem{drijvers19security}
Manu Drijvers, Kasra Edalatnejad, Bryan Ford, Eike Kiltz, Julian Loss, Gregory
  Neven, and Igors Stepanovs.
\newblock On the security of two-round multi-signatures.
\newblock In {\em \bibconf[40th]{SP}{IEEE Symposium on Security and Privacy}},
  May 2019.

\bibitem{drummond03reducing}
L\'{u}cia M.~A. Drummond and Valmir~C. Barbosa.
\newblock \href{https://arxiv.org/pdf/cs/0309042.pdf}{On reducing the
  complexity of matrix clocks}.
\newblock {\em Parallel Computing}, 29(7):895--905, July 2003.

\bibitem{duan18beat}
Sisi Duan, Michael~K. Reiter, and Haibin Zhang.
\newblock \href{https://dl.acm.org/citation.cfm?id=3243812}{BEAT: Asynchronous
  BFT Made Practical}.
\newblock In {\em \bibconf{CCS}{Computer and Communications Security}}, October
  2018.

\bibitem{eskandari19transparent}
Shayan Eskandari, Seyedehmahsa Moosavi, and Jeremy Clark.
\newblock \href{http://fc19.ifca.ai/wtsc/TransparentDishonesty.pdf}{Transparent
  Dishonesty: front-running attacks on Blockchain}.
\newblock In {\em \bibconf[3rd]{WTSC}{Workshop on Trusted Smart Contracts}},
  February 2019.

\bibitem{feldman88optimal}
Paul Feldman and Silvio Micali.
\newblock \href{https://dl.acm.org/citation.cfm?id=62225}{Optimal Algorithms
  for Byzantine Agreement}.
\newblock In {\em \bibconf[20th]{STOC}{Symposium on Theory of Computing}},
  pages 148--161, May 1988.

\bibitem{fidge91logical}
Colin Fidge.
\newblock Logical time in distributed computing systems.
\newblock {\em IEEE Computer}, 24(8):28--33, August 1991.

\bibitem{fienberg71randomization}
Stephen~E. Fienberg.
\newblock
  \href{https://science.sciencemag.org/content/171/3968/255}{Randomization and
  Social Affairs: The 1970 Draft Lottery}.
\newblock {\em Science}, 171(3968):255--261, January 1971.

\bibitem{fischer85impossibility}
Michael~J Fischer, Nancy~A Lynch, and Michael~S Paterson.
\newblock
  \href{https://groups.csail.mit.edu/tds/papers/Lynch/jacm85.pdf}{Impossibility
  of distributed consensus with one faulty process}.
\newblock {\em Journal of the ACM (JACM)}, 32(2):374--382, 1985.

\bibitem{fischer82sacrificing}
Michael~J. Fischer and Alan Michael.
\newblock Sacrificing serializability to attain high availability of data in an
  unreliable network.
\newblock In {\em Symposium on Principles of Database Systems (PODS)}, pages
  70--75, March 1982.

\bibitem{fishkin05experimenting}
James~S Fishkin and Robert~C Luskin.
\newblock
  \href{https://link.springer.com/article/10.1057%2Fpalgrave.ap.5500121}{Experimenting
  with a Democratic Ideal: Deliberative Polling and Public Opinion}.
\newblock {\em Acta Politica}, 40(3):284–298, September 2005.

\bibitem{ford16apple}
Bryan Ford.
\newblock
  \href{https://freedom-to-tinker.com/blog/bford/apple-fbi-and-software-transparency/}{Apple,
  FBI, and Software Transparency}.
\newblock {\em \href{https://freedom-to-tinker.com/}{Freedom to Tinker}}, March
  2016.

\bibitem{forgrave18man}
Reid Forgrave.
\newblock
  \href{https://www.nytimes.com/interactive/2018/05/03/magazine/money-issue-iowa-lottery-fraud-mystery.html}{The
  Man Who Cracked the Lottery}.
\newblock {\em The New York Times Magazine}, May 2018.

\bibitem{friedman05simple}
Roy Friedman, Achour Mostefaoui, and Michel Raynal.
\newblock Simple and efficient oracle-based consensus protocols for
  asynchronous {Byzantine} systems.
\newblock {\em IEEE Transactions on Dependable and Secure Computing}, 2(1),
  January 2005.

\bibitem{ganeriwal08secure}
Saurabh Ganeriwal, Christina P\"opper, Srdjan \v{C}apkun, and Mani~B.
  Srivastava.
\newblock \href{https://dl.acm.org/citation.cfm?doid=1380564.1380571}{Secure
  Time Synchronization in Sensor Networks}.
\newblock {\em ACM Transactions on Information and System Security (TISSEC)},
  11(4), July 2008.

\bibitem{geambasu09vanish}
Roxana Geambasu, Tadayoshi Kohno, Amit~A Levy, and Henry~M Levy.
\newblock
  \href{https://www.usenix.org/legacy/events/sec09/tech/full_papers/sec09_crypto.pdf}{Vanish:
  Increasing Data Privacy with Self-Destructing Data.}
\newblock In {\em USENIX Security Symposium}, pages 299--316, 2009.

\bibitem{gennaro07secure}
Rosario Gennaro, Stanis\l{}aw Jarecki, Hugo Krawczyk, and Tal Rabin.
\newblock
  \href{https://link.springer.com/article/10.1007/s00145-006-0347-3}{Secure
  Distributed Key Generation for Discrete-Log Based Cryptosystems}.
\newblock 20(1):51--83, January 2007.

\bibitem{gennaro98simplified}
Rosario Gennaro, Michael~O. Rabin, and Tal Rabin.
\newblock \href{https://dl.acm.org/citation.cfm?doid=277697.277716}{Simplified
  VSS and Fast-track Multiparty Computations with Applications to Threshold
  Cryptography}.
\newblock In {\em \bibconf[17th]{PODC}{Principles of Distributed Computing}},
  June 1998.

\bibitem{gervais16security}
Arthur Gervais, Ghassan~O Karame, Karl W\"ust, Vasileios Glykantzis, Hubert
  Ritzdorf, and Srdjan \v{C}apkun.
\newblock \href{https://dl.acm.org/citation.cfm?id=2976749.2978341}{On the
  Security and Performance of Proof of Work Blockchains}.
\newblock October 2016.

\bibitem{gilad17algorand}
Yossi Gilad, Rotem Hemo, Silvio Micali, Georgios Vlachos, and Nickolai
  Zeldovich.
\newblock \href{https://dl.acm.org/authorize?N47148}{Algorand: Scaling
  Byzantine Agreements for Cryptocurrencies}, October 2017.

\bibitem{golang}
\href{http://golang.org/}{The {Go} Programming Language}, February 2018.

\bibitem{haeberlen10accountable}
Andreas Haeberlen, Paarijaat Aditya, Rodrigo Rodrigues, and Peter Druschel.
\newblock
  \href{https://www.usenix.org/conference/osdi10/accountable-virtual-machines}{Accountable
  Virtual Machines}.
\newblock In {\em \bibconf[9th]{OSDI}{USENIX Symposium on Operating Systems
  Design and Implementation}}, October 2010.

\bibitem{haeberlen07peerreview}
Andreas Haeberlen, Petr Kouznetsov, and Peter Druschel.
\newblock
  \href{http://www.sosp2007.org/papers/sosp118-haeberlen.pdf}{{PeerReview}:
  Practical Accountability for Distributed Systems}.
\newblock In {\em \bibconf[21st]{SOSP}{{ACM} Symposium on Operating Systems
  Principles}}, October 2007.

\bibitem{hanke18dfinity}
Timo Hanke, Mahnush Movahedi, and Dominic Williams.
\newblock \href{https://arxiv.org/abs/1805.04548}{DFINITY Technology Overview
  Series: Consensus System}, May 2018.

\bibitem{herzberg95proactive}
Amir Herzberg, Stanis{\l}aw Jarecki, Hugo Krawczyk, and Moti Yung.
\newblock
  \href{https://link.springer.com/chapter/10.1007/3-540-44750-4_27}{Proactive
  Secret Sharing Or: How to Cope With Perpetual Leakage}.
\newblock pages 339--352, August 1995.

\bibitem{howard15raft}
Heidi Howard, Malte Schwarzkopf, Anil Madhavapeddy, and Jon Crowcroft.
\newblock {Raft} refloated: Do we have consensus?
\newblock {\em ACM SIGOPS Operating Systems Review}, 49(1):12--21, January
  2015.

\bibitem{hughes17radix}
Dan Hughes.
\newblock \href{https://papers.radixdlt.com/tempo/}{Radix -- Tempo}, September
  2017.

\bibitem{intel14software}
{Intel}.
\newblock
  \href{https://software.intel.com/sites/default/files/managed/48/88/329298-002.pdf}{Software
  Guard Extensions Programming Reference}, October 2014.

\bibitem{jefferson85virtual}
David~R. Jefferson.
\newblock Virtual time.
\newblock {\em ACM Transactions on Programming Languages and Systems}, 7(3),
  July 1985.

\bibitem{johnson77urn}
Norman~Lloyd Johnson.
\newblock {\em Urn models and their application: An approach to modern discrete
  probability theory}.
\newblock Wiley, 1977.

\bibitem{kate10distributed}
Aniket Kate and Ian Goldberg.
\newblock
  \href{https://link.springer.com/chapter/10.1007/978-3-642-15317-4_27}{Distributed
  Private-Key Generators for Identity-Based Cryptography}.
\newblock In {\em \bibconf[7th]{SCN}{Security and Cryptography for Networks}},
  September 2010.

\bibitem{kate12distributed}
Aniket Kate, Yizhou Huang, and Ian Goldberg.
\newblock \href{https://eprint.iacr.org/2012/377.pdf}{Distributed Key
  Generation in the Wild}.
\newblock Cryptology ePrint Archive, Report 2012/377, July 2012.

\bibitem{nist19reference}
John Kelsey, Lu\'is T. A. N.~Brand\ ao, Rene Peralta, and Harold Booth.
\newblock \href{https://csrc.nist.gov/publications/detail/nistir/8213/draft}{A
  Reference for Randomness Beacons: Format and Protocol Version 2}.
\newblock Technical Report NISTIR 8213 (DRAFT), {National Institute of
  Standards and Technology}, May 2019.

\bibitem{kokoris19secure}
Eleftherios Kokoris-Kogias.
\newblock {\em
  \href{https://drive.google.com/open?id=1tl1nNqLUhgubzxlRkNfMBcWZVl_N0WT3}{Secure,
  Confidential Blockchains Providing High Throughput and Low Latency}}.
\newblock PhD thesis, \'Ecole Polytechnique F\'ed\'erale de Lausanne (EPFL),
  May 2019.

\bibitem{kokoris18calypso}
Eleftherios Kokoris-Kogias, Enis~Ceyhun Alp, Sandra~Deepthy Siby, Nicolas
  Gailly, Linus Gasser, Philipp Jovanovic, Ewa Syta, and Bryan Ford.
\newblock \href{https://eprint.iacr.org/2018/209}{CALYPSO: Auditable Sharing of
  Private Data over Blockchains}.
\newblock Cryptology ePrint Archive, Report 2018/209, 2018.

\bibitem{kokoris16enhancing}
Eleftherios Kokoris-Kogias, Philipp Jovanovic, Nicolas Gailly, Ismail Khoffi,
  Linus Gasser, and Bryan Ford.
\newblock \href{http://arxiv.org/abs/1602.06997}{Enhancing Bitcoin Security and
  Performance with Strong Consistency via Collective Signing}.
\newblock In {\em Proceedings of the 25th USENIX Conference on Security
  Symposium}, 2016.

\bibitem{kokoris17omniledger}
Eleftherios Kokoris-Kogias, Philipp Jovanovic, Linus Gasser, Nicolas Gailly,
  Ewa Syta, and Bryan Ford.
\newblock \href{https://eprint.iacr.org/2017/406.pdf}{OmniLedger: A Secure,
  Scale-Out, Decentralized Ledger via Sharding}.
\newblock In {\em \bibconf[39th]{SP}{IEEE Symposium on Security and Privacy}},
  pages 19--34. IEEE, 2018.

\bibitem{kopetz11real-time}
Hermann Kopetz.
\newblock {\em Real-Time Systems: Design Principles for Distributed Embedded
  Applications}.
\newblock Springer, April 2011.

\bibitem{kotla09zyzzyva}
Ramakrishna Kotla, Lorenzo Alvisi, Mike Dahlin, Allen Clement, and Edmund Wong.
\newblock {Zyzzyva}: Speculative {Byzantine} fault tolerance.
\newblock {\em ACM Transactions on Computer Systems (TOCS)}, 27(4), December
  2009.

\bibitem{kozlov19league}
Dina Kozlov.
\newblock \href{https://blog.cloudflare.com/league-of-entropy/}{League of
  Entropy: Not All Heroes Wear Capes}.
\newblock CloudFlare Blog, June 2019.

\bibitem{lamport78time}
Leslie Lamport.
\newblock Time, clocks, and the ordering of events in a distributed system.
\newblock {\em Communications of the ACM}, 21(7):558--565, July 1978.

\bibitem{lamport98parttime}
Leslie Lamport.
\newblock \href{https://dl.acm.org/citation.cfm?id=279229}{The Part-Time
  Parliament}.
\newblock {\em ACM Transactions on Computer Systems}, 16(2):133--169, May 1989.

\bibitem{lamport01paxos}
Leslie Lamport.
\newblock Paxos made simple.
\newblock {\em ACM SIGACT News}, 32(4):51--58, December 2001.

\bibitem{lenstra15random}
Arjen~K. Lenstra and Benjamin Wesolowski.
\newblock \href{https://eprint.iacr.org/2015/366.pdf}{A random zoo: sloth,
  unicorn, and trx}.
\newblock {IACR} eprint archive, April 2015.

\bibitem{lewandowski91gps}
Wlodzimierz Lewandowski and Claudine Thomas.
\newblock \href{https://ieeexplore.ieee.org/document/84976/authors#authors}{GPS
  Time Transfer}.
\newblock {\em Proceedings of the IEEE}, 79(7):991--1000, July 1991.

\bibitem{lewenberg15inclusive}
Yoad Lewenberg, Yonatan Sompolinsky, and Aviv Zohar.
\newblock
  \href{https://link.springer.com/chapter/10.1007/978-3-662-47854-7_33}{Inclusive
  block chain protocols}.
\newblock In {\em International Conference on Financial Cryptography and Data
  Security}, pages 528--547. Springer, 2015.

\bibitem{li06bar}
Harry~C. Li, Allen Clement, Edmund~L. Wong, Jeff Napper, Indrajit Roy, Lorenzo
  Alvisi, and Michael Dahlin.
\newblock \href{https://dl.acm.org/citation.cfm?id=1298474}{BAR Gossip}.
\newblock In {\em \bibconf[7th]{OSDI}{Operating Systems Design and
  Implementation}}, November 2006.

\bibitem{lindeman12gentle}
Mark Lindeman and Philip~B. Stark.
\newblock \href{https://ieeexplore.ieee.org/document/6175884}{A Gentle
  Introduction to Risk-Limiting Audits}.
\newblock {\em IEEE Security \& Privacy}, 10(5), September 2012.

\bibitem{liskov86highly}
Barbara Liskov and Rivka Ladin.
\newblock Highly-available distributed services and fault-tolerant distributed
  garbage collection.
\newblock In {\em Principles of Distributed Computing}, pages 29--39, August
  1986.

\bibitem{luu16secure}
Loi Luu, Viswesh Narayanan, Chaodong Zheng, Kunal Baweja, Seth Gilbert, and
  Prateek Saxena.
\newblock \href{{http://doi.acm.org/10.1145/2976749.2978389}}{{A Secure
  Sharding Protocol For Open Blockchains}}.
\newblock In {\em Proceedings of the 2016 ACM SIGSAC Conference on Computer and
  Communications Security}, CCS '16, pages 17--30, New York, NY, USA, 2016.
  ACM.

\bibitem{lynch96distributed}
Nancy~A. Lynch.
\newblock {\em Distributed Algorithms}.
\newblock Morgan Kaufmann, March 1996.

\bibitem{mahmoody11time}
Mohammad Mahmoody, Tal Moran, and Salil Vadhan.
\newblock
  \href{https://link.springer.com/chapter/10.1007%2F978-3-642-22792-9_3}{Time-Lock
  Puzzles in the Random Oracle Model}.
\newblock In {\em \bibconf{CRYPTO}{Advances in Cryptology}}, pages 39--50.
  Springer, 2011.

\bibitem{mahmoud08polya}
Hosam Mahmoud.
\newblock {\em P\'olya Urn Models}.
\newblock Chapman and Hall/CRC, June 2008.

\bibitem{malhotra16attacking}
Aanchal Malhotra, Isaac~E. Cohen, Erik Brakke, and Sharon Goldberg.
\newblock
  \href{https://www.ndss-symposium.org/wp-content/uploads/2017/09/attacking-network-time-protocol.pdf}{Attacking
  the Network Time Protocol}.
\newblock In {\em \bibconf{NDSS}{Network and Distributed System Security
  Symposium}}, February 2016.

\bibitem{malhotra16broadcast}
Aanchal Malhotra and Sharon Goldberg.
\newblock \href{http://www.sigcomm.org/node/3899}{Attacking NTP’s
  Authenticated Broadcast Mode}.
\newblock {\em ACM SIGCOMM Computer Communication Review}, 46(2), April 2016.

\bibitem{malhotra17security}
Aanchal Malhotra, Matthew~Van Gundy, Mayank Varia, Haydn Kennedy, Jonathan
  Gardner, and Sharon Goldberg.
\newblock \href{https://fc17.ifca.ai/preproceedings/paper_29.pdf}{The Security
  of NTP’s Datagram Protocol}.
\newblock In {\em \bibconf{FC}{Financial Cryptography and Data Security}},
  April 2017.

\bibitem{maniatis02secure}
Petros Maniatis and Mary Baker.
\newblock Secure history preservation through timeline entanglement.
\newblock In {\em 11th USENIX Security Symposium}, August 2002.

\bibitem{mattern89virtual}
Friedemann Mattern.
\newblock
  \href{http://citeseerx.ist.psu.edu/viewdoc/download?doi=10.1.1.63.4399&rep=rep1&type=pdf}{Virtual
  Time and Global States of Distributed Systems}.
\newblock In {\em International Workshop on Parallel and Distributed
  Algorithms}, page 215–226, 1989.

\bibitem{merkle79secrecy}
Ralph~Charles Merkle.
\newblock {\em \href{http://www.merkle.com/papers/Thesis1979.pdf}{Secrecy,
  Authentication, and Public Key Systems}}.
\newblock PhD thesis, Stanford University, June 1979.

\bibitem{miller16honey}
Andrew Miller, Yu~Xia, Kyle Croman, Elaine Shi, and Dawn Song.
\newblock \href{https://eprint.iacr.org/2016/199.pdf}{The Honey Badger of BFT
  Protocols}.
\newblock In {\em \bibconf{CCS}{Computer and Communications Security}}, pages
  31--42, New York, NY, USA, October 2016. ACM.

\bibitem{rfc5905}
D.~Mills, J.~{Martin, Ed.}, J.~Burbank, and W.~Kasch.
\newblock Network time protocol version 4: Protocol and algorithms
  specification, June 2010.
\newblock RFC 5905.

\bibitem{mills91internet}
David~L. Mills.
\newblock \href{https://ieeexplore.ieee.org/document/103043}{Internet Time
  Synchronization: The Network Time Protocol}.
\newblock {\em IEEE Transactions on Communications}, 39(10):1482--1493, October
  1991.

\bibitem{moran58random}
{P. A. P.} Moran.
\newblock Random processes in genetics.
\newblock {\em \href{https://doi.org/10.1017/S0305004100033193}{Mathematical
  Proceedings of the Cambridge Philosophical Society}}, 54(1):60--71, January
  1958.

\bibitem{mostefaoui14signature}
Achour Most\'efaoui, Hamouma Moumen, and Michel Raynal.
\newblock \href{https://dl.acm.org/citation.cfm?id=2611468}{Signature-Free
  Asynchronous Byzantine Consensus with $t < n/3$ and $O(n^2)$ Messages}.
\newblock In {\em \bibconf{PODC}{Principles of Distributed Computing}}, July
  2014.

\bibitem{nakamoto08bitcoin}
Satoshi Nakamoto.
\newblock \href{https://bitcoin.org/bitcoin.pdf}{Bitcoin: A Peer-to-Peer
  Electronic Cash System}, 2008.

\bibitem{nikitin17chainiac}
Kirill Nikitin, Eleftherios Kokoris-Kogias, Philipp Jovanovic, Nicolas Gailly,
  Linus Gasser, Ismail Khoffi, Justin Cappos, and Bryan Ford.
\newblock
  \href{https://www.usenix.org/conference/usenixsecurity17/technical-sessions/presentation/nikitin}{{CHAINIAC}:
  Proactive Software-Update Transparency via Collectively Signed Skipchains and
  Verified Builds}.
\newblock In {\em 26th {USENIX} Security Symposium ({USENIX} Security 17)},
  pages 1271--1287. {USENIX} Association, 2017.

\bibitem{nowak06evolutionary}
Martin~A. Nowak.
\newblock {\em Evolutionary Dynamics: Exploring the Equations of Life}.
\newblock Belknap Press, September 2006.

\bibitem{ongaro14search}
Diego Ongaro and John Ousterhout.
\newblock In search of an understandable consensus algorithm.
\newblock In {\em \bibconf{USENIX ATC}{USENIX Annual Technical Conference}},
  June 2014.

\bibitem{pease80reaching}
Marshall Pease, Robert Shostak, and Leslie Lamport.
\newblock \href{http://dl.acm.org/citation.cfm?id=322188}{Reaching Agreement in
  the Presence of Faults}.
\newblock {\em Journal of the ACM (JACM)}, 27(2):228--234, April 1980.

\bibitem{plakal98lamport}
Manoj Plakal, Daniel~J. Sorin, Anne~E. Condon, and Mark~D. Hill.
\newblock \href{https://dl.acm.org/citation.cfm?doid=277651.277672}{Lamport
  Clocks: Verifying a Directory Cache-Coherence Protocol}.
\newblock In {\em \bibconf[10th]{SPAA}{Symposium on Parallel Algorithms and
  Architectures}}, June 1998.

\bibitem{popov18tangle}
Serguei Popov.
\newblock \href{https://www.iota.org/research/academic-papers}{The Tangle},
  April 2018.

\bibitem{propp15polyas}
James Propp.
\newblock
  \href{https://mathenchant.wordpress.com/2015/10/16/polyas-urn/}{P\'olya's
  Urn}, October 2015.

\bibitem{psiaki16gnss}
Mark~L. Psiaki and Todd~E. Humphreys.
\newblock \href{https://ieeexplore.ieee.org/document/7445815}{GNSS Spoofing and
  Detection}.
\newblock {\em Proceedings of the IEEE}, 104(6), April 2016.

\bibitem{rabin83randomized}
Michael~O. Rabin.
\newblock Randomized {Byzantine} generals.
\newblock In {\em Symposium on Foundations of Computer Science (SFCS)},
  November 1983.

\bibitem{rabin98simplified}
Tal Rabin.
\newblock \href{https://doi.org/10.1007/BFb0055722}{A Simplified Approach to
  Threshold and Proactive {RSA}}.
\newblock In {\em \bibconf{CRYPTO}{Advances in Cryptology}}, August 1998.

\bibitem{raynal92about}
Michel Raynal.
\newblock About logical clocks for distributed systems.
\newblock {\em ACM SIGOPS Operating Systems Review}, 26(1), January 1992.

\bibitem{reiter94secure}
Michael~K. Reiter.
\newblock \href{https://dl.acm.org/citation.cfm?id=191194}{Secure Agreement
  Protocols: Reliable and Atomic Group Multicast in Rampart}.
\newblock In {\em \bibconf[2nd]{CCS}{Computer and Communications Security}},
  November 1994.

\bibitem{renesse15paxos}
Robbert~Van Renesse and Deniz Altinbuken.
\newblock {Paxos} made moderately complex.
\newblock {\em ACM Computing Surveys (CSUR)}, 47(3), April 2015.

\bibitem{rfc8446}
E.~Rescorla.
\newblock The transport layer security {(TLS)} protocol version 1.3, August
  2018.
\newblock RFC 8446.

\bibitem{rivest96time}
Ronald~L. Rivest, Adi Shamir, and David~A. Wagner.
\newblock \href{https://dl.acm.org/citation.cfm?id=888615}{Time-lock puzzles
  and timed-release crypto}.
\newblock Technical report, Cambridge, MA, USA, March 1996.

\bibitem{robinson17using}
David~``Karit'' Robinson.
\newblock
  \href{https://zxsecurity.co.nz/presentations/201707_Defcon-ZXSecurity-GPSSpoofing.pdf}{Using
  GPS Spoofing to Control Time}.
\newblock DEFCON 25, July 2017.

\bibitem{robinson50bias}
W.~S. Robinson.
\newblock \href{https://www.jstor.org/stable/2086402}{Bias, Probability, and
  Trial by Jury}.
\newblock {\em American Sociological Review}, 15(1):73--78, February 1950.

\bibitem{roosta07time}
Tanya Roosta, Mike Manzo, and Shankar Sastry.
\newblock
  \href{https://link.springer.com/chapter/10.1007%2F978-0-387-46276-9_14}{Time
  Synchronization Attacks in Sensor Networks}.
\newblock In Radha Poovendran, Cliff Wang, and Sumit Roy, editors, {\em Secure
  Localization and Time Synchronization for Wireless Sensor and Ad Hoc
  Networks}, pages 325--345. Springer, 2007.

\bibitem{ruget94cheaper}
Fr\'{e}d\'{e}ric Ruget.
\newblock
  \href{https://pdfs.semanticscholar.org/e060/433cd977fc11c29919bb01bc7c8b9500a144.pdf}{Cheaper
  matrix clocks}.
\newblock In {\em International Workshop on Distributed Algorithms (WDAG)},
  pages 355--369, September 1994.

\bibitem{sarin87discarding}
Sunil~K. Sarin and Nancy~A. Lynch.
\newblock Discarding obsolete information in a replicated database system.
\newblock {\em IEEE Transactions on Software Engineering}, SE-13(1), January
  1987.

\bibitem{schneider90implementing}
Fred~B. Schneider.
\newblock Implementing fault-tolerant services using the state machine
  approach: A tutorial.
\newblock {\em {ACM} Computing Surveys}, 22(4):299--319, December 1990.

\bibitem{schneier99secure}
Bruce Schneier and John Kelsey.
\newblock Secure audit logs to support computer forensics.
\newblock {\em ACM Transactions on Information and System Security},
  2(2):159--176, May 1999.

\bibitem{schoenmakers99simple}
Berry Schoenmakers.
\newblock \href{https://link.springer.com/chapter/10.1007/3-540-48405-1_10}{A
  Simple Publicly Verifiable Secret Sharing Scheme and Its Application to
  Electronic Voting}.
\newblock In {\em \bibconf{CRYPTO}{IACR International Cryptology Conference}},
  pages 784--784, August 1999.

\bibitem{shamir79share}
Adi Shamir.
\newblock \href{https://cs.jhu.edu/~sdoshi/crypto/papers/shamirturing.pdf}{How
  to Share a Secret}.
\newblock {\em Communications of the ACM}, 22(11):612--613, 1979.

\bibitem{shamir84identity}
Adi Shamir.
\newblock
  \href{https://link.springer.com/chapter/10.1007/3-540-39568-7_5}{Identity-Based
  Cryptosystems and Signature Schemes}.
\newblock In {\em \bibconf{CRYPTO}{Advances in Cryptology}}, pages 47--53,
  August 1984.

\bibitem{shoup00practical}
Victor Shoup.
\newblock
  \href{https://link.springer.com/content/pdf/10.1007/3-540-45539-6_15.pdf}{Practical
  Threshold Signatures}.
\newblock In {\em Eurocrypt}, May 2000.

\bibitem{shoup98securing}
Victor Shoup and Rosario Gennaro.
\newblock \href{https://link.springer.com/chapter/10.1007/BFb0054113}{Securing
  threshold cryptosystems against chosen ciphertext attack}.
\newblock {\em Advances in Cryptology --- EUROCRYPT'98}, pages 1--16, 1998.

\bibitem{song18guess}
Jonghyuk Song.
\newblock
  \href{https://medium.com/coinmonks/attack-on-pseudo-random-number-generator-prng-used-in-1000-guess-an-ethereum-lottery-game-7b76655f953d}{Attack
  on Pseudo-random number generator (PRNG) used in 1000 Guess, an Ethereum
  lottery game (CVE-2018–12454)}, July 2018.

\bibitem{sonnino18coconut}
Alberto Sonnino, Mustafa Al-Bassam, Shehar Bano, and George Danezis.
\newblock \href{https://arxiv.org/abs/1802.07344}{Coconut: Threshold Issuance
  Selective Disclosure Credentials with Applications to Distributed Ledgers}.
\newblock {\em arXiv preprint arXiv:1802.07344}, 2018.

\bibitem{beacon14se}
How useful is {NIST}'s {R}andomness {B}eacon for cryptographic use?
\newblock
  \url{http://crypto.stackexchange.com/questions/15225/how-useful-is-nists-randomness-beacon-for-cryptographic-use}.

\bibitem{stadler96publicly}
Markus Stadler.
\newblock
  \href{https://link.springer.com/content/pdf/10.1007/3-540-68339-9_17.pdf}{Publicly
  Verifiable Secret Sharing}.
\newblock In {\em Eurocrypt}, May 1996.

\bibitem{sundararaman05clock}
Bharath Sundararaman, Ugo Buy, and Ajay~D. Kshemkalyani.
\newblock
  \href{https://www.sciencedirect.com/science/article/pii/S1570870505000144?via%3Dihub}{Clock
  synchronization for wireless sensor networks: a survey}.
\newblock {\em Ad Hoc Networks}, 3(3), May 2005.

\bibitem{syta17scalable}
Ewa Syta, Philipp Jovanovic, Eleftherios Kokoris-Kogias, Nicolas Gailly, Linus
  Gasser, Ismail Khoffi, Michael~J. Fischer, and Bryan Ford.
\newblock
  \href{https://www.ieee-security.org/TC/SP2017/papers/413.pdf}{Scalable
  Bias-Resistant Distributed Randomness}.
\newblock In {\em 38th IEEE Symposium on Security and Privacy}, May 2017.

\bibitem{syta16keeping}
Ewa Syta, Iulia Tamas, Dylan Visher, David~Isaac Wolinsky, Philipp Jovanovic,
  Linus Gasser, Nicolas Gailly, Ismail Khoffi, and Bryan Ford.
\newblock \href{http://dedis.cs.yale.edu/dissent/papers/witness-abs}{Keeping
  Authorities ``Honest or Bust'' with Decentralized Witness Cosigning}.
\newblock In {\em 37th IEEE Symposium on Security and Privacy}, May 2016.

\bibitem{szalachowski18towards}
Pawel Szalachowski.
\newblock \href{https://ieeexplore.ieee.org/document/8525399}{Towards More
  Reliable Bitcoin Timestamps}.
\newblock In {\em Crypto Valley Conference}, June 2018.

\bibitem{rfc793}
Transmission control protocol, September 1981.
\newblock RFC 793.

\bibitem{topalovic12towards}
Emin Topalovic, Brennan Saeta, Lin-Shung Huang, Collin Jackson, and Dan Boneh.
\newblock
  \href{http://www.ieee-security.org/TC/W2SP/2012/papers/w2sp12-final9.pdf}{Towards
  Short-Lived Certificates}.
\newblock In {\em \bibconf{W2SP}{Web 2.0 Security \& Privacy}}, May 2012.

\bibitem{vratonjic11inconvenient}
Nevena Vratonjic, Julien Freudiger, Vincent Bindschaedler, and Jean-Pierre
  Hubaux.
\newblock
  \href{https://link.springer.com/chapter/10.1007%2F978-1-4614-1981-5_5}{The
  Inconvenient Truth about Web Certificates}.
\newblock In {\em \bibconf[10th]{WEIS}{Workshop on Economics of Information
  Security}}, pages 79--117, June 2011.

\bibitem{waters05efficient}
Brent Waters.
\newblock
  \href{https://link.springer.com/chapter/10.1007%2F11426639_7}{Efficient
  Identity-Based Encryption Without Random Oracles}.
\newblock In {\em Eurocrypt}, May 2005.

\bibitem{weissman15how}
Cale~Guthrie Weissman.
\newblock
  \href{https://www.businessinsider.in/How-a-man-who-worked-for-the-Lottery-Association-may-have-hacked-the-system-for-a-winning-ticket/articleshow/46913403.cms}{How
  a man who worked for the Lottery Association may have hacked the system for a
  winning ticket}.
\newblock {\em Business Insider}, April 2015.

\bibitem{wesolowski19efficient}
Benjamin Wesolowski.
\newblock
  \href{https://link.springer.com/chapter/10.1007%2F978-3-030-17659-4_13}{Efficient
  Verifiable Delay Functions}.
\newblock In {\em Eurocrypt}, May 2019.

\bibitem{wolinsky12scalable}
David~Isaac Wolinsky, Henry Corrigan-Gibbs, Bryan Ford, and Aaron Johnson.
\newblock Scalable anonymous group communication in the anytrust model.
\newblock In {\em \bibconf{EuroSec}{European Workshop on System Security}},
  April 2012.

\bibitem{wood14ethereum}
Gavin Wood.
\newblock \href{https://github.com/ethereum/wiki/wiki/White-Paper}{Ethereum: A
  Secure Decentralised Generalised Transaction Ledger}.
\newblock {\em Ethereum Project Yellow Paper}, 2014.

\bibitem{wuu84efficient}
Gene~T.J. Wuu and Arthur~J. Bernstein.
\newblock Efficient solutions to the replicated log and dictionary problems.
\newblock In {\em Principles of Distributed Computing}, pages 232--242, August
  1984.

\bibitem{yin19hotstuff}
Maofan Yin, Dahlia Malkhi, Michael~K. Reiter, Guy~Golan Gueta, and Ittai
  Abraham.
\newblock
  \href{https://research.vmware.com/files/attachments/0/0/0/0/0/7/7/podc.pdf}{HotStuff:
  BFT Consensus with Linearity and Responsiveness}.
\newblock In {\em \bibconf{PODC}{Principles of Distributed Computing}}, July
  2019.

\bibitem{zhao04wireless}
Feng Zhao and Leonidas Guibas.
\newblock {\em Wireless Sensor Networks: An Information Processing Approach}.
\newblock Morgan Kaufmann, July 2004.

\bibitem{zimmermann80osi}
Hubert Zimmermann.
\newblock {OSI} reference model---the {ISO} model of architecture for open
  systems interconnection.
\newblock {\em IEEE Transactions on Communications}, 28(4):425--432, April
  1980.

\end{thebibliography}
